\documentclass[twocolumn,tighten]{aastex62}
\hypersetup{linkcolor=red,citecolor=blue,filecolor=cyan,urlcolor=magenta}
\usepackage{natbib}
\usepackage{color}
\usepackage{hyperref}
\usepackage{url}
\usepackage{comment}
\usepackage{graphicx}	
\usepackage{mathrsfs} 

\newcommand\ion[2]{\text{#1\,\textsc{\lowercase{#2}}}}	

\usepackage[T1]{fontenc}
\usepackage{ae,aecompl}
\usepackage{amsmath,amssymb}
\usepackage{textgreek}
\usepackage{booktabs}

\newcommand{\ur}[1]{\,\mathrm{#1}}
\usepackage{siunitx}
\usepackage{mhchem}

\makeatletter
\def\tabletypesize#1{\gdef\currtabletypesize{#1}
\def\@table@type@size{#1}}%
\tabletypesize{\small}
\global\def\tablenotemark#1{{\normalfont\textsuperscript{\normalsize\it #1}}}
\def\tablecomments#1{\vskip1pt{\small\vskip1sp\indent\vrule height 11pt depth 2pt
width 0pt\currtabletypesize{\sc Note}---{#1}\vskip1pt}}
\def\tablenotetext#1#2{\vskip1pt{\currtabletypesize\vskip1pt\indent\vrule
height 11pt depth
2pt width0pt\relax$^{\hbox to 5pt{$#1$}}$#2\vskip1pt}}
\def\tablerefs#1{{\small\vskip3pt\indent\vrule height 11pt depth 2pt
width 0pt\currtabletypesize{\bf References}---{#1}\vskip1sp}}
\makeatother

\usepackage{multirow}
\usepackage{xspace}
\newcommand{\targetstar}{WINERED-HVS1\xspace}
\newcommand{\targetstarGaiaSID}{Gaia DR3 6434696676903108864\xspace}

\newcommand{\msun}{\mbox{$M_{\odot}$}}

\newcommand{\mhxp}{\mbox{[M/H]$_\mathrm{XP}$}}
\newcommand{\amxp}{\mbox{[$\alpha$/M]$_\mathrm{XP}$}}

\def\vector#1{\mbox{\boldmath $#1$}}
\newcommand{\AU}{\ensuremath{\,\mathrm{AU}}}
\newcommand{\pc}{\ensuremath{\,\mathrm{pc}}}
\newcommand{\kpc}{\ensuremath{\,\mathrm{kpc}}}
\newcommand{\yr}{\ensuremath{\,\mathrm{yr}}}
\newcommand{\Myr}{\ensuremath{\,\mathrm{Myr}}}
\newcommand{\Gyr}{\ensuremath{\,\mathrm{Gyr}}}
\newcommand{\kms}{\ensuremath{\,\mathrm{km\ s}^{-1}}}

\newcommand{\masyr}{\ensuremath{\,\mathrm{mas\ yr}^{-1}}}

\newcommand{\vlos}{v_{\ensuremath{\mathrm{los}}}}

\newcommand{\mualpha}{\mu_{\alpha*}}
\newcommand{\mudelta}{\mu_\delta}
\newcommand{\pmra}{\mbox{$\mu_{\alpha *}$}}
\newcommand{\pmdec}{\mbox{$\mu_{\delta}$}}

\newcommand{\rperi}{r_\ensuremath{\mathrm{peri}}}

\newcommand{\StarHorse}{\texttt{{StarHorse}}}

\newcommand{\eq}[1]{\begin{align}#1\end{align}}

\newcommand{\revise}[1]{#1}

\begin{document}

\title{%
Discovery of WINERED-HVS1: A metal-rich hyper-velocity star candidate ejected from the Galactic center
}

\shorttitle{
Chemistry of a hyper-velocity star
}
\shortauthors{Hattori et al.}

\author[0000-0001-6924-8862]{Kohei~Hattori}
\affiliation{National Astronomical Observatory of Japan, 2-21-1 Osawa, Mitaka, Tokyo 181-8588, Japan}
\affiliation{The Institute of Statistical Mathematics, 10-3 Midoricho, Tachikawa, Tokyo 190-8562, Japan}
\affiliation{Department of Astronomy, University of Michigan,
1085 S.\ University Avenue, Ann Arbor, MI 48109, USA}
\email{Email:\ khattori@ism.ac.jp}

\author[0000-0002-2861-4069]{Daisuke~Taniguchi}
\affiliation{National Astronomical Observatory of Japan, 2-21-1 Osawa, Mitaka, Tokyo 181-8588, Japan}

\author[0000-0002-9397-3658]{Takuji~Tsujimoto}
\affiliation{National Astronomical Observatory of Japan, 2-21-1 Osawa, Mitaka, Tokyo 181-8588, Japan}

\author{Noriyuki~Matsunaga}
\affiliation{Department of Astronomy, Graduate School of Science, The University of Tokyo, 7-3-1 Hongo, Bunkyo-ku, Tokyo 113-0033, Japan}
\affiliation{Laboratory of Infrared High-resolution spectroscopy (LiH), Koyama Astronomical Observatory, Kyoto Sangyo University, Motoyama, Kamigamo, Kita-ku, Kyoto 603-8555, Japan}

\author[0000-0001-6401-723X]{Hiroaki~Sameshima}
\affiliation{Institute of Astronomy, Graduate School of Science, The University of Tokyo, 2-21-1 Osawa, Mitaka, Tokyo 181-0015, Japan}

\author[0000-0001-5642-2569]{Scarlet~S.~Elgueta}
\affiliation{Departamento de Física, Universidad de Santiago de Chile, Av. Victor Jara 3659, Santiago, Chile}
\affiliation{Millenium Nucleus ERIS, Instituto de Estudios Astrof\'isicos, Universidad Diego Portales, Av. Ej\'ercito Libertador 441, Santiago, Chile}

\author{Shogo~Otsubo}
\affiliation{Laboratory of Infrared High-resolution spectroscopy (LiH), Koyama Astronomical Observatory, Kyoto Sangyo University, Motoyama, Kamigamo, Kita-ku, Kyoto 603-8555, Japan}

\correspondingauthor{Kohei Hattori}

\begin{abstract}

We report the discovery of a metal-rich red giant star, WINERED-HVS1, which is a candidate for a hyper-velocity star (HVS). Its past trajectory suggests that this star may have been ejected by the Galactic supermassive black hole (SMBH; Sgr A*),  
with a modest ejection velocity of at least $500\; \mathrm{km\;s^{-1}}$.
Since WINERED-HVS1 is gravitationally bound to the Milky Way and is not necessarily young, 
it is not an unambiguous HVS candidate from a kinematic perspective, unlike the previously confirmed A-type main-sequence HVS known as S5-HVS1. Therefore, the chemical characterization of this star is essential for understanding its origin. 
Through a spectroscopic follow-up observation with Magellan/WINERED, we determined its metallicity $\mathrm{[Fe/H]}=-0.147^{+0.046}_{-0.045}$, as well as the abundances for Na, Mg, Si, S, K, Ca, Ti, Cr, Ni, and Sr. The high metallicity value suggests that this star was ejected from the Galactic center and is unlikely to be a halo star with a radial orbit. 
The abundance pattern of this star---including the pattern of 
[$\alpha$/Fe], [Mg/Fe], [Si/Mg], and [Ca/Mg]---is 
consistent with that of the nuclear star cluster 
surrounding the SMBH, 
further supporting our view. 
This discovery opens a new window to look through the Galactic center environment without the hindrance of dust extinction, by using the HVSs  moving in the halo region where dust extinction is minimal.


\end{abstract}
 
\section{Introduction}
\label{section:introduction}

\subsection{Environment near the supermassive black hole in the Milky Way}

The center of the Milky Way is home to a supermassive black hole (SMBH) 
known as Sgr A*, which has a mass of $\sim 4 \times 10^6 \msun$
\citep{Ghez2003ApJ...586L.127G,Gillessen2017ApJ...837...30G}. 
Although the intense tidal forces from this SMBH hinder star formation 
in its immediate vicinity \citep{Levin2007MNRAS.374..515L}, 
observations have revealed that it is surrounded by 
a group of stars referred to as the S-stars. 
These stars orbit within a radius of $0.04 \pc$ 
and include young massive stars. 
If we look further out, 
both the SMBH and the S-stars are 
contained within a nuclear star cluster (NSC) 
that has a half light radius of $\sim 4 \pc$ and a total mass of $2$--$4 \times 10^7 \msun$
\citep{Neumayer2020A&ARv..28....4N}. 
The origins of these stellar components in the Galactic center 
remain a subject of ongoing debate 
(e.g., \citealt{Guillard2016MNRAS.461.3620G, Akiba2024arXiv241019881A}). 
Studying this region is particularly challenging 
due to high levels of interstellar dust toward the Galactic center\footnote{
When the manuscript of this paper was nearly ready for submission, we became aware of a study by \cite{Nandakumar2025ApJ...982L..14N} presenting the first results on the detailed chemical abundances of NSC stars. 
To maintain the independence of our analysis, we have not incorporated their findings in this paper. However, we fully acknowledge their pioneering contribution to this field.
In our forthcoming paper (Taniguchi et al. in prep), we will discuss the implications from \cite{Nandakumar2025ApJ...982L..14N} in some detail. 
}  
\citep{
Carr2000ApJ...530..307C, 
Ramirez2000ApJ...537..205R,
Cunha2007ApJ...669.1011C,
Davies2009ApJ...694...46D, 
Pfuhl2011ApJ...741..108P,
Rich2017AJ....154..239R, 
Ryde2025ApJ...979..174R, 
Nandakumar2025ApJ...982L..14N}. 
Motivated by these difficulties, 
here we explore an alternative approach to infer the Galactic center environment, 
by using stars ejected from the SMBH, known as hyper-velocity stars (HVSs).

\subsection{Hyper-velocity stars as a new window to see the Galactic center environment}
\label{section:introduction_HVS}

\cite{Hills1988Natur.331..687H} 
predicted that when the SMBH disrupts a binary star system, 
it gravitationally captures one star (forming an S-star\footnote{
The formation process of S-stars is not fully understood, 
and there may be multiple channels to form these stars. 
Binary disruption is one of the possible mechanisms to form S-stars. 
}), 
while ejecting the other star with an ejection velocity of $\sim 10^3 \kms$, which is called an HVS.\footnote{
In this paper, we define HVSs as those stars ejected from the Galactic center, 
regardless of their velocities. 
} 
If such an HVS exists, 
it would convey 
information on the 
chemical abundances of stars 
\revise{near} the SMBH 
\citep{Koposov2020MNRAS.491.2465K}. 
HVSs are 
currently moving in the halo region, 
where the dust extinction is negligible. 
Therefore, finding HVSs in the halo 
and characterizing their chemical abundances 
may provide an access to the 
detailed chemical properties of the Galactic center.

\cite{Brown2005ApJ...622L..33B} 
discovered an unbound star named HVS1, 
which is a B-type star traveling at the heliocentirc distance of $\sim 39$--$71 \kpc$ in the halo 
with a velocity of $709 \kms$. 
Its metallicity is [Fe/H]$\simeq 0$, 
consistent with the Galactic center environment. 
Given the large velocity and the solar metallicity of this star, 
it is probably an HVS ejected by the SMBH, 
although the heliocentric distance is too uncertain that it is difficult to unambiguously trace back the orbit to the Galactic center. 
Later, 
\cite{Koposov2020MNRAS.491.2465K} 
discovered an unambiguous HVS dubbed S5-HVS1, 
which is an A-type main-sequence star 
moving at $9 \kpc$ from the Sun 
with a velocity of $1755 \kms$. 
Due to its 
proximity from the Sun, 
its orbit can be traced back to the Galactic center 
with a small uncertainty. 
S5-HVS1 is currently the only unambiguous HVS from the Galactic center. 
Although the metallicity is estimated to be 
[Fe/H]$\simeq 0.29$ \citep{Koposov2020MNRAS.491.2465K}, 
its detailed chemical abundances 
are yet to be determined 
due to its high temperature.

The ejection rate of HVSs is estimated to be 
one per $\sim 10^4$ years (\citealt{Zhang2013ApJ...768..153Z, Brown2015ARAA, Marchetti2018MNRAS.476.4697M}; 
see also \citealt{Verberne2024MNRAS.533.2747V}). 
If HVSs travel with a velocity of $\sim 10^3 \kms$, 
a simple calculation suggests that there are $\sim 10^4$ HVSs within 100 kpc from the Galactic center 
(see Section 4.5 of \citealt{Brown2015ARAA}; see also \citealt{Marchetti2018MNRAS.476.4697M}). 
Considering that the stellar halo contains $\sim 10^{9}$ stars 
\citep{Deason2019MNRAS.490.3426D}, 
only one in $10^{5}$ of stars in the halo region is an HVS.
The rarity of HVSs means that we need to search for HVSs from a large catalog. 
Following the major data releases from Gaia, namely Gaia DR2, EDR3, and DR3 
\citep{Gaia2016A&A...595A...1G, Gaia2018A&A...616A...1G, Gaia2021A&A...649A...1G, GaiaVallenari2023A&A...674A...1G},  
many authors have attempted to find HVSs from 
Gaia's large 6D position-velocity catalog 
\citep{Hattori2018ApJ...866..121H, Marchetti2019MNRAS.490..157M, Marchetti2021MNRAS.503.1374M, Marchetti2022MNRAS.515..767M, Li2021ApJS..252....3L}. 
\revise{None of their fast-moving stars} are unambiguous HVSs. 
This is 
partly because the stellar halo is dominated by 
halo stars with nearly radial orbits 
\citep{Belokurov2018MNRAS.478..611B, Helmi2018Natur.563...85H}.

Although previous kinematical searches for HVSs in the Gaia catalog 
did not result in discoveries of unambiguous HVSs, 
we still have a hope to find HVSs. 
Since typical stars near the Galactic center are  
much more metal-rich than field halo stars, 
HVSs should stand out 
when viewed in the chemistry space. 
Recently, 
extremely low-resolution stellar spectra named Gaia XP spectra 
were published for 219 million stars as part of the Gaia DR3 \citep{DeAngeli2023A&A...674A...2D, Montegriffo2023A&A...674A...3M}.  
By using the Gaia XP spectra, 
\cite{Hattori2025ApJ...980...90H} 
estimated the metallicity $\mhxp$ and $\alpha$-abundance $\amxp$ 
for 48 million stars in the low dust extinction regions. 
In this paper, 
we combine the chemical data from \cite{Hattori2025ApJ...980...90H} 
and the position/velocity data from Gaia DR3 
to hunt for HVSs ejected from the Galactic center. 
By using Magellan/WINERED, we conduct a follow-up observation for the most metal-rich candidate, \targetstar, and we conclude that it is the most promising HVS candidate 
based on its \revise{chemistry}.

\subsection{Structure of this paper}

This paper is organized as follows. 
\autoref{section:strategy} \revise{describes our strategy.} 
\autoref{section:data} \revise{describes the data}. 
\autoref{section:Solar_pos_vel} \revise{describes} our assumptions on the Solar position and velocity. 
In \autoref{section:search_HVS}, 
we first 
\revise{kinematically select 74 HVS candidates.}
Then we select 22 metal-rich HVS candidates 
by additionally using their \revise{chemistry}. 
A follow-up observation for the most metal-rich HVS candidate 
(\targetstar) with Magellan/WINERED  
\revise{is} described in \autoref{section:WINERED}. 
In \autoref{section:discussion}, 
\revise{we demonstrate} that \targetstar is the most promising HVS \revise{candidate}. 
In \autoref{section:conclusion}, we summarize our paper.

\section{Strategy to find HVS candidates} 
\label{section:strategy}

The aim of this paper is to find \revise{candidates of bound or unbound} HVSs that were 
\revise{ejected from the Galactic center}. 
As HVSs are no longer located in the Galactic center region, 
finding candidates of HVSs in the Gaia catalog 
(or any given \revise{stellar catalog}) 
requires us to have a reliable model of the stellar orbits in the past. 
To build such a model, we need to know 
(i) the Galactic potential; 
(ii) the Solar position/velocity with respect to the Galactic center; and 
(iii) the stellar position/velocity relative to the Sun.  
If we had the perfect knowledge on (i)--(iii), 
finding HVSs would be straightforward because we would just need to trace the orbit backward in time 
and to find stars that `penetrate' through the Galactic center. 
In reality, however, we have a limited understanding of (i)--(iii) mentioned above. 
To handle these uncertainties, 
\revise{we strategically design our HVS search as described below.}

Regarding the uncertainty associated with (i), 
we use two widely-used Galactic potential models
(see \autoref{table:4methods}). 
The fiducial model is an axisymmetric model explored by \cite{McMillan2017}, 
and the secondary model is 
\revise{a potential with a rotating bar} 
\citep{Portail2017MNRAS.465.1621P}. 
By using these two potentials, 
we can search for HVS candidates 
that are robust against the specific choice of the adopted potential. 
Also, 
\revise{we integrate the stellar orbits backward in time for only 200 Myr, 
and search for candidates of 
HVSs that have passed near the Galactic center 
within the last 200 Myr. 
Due to this short integration time, 
we essentially search for  
unbound HVSs; 
bound, recently-ejected HVSs that have not experienced disk crossing after ejection; 
or
bound HVSs whose last disk crossings after the ejection were close to the Galactic center. 
(We may miss bound HVSs 
whose last disk crossings 
were not close to the Galactic center.) 
}
Because our inference on the stellar orbits 
becomes less reliable 
if we adopt longer integration time, 
our choice of short integration time 
minimizes the mis-classification of non-HVSs as HVS candidates.\footnote{
We choose a short integration time of $200 \Myr$,  
which makes our work distinct from 
\cite{Marchetti2022MNRAS.515..767M}, 
who integrated the orbits for as long as $1 \Gyr$ to search for HVS candidates. 
}

Regarding the uncertainty associated with (ii), 
we try two assumptions on the Solar position and velocity, $(\vector{x}_\odot, \vector{v}_\odot)$ 
\revise{(see \autoref{table:4methods}).} 
For Methods A and C, 
we fix 
\revise{$(\vector{x}_\odot, \vector{v}_\odot)$};
while for Methods B and D, 
we try different set of 
\revise{$(\vector{x}_\odot, \vector{v}_\odot)$} 
each time we compute the stellar orbits
(see \autoref{section:Solar_pos_vel}). 
This is the first time the uncertainties in 
\revise{$(\vector{x}_\odot, \vector{v}_\odot)$} 
are taken into account in finding HVSs.\footnote{
There has been an attempt to determine 
\revise{$(\vector{x}_\odot, \vector{v}_\odot)$} 
by using an HVS \citep{Koposov2020MNRAS.491.2465K}
with a method explored by \cite{Hattori2018b}.
} 
Our investigation allows us 
to check how our search of HVSs are sensitive to the 
assumptions on 
$\vector{x}_\odot$ and $\vector{v}_\odot$.

Regarding the uncertainty associated with (iii), 
for each star, 
we draw $N_\mathrm{MC}=10^4$ Monte Carlo samples of  
the stellar position and velocity 
that reflect the observational uncertainty 
(see \autoref{section:UncertaintySet}). 
For each star, 
\revise{we construct $N_\mathrm{MC}$ orbits 
in the last $200 \Myr$, 
which are used to select 
candidates of HVSs.}

\begin{deluxetable}{l rrr }
\tablecaption{Four methods to kinematically search for HVS candidates
\label{table:4methods}}
\tablewidth{0pt}
\tabletypesize{\scriptsize}
\tablehead{
\colhead{(1)} &
\colhead{(2)} &
\colhead{(3)} \\
\colhead{Search method} &
\colhead{Potential model} &
\colhead{Assumed $(R_0, V_\odot)$} 
}
\startdata
{Method A (fiducial)} & {\cite{McMillan2017}} & {Fix $(R_0, V_\odot)$} \\
{Method B} & {\cite{McMillan2017}} & {Vary $(R_0, V_\odot)$} \\
{Method C} & {\cite{Portail2017MNRAS.465.1621P}} & {Fix $(R_0, V_\odot)$} \\
{Method D} & {\cite{Portail2017MNRAS.465.1621P}} & {Vary $(R_0, V_\odot)$} \\
\hline 
\enddata
\tablecomments{
(1) Names of the search methods. 
Method A is the fiducial \revise{method}. 
(2) Assumed Galactic potential. 
The model in \cite{McMillan2017} is axisymmetric.  
The model in \cite{Portail2017MNRAS.465.1621P} is 
\revise{a potential with a rotating bar 
(with a constant pattern speed $\Omega_\mathrm{bar}=39 \kpc\kms$). 
(3) The assumptions on the Solar position and velocity. 
(See Section \ref{section:Solar_pos_vel} for details.)
}}
\end{deluxetable}

\section{Stellar data}
\label{section:data}

\revise{Here we describe the data set used in this paper. 
Our notations are summarized in \autoref{table:notation}.}

\subsection{Selection of data}
\label{section:data_selection}

We construct the catalog of metal-rich stars 
by using the position-velocity data in Gaia DR3 catalog 
\citep{GaiaVallenari2023A&A...674A...1G}, 
the photometric distance from \StarHorse\ catalog \citep{Anders2022AandA...658A..91A}, 
and the chemical abundance catalog from \cite{Hattori2025ApJ...980...90H}.

First, from Gaia DR3 catalog, 
we select stars with 
\texttt{parallax\_over\_error}$>5$, 
with valid \texttt{radial\_velocity} ($\vlos$), 
and with $v_\mathrm{tot} > 200 \kms$. 
Here, 
$v_\mathrm{tot}$ 
is the point-estimate of the 
total velocity in the Galactocentric rest frame, 
\revise{derived with assumptions in \autoref{section:fixedR0V0}. 
The selection by $v_\mathrm{tot}$ is not stringent, 
and it is set to reduce the computational cost 
in finding HVS candidates. 
}
The lower limit on $v_\mathrm{tot}$ 
is set for 
\revise{
excluding some disk stars 
}
and 
excluding halo stars with low velocity. 
In computing 
$v_\mathrm{tot}$, 
\revise{we use the point-estimate of the 
heliocentric stellar distance ($d=1/\varpi$).}

Secondly, 
we cross-match the catalog with 
the \StarHorse\ catalog 
to obtain photometric distance information. 
As described in \autoref{section:UncertaintySet}, 
we will use the 5th and 95th percentile values 
from the probability distribution of the photometric distance 
for each star, 
which are denoted as \texttt{dist05} and \texttt{dist95}.

Finally, 
we cross-match the sample with 
the chemical abundance catalog 
(\url{https://zenodo.org/records/10902172}) 
in  
\cite{Hattori2025ApJ...980...90H}. 
To reduce the contamination from halo stars, 
we select stars with $\mathrm{[M/H]_{XP}}>-1$ 
and \texttt{bool\_flag\_cmd\_good=True}. 
The remaining $6.6\times10^6$ stars will be analyzed.

\subsection{Monte Carlo realizations of the observed quantities of the star}
\label{section:UncertaintySet}

Here we describe how we generate Monte Carlo realizations of 
the stellar position/velocity 
that reflect the uncertainties in the observed quantities. 
In the following, we focus on the procedure 
for $i$th star in our catalog.

For $i$th star,
we draw $N_\mathrm{MC}=10^4$ realizations of the observable vector 
$(\alpha, \delta, d, \mualpha, \mudelta, \vlos)$ 
by taking into account the associated error distribution, 
in a similar manner as in 
\revise{\cite{Okuno2022arXiv220408205O} and} \cite{Hattori2023ApJ...946...48H}. 
We denote the $j$th realization of the observable vector (for star $i$) as 
\eq{\label{eq:qij}
\vector{q}_{ij} = (\alpha^{(ij)}, \delta^{(ij)}, d^{(ij)}, \mualpha^{(ij)}, \mudelta^{(ij)}, \vlos^{(ij)}).
}
We neglect the tiny observational uncertainty in the sky coordinate, 
and thus we set $(\alpha^{(ij)}, \delta^{(ij)})$ to the observed 
\revise{values in Gaia DR3 catalog.}
We draw the heliocentric distance $d^{(ij)}$ 
from the uniform distribution between 
$\texttt{dist05} < d < \texttt{dist95}$, 
where \texttt{dist05} and \texttt{dist95} are taken from the \StarHorse\ catalog.\footnote{
\revise{
We have confirmed that the most important HVS candidate in this paper (\targetstar) was also identified with our fiducial method 
(Method A), 
even if we set the upper and lower limits on the distance 
based on the zero-point-corrected Gaia parallax and its uncertainty. 
}
} 
We draw the proper motion vector 
$(\mualpha^{(ij)}, \mudelta^{(ij)})$ from the 
error distribution in Gaia DR3 catalog, 
by taking into account the correlation between the two proper motion components. 
We draw the line-of-sight velocity $\vlos^{(ij)}$ 
from the error distribution in Gaia DR3 catalog. 
After these procedures, 
for each star $i$, 
we obtain $N_\mathrm{MC}=10^4$ sets of 
phase-space vector $\vector{q}_{ij}$. 
The same sets of $\vector{q}_{ij}$ will be analyzed with 
\revise{the four Methods A--D in \autoref{table:4methods}.}

\section{The position and velocity of the Sun} 
\label{section:Solar_pos_vel}

We use a right-handed Cartesian coordinate 
\revise{where} 
the Galactic plane is the $(x,y)$-plane 
and $z$ is directed to the North Galactic Pole. 
The Sun is located at $\vector{x}_\odot$ $=(x,y,z)_\odot = (-R_0, 0, z_0)$. 
Its velocity is $\vector{v}_\odot = (v_x,v_y,v_z)_\odot=(U_\odot, V_\odot, W_\odot)$. 
We fix $(\vector{x}_\odot, \vector{v}_\odot)$ 
in Methods A and C; 
we vary \revise{them} in Methods B and D.

\subsection{Fixed Solar position and velocity: 
Methods A/C} 
\label{section:fixedR0V0}

For Methods A and C (see \autoref{table:4methods}), 
we fix the Solar position and velocity. 
We assume 
$z_0 = 0.0208 \kpc$ \citep{Bennett2019MNRAS.482.1417B} and 
$(U_\odot, W_\odot)=(11.1, 7.25) \kms$ \citep{Schonrich2010}. 
Also, we fix $(R_0, V_\odot) = (8.277 \kpc, 251.5 \kms)$  
\citep{GRAVITY2022A&A...657L..12G, Reid2020ApJ...892...39R}.

\subsection{Varied Solar position and velocity: 
Methods B/D} 

For Methods B and D (see \autoref{table:4methods}), 
we take into account the uncertainties 
in \revise{$(\vector{x}_\odot, \vector{v}_\odot)$}.
We prepare $N_\mathrm{MC}=10^4$ combinations of 
$(\vector{x}_\odot^{(j)}, \vector{v}_\odot^{(j)})$, 
where $j=1, \cdots, N_\mathrm{MC}$,
in the following manner. 
We assume the same fixed values for $(z_0, U_\odot, W_\odot)$ 
as before. 
For $(R_0, V_\odot)$, 
we use $N_\mathrm{MC}=10^4$ pairs of values, 
$\{ (R_0^{(j)}, V_\odot^{(j)}) \mid j = 1, \cdots, N_\mathrm{MC}\}$,  
which are drawn from:
\eq{
R_0^{(j)} &\sim \mathscr{N}(8.277, 0.1) \kpc, \\
\Omega_\mathrm{SgrA}^{(j)} &\sim \mathscr{N}(-6.411, 0.008) \masyr, \\
V_\odot^{(j)} &= k R_0^{(j)} (-\Omega_\mathrm{SgrA}^{(j)}) , \label{eq:Vodotj}
}
with $k=4.74047 \kms \kpc^{-1} (\masyr)^{-1}$. 
Here, $a \sim \mathscr{N}(m, s)$ denotes 
that $a$ is drawn from a Gaussian distribution with 
mean $m$ and standard deviation $s$. 
\revise{We assume the uncertainty in $R_0$ to be $0.1 \kpc$ 
\citep{BlandHawthorn2016ARA&A..54..529B}.
}
We assume that the 
angular motion of Sgr A* 
in the Galactic longitude direction, 
$\Omega_\mathrm{SgrA} = -6.411 \pm 0.008 \masyr$ \citep{Reid2020ApJ...892...39R}, 
corresponds to the Solar reflex motion.

\section{Search for HVS candidates }
\label{section:search_HVS}

\subsection{Backward orbit integration}
\label{section:backward_orbit}

Here we describe how we judge whether 
$i$th star in our catalog has a non-trivial probability of having been ejected by the SMBH.

From \autoref{section:UncertaintySet}, 
we have the position-velocity data of $i$th star relative to the Sun, 
represented by $N_\mathrm{MC}$ Monte Carlo samples. 
For $j$th Monte Carlo sample, 
we have the 6D observable vector $\vector{q}_{ij}$ defined in equation (\ref{eq:qij}). 
We convert $\vector{q}_{ij}$ to 
the Galactocentric Cartesian position and velocity vector 
$(\vector{x}_{ij}, \vector{v}_{ij})$, 
by using the assumed Solar position and velocity 
$(\vector{x}_\odot, \vector{v}_\odot)$. 
Here, we use the fixed value of $(\vector{x}_\odot, \vector{v}_\odot)$ 
in Methods A and C, 
while we use $(\vector{x}_\odot, \vector{v}_\odot)=(\vector{x}_\odot^{(j)}, \vector{v}_\odot^{(j)})$ 
in Methods B and D.  
Then, from the current-day position-velocity 
$(\vector{x}_{ij}, \vector{v}_{ij})$, 
we integrate the orbit 
backward in time for $200 \Myr$ 
in the model potential of the Milky Way, 
\revise{as} shown in \autoref{table:4methods}. 


If $j$th orbit of $i$th star crosses the disk plane at $-200 \Myr < t < 0 \Myr$, 
\revise{
we record 
the last disk-crossing time 
$t=-\tau$ (0 Myr $<\tau < 200$ Myr). 
We record the position 
$\vector{x}_\mathrm{ej} = (x_\mathrm{ej}, y_\mathrm{ej}, 0)$ 
at $t=-\tau$  
and its Galactocentric radius $r_\mathrm{ej} = (x_\mathrm{ej}^2 + y_\mathrm{ej}^2)^{1/2}$. 
Also, we record the velocity 
$\vector{v}_\mathrm{ej} = (v_{x,\mathrm{ej}}, v_{y,\mathrm{ej}}, v_{z,\mathrm{ej}})$ 
at $t=-\tau$ and its magnitude 
$v_\mathrm{ej} = (v_{x,\mathrm{ej}}^2 + v_{y,\mathrm{ej}}^2 + v_{z,\mathrm{ej}}^2)^{1/2}$.}

\revise{
For brevity, 
we call the disk-crossing position $\vector{x}_\mathrm{ej}$ 
and velocity $\vector{v}_\mathrm{ej}$ 
as the 
{\it ejection} position and 
{\it ejection} velocity, respectively, 
although their interpretation 
depends on the actual nature of $i$th star: 
}
\revise{
\begin{itemize}
\item 
If the star is an unbound HVS 
or a bound, recently-ejected HVS (ejected in the last 200 Myr), 
($-\tau, \vector{x}_\mathrm{ej}, \vector{v}_\mathrm{ej}$) 
in our analysis correspond to  
the ejection time, ejection position, and ejection velocity of the HVS, respectively. 
(This is why we put the subscript `ej' in these quantities.)
The HVS's flight time since the ejection is $\tau$. 
\item 
If the star is a bound HVS that has experienced one or more disk crossings after the ejection 
(e.g., bound HVSs ejected more than 200 Myr ago), 
($-\tau, \vector{x}_\mathrm{ej}, \vector{v}_\mathrm{ej}$) 
indicates the orbital properties of the HVS 
at the {\it last} disk crossing. 
If we assume that the Galactic potential has been approximately static, 
then the disk-crossing velocity should have been nearly the same 
whenever the star crossed the disk plane near the Galactic center.  
Therefore, the value of $v_\mathrm{ej}$ 
can be interpreted as the ejection velocity 
when it was originally ejected from the SMBH. 
Since we only consider the stellar orbit 
in the last 200 Myr, 
$\tau$ is not the flight time since the ejection, 
but is a lower limit on the flight time. 
\item 
If the star is not an HVS but an ordinary Galactic star, 
($-\tau, \vector{x}_\mathrm{ej}, \vector{v}_\mathrm{ej}$) 
are their properties at the last disk crossing. 
\end{itemize}
}

\revise{
Although we trace the stellar orbit only in the last 200 Myr, 
the quantities such as $(\tau, r_\mathrm{ej}, v_\mathrm{ej})$ are still useful to find candidates of HVSs, as we will describe in \autoref{section:kinematic_conditions}. 
(We admit, however, that 
the subscript `ej' 
in $r_\mathrm{ej}$ or $\vector{x}_\mathrm{ej}$ 
may be misleading for bound HVSs ejected long time ago or non-HVSs.) 
}

\subsection{Kinematically-selected HVS candidates} 
\label{section:kinematic_conditions}

For each Method (A--D), 
we conduct the backward orbit integration and 
derive \revise{$\vector{x}_\mathrm{ej}$ and $\vector{v}_\mathrm{ej}$.} 
Then we consider a star as a `kinematically-selected HVS candidate' if at least one of the following conditions are satisfied:
\begin{itemize}
\item (Condition 1) 
For star $i$, count the number of Monte Carlo orbits that satisfy  
$r_\mathrm{ej} < 50 \pc$ 
and $v_\mathrm{ej} > 500 \kms$. 
If more than 1\% of Monte Carlo orbits satisfy these conditions, we regard star $i$ as an HVS candidate. 
\item (Condition 2) 
For star $i$, count the number of Monte Carlo orbits that satisfy 
$r_\mathrm{ej} < 200 \pc$
and 
$v_\mathrm{ej} > 500 \kms$. 
If more than 25\% of Monte Carlo orbits satisfy these conditions, we regard star $i$ as an HVS candidate. 
\end{itemize}
These conditions are not mutually exclusive, so a star may satisfy both (Condition 1) and (Condition 2). 
(The number of kinematically-selected HVS candidates will be described in 
\autoref{section:number_of_kinematically_selected_HVS}.)

\subsubsection{Justification for (Condition 1) and (Condition 2)}
\label{section:justification_for_condition_1_2}

\revise{The purpose of (Condition 1) and (Condition 2) is to ensure that genuine, recently-ejected HVSs are not missed, including both unbound HVSs and bound ones that have not yet crossed the Galactic disk plane since their ejection.}\footnote{
\label{footnote:caveat_on_condition_1_2}
\revise{We designed (Condition 1) and (Condition 2) to 
search for HVSs that have been recently ejected from the Galactic center.}\revise{
However, with these conditions, we may also detect 
bound HVSs whose last disk crossings were close to the Galactic center region.}
} 
For example, 
suppose that we are confident on the assumed gravitational potential model and the Solar position and velocity. 
In this case, the uncertainty in the stellar orbit 
arises mainly from the uncertainty in the current-day position and velocity of the star. 
Since we have prepared a large number of Monte Carlo orbits per star, 
we believe that 
at least a small fraction of of the Monte Carlo orbits 
are close to the genuine orbit and have small $r_\mathrm{ej}$. 
Therefore, (Condition 1) works well to select HVS candidates.

In contrast, 
suppose that we are not very confident on the assumed gravitational potential model or the Solar position and velocity. 
In this case, a genuine, recently-ejected HVS may not have very small $r_\mathrm{ej}$ if the assumed potential is not a good approximation 
to the actual Milky Way potential. 
In order to investigate how the wrongly-assumed potential models affect the value of $r_\mathrm{ej}$, we conduct a simple test. 
Here, we eject HVSs from the Galactic center (with ejection velocity of $\sim 600 \kms$ and flight time of $\sim 30 \Myr$) assuming a realistic potential model (the model in \citealt{McMillan2017}) and rewinding the orbit backward in time in another realistic potential 
(the model in \citealt{Portail2017MNRAS.465.1621P}) 
until the disk crossing. 
\revise{
In this test, we do not assign observational errors, 
because we are interested in the systematic uncertainty in the ejection position arising purely from the choice of the gravitational potential. 
}
As a result, we find that $r_\mathrm{ej} < 200 \pc$ is satisfied for 
the majority ($\sim 80\%$) of simulated HVSs. 
Our simplified test suggests 
that a genuine, recently-ejected HVS probably satisfy 
$r_\mathrm{ej} < 200 \pc$. 
Therefore, (Condition 2) works well to select HVS candidates.

\begin{figure*}
\centering
\includegraphics[width=0.7\textwidth]{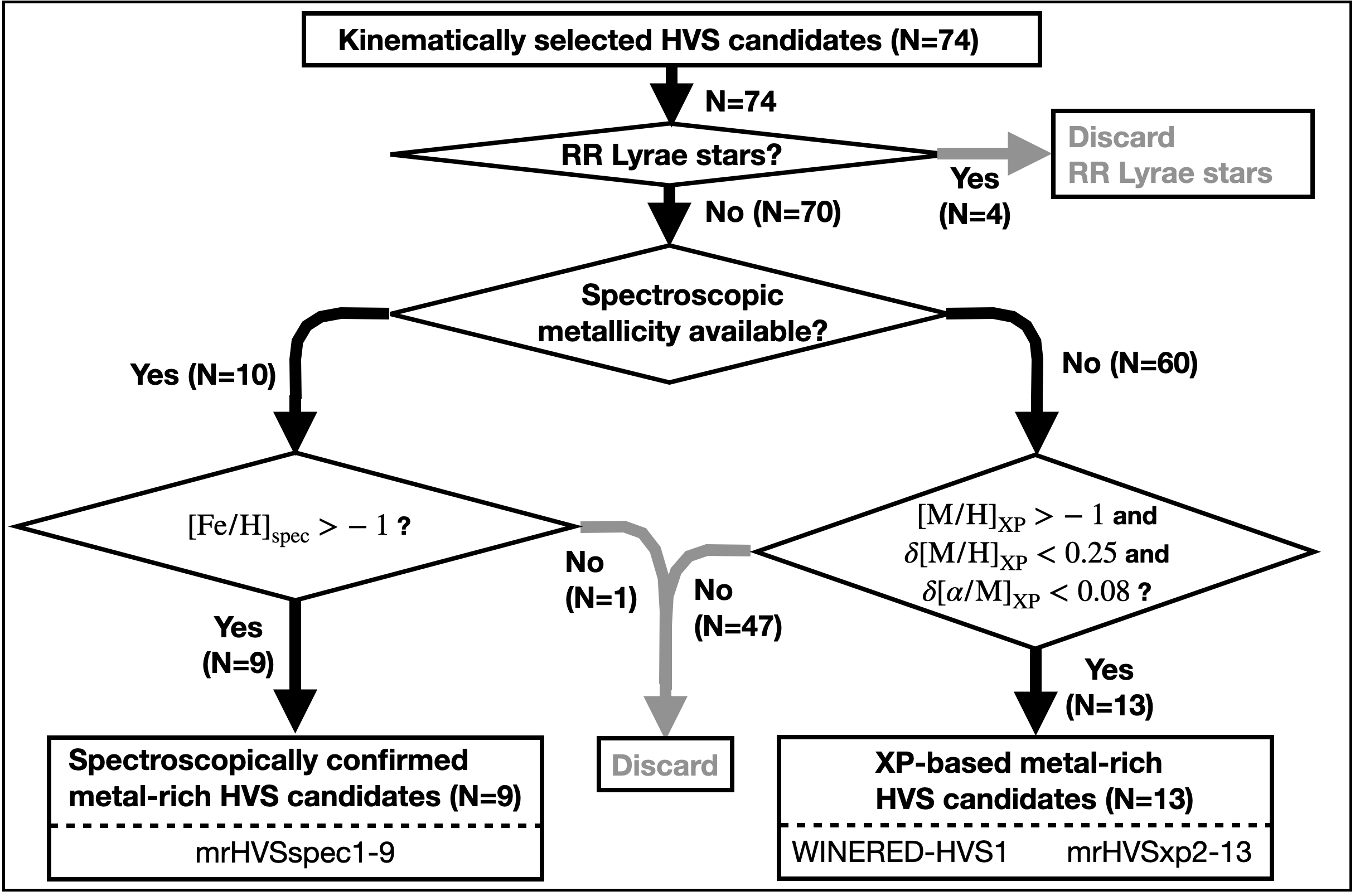}
\caption{
Flow chart to chemically select 22 metal-rich HVS candidates 
from the 74 kinematically selected HVS candidates. 
We refer to the 9 candidates with spectroscopic chemistry from literature 
as `spectroscopically confirmed metal-rich HVS candidates.' 
We refer to the remaining 13 candidates  
as `XP-based meta-rich HVS candidates.' 
The most metal-rich candidate, \targetstar, 
is among the XP-based metal-rich HVS candidates. 
}
\label{fig:flow_chart_chemistry}
\end{figure*}
\subsection{Number of kinematically-selected HVS candidates}
\label{section:number_of_kinematically_selected_HVS}

After searching for the kinematically-selected HVS candidates 
with Methods A--D, 
we compared the list of candidates for each method. 
It turned out that 
the kinematically-selected HVS candidates mainly depend on the assumed potential and only weakly depend on the Solar position and velocity.  
Given that the list of kinematically-selected HVS candidates 
depends on the assumed potential, 
it is reassuring to note that 
74 stars are regarded as kinematically-selected HVS candidates 
for all the methods we explored. 
It turned out that all of these 74 candidates are bound to the Milky Way. 
In the following,
we will analyze these 74 kinematically-selected HVS candidates 
in more detail.

\subsection{Further pruning of HVS candidates with chemistry}

From the 74 kinematically-selected HVS candidates, 
we further select 22 metal-rich HVS candidates 
as summarized in Fig.~\ref{fig:flow_chart_chemistry}.

First, we remove 4 RR Lyrae stars by using an RR Lyrae catalog 
\cite{Clementini2023A&A...674A..18C} 
and SIMBAD. 
Among the remaining 70 stars, 10 stars have 
reliable 
\revise{chemistry} 
from either 
APOGEE DR17 \citep{Abdurrouf2022ApJS..259...35A}, 
GALAH DR4 \citep{Buder2024arXiv240919858B}, 
RAVE DR6  \citep{Steinmetz2020AJ....160...83S}, 
LAMOST DR9 v2.0 \citep{Zhao2012}. 
Among these stars, 9 stars satisfy\footnote{
Throughout this paper, 
$\mathrm{[X/Y]_{spec}}$ denotes the spectroscopic measurements 
of [X/H]. See Table~\ref{table:notation}. 
}
$\mathrm{[Fe/H]_{spec}} > -1$. 
We refer to these stars as 
`spectroscopically confirmed metal-rich HVS candidates' 
and label them as mrHVSspec1-9.
All of them are red giants.

For the 60 stars without spectroscopic chemical abundances,
we use the chemical abundances 
$(\mhxp$, $\amxp)$ 
\revise{
taken from \cite{Hattori2025ApJ...980...90H} (see \autoref{table:notation}).} 
We select 13 stars 
that satisfy the quality criteria in \cite{Hattori2025ApJ...980...90H}, 
while showing $\mhxp > -1$.\footnote{
We have already applied the cut $\mhxp > -1$ 
in Section \ref{section:data}. 
We show this cut also in Fig.\ref{fig:flow_chart_chemistry} 
just to emphasize that XP-based metal-rich HVS candidates 
do not include metal-poor stars. 
} 
We refer to these stars as 
`XP-based metal-rich HVS candidates.' 
We label the most metal-rich candidate 
as WINERED-HVS1, 
while we label the rest of them 
as mrHVSxp2-13. 
Again, all of these candidates are red giants.

As can be inferred from the unique name convention for \targetstar, 
we regard this star different from other HVS candidates 
because it has intriguing chemical properties. 
Its XP-based metallicity, $\mhxp = -0.23$, 
is higher than any other HVS candidates analyzed in this paper. 
Also, 
its $\alpha$-abundance, $\amxp = 0.06$, 
is one of the least $\alpha$-enhanced stars in our HVS candidates. 
At face value, the metal-rich and $\alpha$-poor nature of \targetstar 
is reminiscent of the stellar population near the Galactic center region. 
We did a follow-up observation of this star with Magellan/WINERED 
and spectroscopically confirmed the metal-rich and $\alpha$-poor nature of this star (see \autoref{section:WINERED}).  
Kinematics and chemistry of the 22 HVS candidates, 
including \targetstar, 
will be discussed in \autoref{section:discussion}.

\begin{deluxetable*}{r rrr rrrr rr rr l}
\tablecaption{Kinematical properties of our HVS candidates
\label{table:kinematics}}
\rotate 
\tablewidth{0pt}
\tabletypesize{\scriptsize}
\tablehead{
\colhead{(1)} &
\colhead{(2)} &
\colhead{(3)} &
\colhead{(4)} &
\colhead{(5)} &
\colhead{(6)} &
\colhead{(7)} &
\colhead{(8)} &
\colhead{(9)} &
\colhead{(10)} &
\colhead{(11)} \\
\colhead{Name} &
\colhead{DR3 source\_id} &
\colhead{$(f_\mathrm{50 \pc}, f_\mathrm{200 \pc})$} &
\colhead{$r_\mathrm{ej}$} &
\colhead{$v_\mathrm{ej}$} &
\colhead{$\tau$} &
\colhead{$(x,y,z)$} &
\colhead{$(r,R)$} &
\colhead{$(v_x,v_y,v_z)$} &
\colhead{$(v_r,v_R,v_\phi)$} &
\colhead{$v$} \\
\colhead{} &
\colhead{} &
\colhead{} &
\colhead{$\pc$} &
\colhead{$\kms$} &
\colhead{$\Myr$} &
\colhead{\kpc} &
\colhead{\kpc} &
\colhead{$\kms$} &
\colhead{$\kms$} &
\colhead{$\kms$} 
}
\startdata
\hline 
{mrHVSspec1} & {1364083369852936576} & {$[0.2525, 0.9708]$} & {0}  & {535}  & {26}  & {$[-7.60, +2.69, +1.60]$} & {$[8.22, 8.06]$} & {$[-175, +62, -7]$} & {$[+181, +186, -0]$} & {$186$} \\
{mrHVSspec2} & {3894254612584073472} & {$[0.0218, 0.1212]$} & {1}  & {578}  & {22}  & {$[-8.04, -1.21, +2.82]$} & {$[8.60, 8.13]$} & {$[-265, -40, +60]$} & {$[+273, +268, -0]$} & {$275$} \\
{mrHVSspec3} & {6508950579775204864} & {$[0.0359, 0.0943]$} & {1}  & {606}  & {15}  & {$[-4.09, -1.67, -5.26]$} & {$[6.87, 4.42]$} & {$[-221, -90, -258]$} & {$[+351, +238, +0]$} & {$351$} \\
{mrHVSspec4} & {2694242193191538816} & {$[0.1053, 0.8001]$} & {1}  & {533}  & {24}  & {$[-7.54, +1.29, -1.09]$} & {$[7.73, 7.65]$} & {$[-197, +34, +16]$} & {$[+196, +200, +0]$} & {$201$} \\
{mrHVSspec5} & {6137389419947219584} & {$[0.0000, 0.2973]$} & {54}  & {521}  & {23}  & {$[-4.44, -5.08, +2.37]$} & {$[7.15, 6.75]$} & {$[-122, -136, +21]$} & {$[+180, +183, -3]$} & {$184$} \\
{mrHVSspec6} & {5467370673479835904} & {$[0.0001, 0.3035]$} & {46}  & {571}  & {21}  & {$[-8.28, -1.63, +0.74]$} & {$[8.47, 8.44]$} & {$[-268, -55, -10]$} & {$[+272, +274, +2]$} & {$274$} \\
{mrHVSspec7} & {6692069426721736832} & {$[0.0504, 0.0536]$} & {5}  & {501}  & {11}  & {$[-2.76, +0.21, -2.83]$} & {$[3.96, 2.77]$} & {$[-186, +14, -153]$} & {$[+239, +186, +1]$} & {$241$} \\
{mrHVSspec8} & {3552388581761193728} & {$[0.0339, 0.1519]$} & {3}  & {556}  & {25}  & {$[-8.26, -2.33, +1.72]$} & {$[8.76, 8.59]$} & {$[-219, -62, +7]$} & {$[+224, +227, +0]$} & {$227$} \\
{mrHVSspec9} & {3657091157063641344} & {$[0.0120, 0.1198]$} & {2}  & {543}  & {21}  & {$[-6.33, -1.04, +3.20]$} & {$[7.17, 6.42]$} & {$[-214, -35, +71]$} & {$[+226, +217, -0]$} & {$228$} \\
\hline
{WINERED-HVS1} & {6434696676903108864} & {$[0.0310, 0.4957]$} & {16}  & {537}  & {14}  & {$[-4.44, -2.30, -2.18]$} & {$[5.45, 5.00]$} & {$[-232, -122, -76]$} & {$[+270, +262, +1]$} & {$273$} \\
{mrHVSxp2} & {4548672495942395264} & {$[0.1131, 0.4637]$} & {0}  & {544}  & {15}  & {$[-4.75, +2.81, +1.89]$} & {$[5.84, 5.52]$} & {$[-234, +138, +54]$} & {$[+274, +271, +0]$} & {$277$} \\
{mrHVSxp3} & {6173750574418020352} & {$[0.0469, 0.0564]$} & {1}  & {502}  & {25}  & {$[-3.60, -3.97, +3.69]$} & {$[6.51, 5.36]$} & {$[-83, -92, +40]$} & {$[+125, +124, -0]$} & {$130$} \\
{mrHVSxp4} & {4572287497444895104} & {$[0.0914, 0.3728]$} & {1}  & {556}  & {17}  & {$[-5.44, +2.72, +2.74]$} & {$[6.67, 6.09]$} & {$[-233, +117, +81]$} & {$[+271, +261, +0]$} & {$273$} \\
{mrHVSxp5} & {6307609625404677376} & {$[0.0147, 0.1091]$} & {7}  & {540}  & {13}  & {$[-2.82, -1.46, +4.01]$} & {$[5.12, 3.18]$} & {$[-164, -84, +200]$} & {$[+272, +185, -0]$} & {$272$} \\
{mrHVSxp6} & {5765626877690171904} & {$[0.0034, 0.2743]$} & {7}  & {518}  & {23}  & {$[-6.41, -2.59, -1.46]$} & {$[7.06, 6.91]$} & {$[-168, -68, +10]$} & {$[+175, +181, -0]$} & {$182$} \\
{mrHVSxp7} & {6773393846499438464} & {$[0.0349, 0.1637]$} & {3}  & {503}  & {77}  & {$[-2.73, +1.92, -1.81]$} & {$[3.80, 3.34]$} & {$[+120, -84, +223]$} & {$[-235, -146, -0]$} & {$267$} \\
{mrHVSxp8} & {6425246443181463680} & {$[0.0188, 0.1082]$} & {2}  & {532}  & {18}  & {$[-3.37, -3.02, -4.16]$} & {$[6.15, 4.53]$} & {$[-136, -122, -133]$} & {$[+225, +183, +0]$} & {$226$} \\
{mrHVSxp9} & {4501932497166815360} & {$[0.0009, 0.2663]$} & {30}  & {547}  & {18}  & {$[-1.69, +6.03, +2.79]$} & {$[6.85, 6.26]$} & {$[-64, +232, +71]$} & {$[+249, +241, -1]$} & {$251$} \\
{mrHVSxp10} & {1497611497903517056} & {$[0.0534, 0.5141]$} & {2}  & {568}  & {23}  & {$[-8.08, +0.86, +2.45]$} & {$[8.48, 8.12]$} & {$[-253, +27, +41]$} & {$[+255, +254, +0]$} & {$257$} \\
{mrHVSxp11} & {1329060763412103808} & {$[0.0159, 0.3891]$} & {8}  & {579}  & {19}  & {$[-6.42, +2.75, +3.42]$} & {$[7.78, 6.98]$} & {$[-251, +107, +101]$} & {$[+289, +273, +1]$} & {$291$} \\
{mrHVSxp12} & {5130376441238310016} & {$[0.0426, 0.2071]$} & {0}  & {575}  & {33}  & {$[-10.16, -0.71, -4.35]$} & {$[11.07, 10.18]$} & {$[-205, -14, -57]$} & {$[+212, +206, +0]$} & {$214$} \\
{mrHVSxp13} & {3929723865560075776} & {$[0.0238, 0.1063]$} & {2}  & {598}  & {25}  & {$[-7.28, -1.31, +6.65]$} & {$[9.95, 7.39]$} & {$[-220, -39, +177]$} & {$[+284, +223, -0]$} & {$285$} \\
\enddata
\tablecomments{
(1) Name of the metal-rich HVS candidates. 
Stars named mrHVSspec1-9 are HVS candidates for which we have spectroscopic abundance information from publicly available data. 
For \targetstar, we have spectroscopic abundance information from our follow-up observations with Magellan/WINERED. 
Stars named mrHVSxp2-15 are HVS candidates for which we have the abundance information from the Gaia XP spectra only. 
(2) Gaia DR3 source id. 
(3) The fraction of orbits that pass within the specified Galactocentric radius 
($50 \pc$ and $200\pc$) in the last $200 \Myr$ 
with a large ejection velocity (disk-crossing velocity) of $v_\mathrm{ej}> 500 \kms$. 
(4)-(5) 
The ejection radius $r_\mathrm{ej} = |\vector{x}_\mathrm{ej}|$ and 
ejection velocity $v_\mathrm{ej} = |\vector{v}_\mathrm{ej}|$ 
in the best orbit. 
Here, the best orbit is chosen among $N_\mathrm{MC}=10001$ Monte Carlo orbits such that the ejection position is closest to the Galactic center 
while having a large ejection velocity ($v_\mathrm{ej}> 500 \kms$). 
(6) 
\revise{
The time since the last disk crossing in the best orbit. 
If our candidates are recently-ejected HVSs, 
this quantity corresponds to the flight time since the ejection 
from the Galactic center. 
}
(7)-(8) The current Galactocentric spherical radius ($r$) 
and cylindrical radius ($R$) of the star in the best orbit, respectively. 
(9)-(11) The current velocity of the star in the best orbit. 
Here, $v = |\vector{v}|$ is the magnitude of the current velocity. 
}
\end{deluxetable*}

\begin{deluxetable*}{r rrr rrrr rr rr l}
\tablecaption{Chemical properties of our HVS candidates
\label{table:chemistry}}
\rotate 
\tablewidth{0pt}
\tabletypesize{\scriptsize}
\tablehead{
\colhead{(1)} &
\colhead{(2)} &
\colhead{(3)} &
\colhead{(4)} &
\colhead{(5)} &
\colhead{(6)} &
\colhead{(7)} &
\colhead{(8)} &
\colhead{(9)} &
\colhead{(10)} &
\colhead{(11)} \\
\colhead{Name} &
\colhead{DR3 source\_id} &
\colhead{[Fe/H]$_\mathrm{spec}$} &
\colhead{[Mg/Fe]$_\mathrm{spec}$} & 
\colhead{[Si/Fe]$_\mathrm{spec}$} & 
\colhead{[Ca/Fe]$_\mathrm{spec}$} & 
\colhead{[Ti/Fe]$_\mathrm{spec}$} &
\colhead{[Al/Fe]$_\mathrm{spec}$} &
\colhead{[$\alpha$/Fe]$_\mathrm{spec}$} &
\colhead{$\mathrm{[M/H]_{XP}}$} &
\colhead{$[\alpha/\mathrm{M}]_\mathrm{XP}$} \\
\colhead{} &
\colhead{} &
\colhead{dex} &
\colhead{dex} &
\colhead{dex} &
\colhead{dex} &
\colhead{dex} &
\colhead{dex} &
\colhead{dex} &
\colhead{dex} 
}
\startdata
{mrHVSspec1} & {1364083369852936576} & {$-0.64 \pm 0.01$} & {$0.35 \pm 0.01$} & {$0.23 \pm 0.02$} & {$0.18 \pm 0.02$} & {$0.12 \pm 0.02$} & {$0.27 \pm 0.02$} & {$0.22 \pm 0.09$} & {$-0.64^{+0.18}_{+0.00}$} & {$+0.29^{+0.00}_{-0.05}$} \\
{mrHVSspec2} & {3894254612584073472} & {$-0.81 \pm 0.01$} & {$0.18 \pm 0.02$} & {$0.08 \pm 0.03$} & {$0.18 \pm 0.03$} & {$0.12 \pm 0.04$} & {$-0.25 \pm 0.03$} & {$0.14 \pm 0.04$} & {$-0.83^{+0.14}_{-0.31}$} & {$+0.26^{+0.05}_{-0.10}$} \\
{mrHVSspec3} & {6508950579775204864} & {$-0.50 \pm 0.06$} & {$0.22 \pm 0.02$} & {$0.34 \pm 0.02$} & {$-0.09 \pm 0.02$} & {$0.02 \pm 0.01$} & {$-0.43 \pm 0.02$} & {$0.12 \pm 0.17$} & {$-0.86^{+0.17}_{-0.07}$} & {$+0.11^{+0.07}_{-0.05}$} \\
{mrHVSspec4} & {2694242193191538816} & {$-0.66 \pm 0.06$} & {$0.39 \pm 0.04$} & {$0.29 \pm 0.03$} & {$0.40 \pm 0.04$} & {$0.28 \pm 0.06$} & {$0.13 \pm 0.11$} & {$0.34 \pm 0.06$} & {$-0.77^{+0.27}_{-0.10}$} & {$+0.28^{+0.03}_{-0.05}$} \\
{mrHVSspec5} & {6137389419947219584} & {$-0.83 \pm 0.06$} & {$0.32 \pm 0.03$} & {$0.32 \pm 0.02$} & {$0.24 \pm 0.04$} & {$0.21 \pm 0.02$} & {$0.09 \pm 0.04$} & {$0.27 \pm 0.05$} & {$-0.97^{+0.16}_{-0.13}$} & {$+0.07^{+0.05}_{-0.04}$} \\
{mrHVSspec6} & {5467370673479835904} & {$-0.98 \pm 0.07$} & {$0.10 \pm 0.07$} & {$0.30 \pm 0.05$} & {$0.34 \pm 0.06$} & {$0.72 \pm 0.17$} & {--} & {$0.37 \pm 0.22$} & {$-1.00^{+0.44}_{-0.44}$} & {$+0.15^{+0.20}_{-0.11}$} \\
{mrHVSspec7} & {6692069426721736832} & {$-0.47 \pm 0.13$} & {--} & {--} & {--} & {--} & {$0.14 \pm 0.13$} & {$0.52 \pm 0.18$} & {$-0.64^{+0.11}_{-0.10}$} & {$+0.31^{+0.01}_{-0.04}$} \\
{mrHVSspec8} & {3552388581761193728} & {$-0.90 \pm 0.13$} & {--} & {--} & {--} & {--} & {$0.23 \pm 0.13$} & {$0.39 \pm 0.25$} & {$-0.95^{+0.17}_{-0.16}$} & {$+0.22^{+0.07}_{-0.08}$} \\
{mrHVSspec9} & {3657091157063641344} & {$-0.39$} & {--} & {--} & {--} & {--} & {--} & {--} & {$-0.45^{+0.09}_{-0.12}$} & {$+0.25^{+0.05}_{-0.12}$} \\
\hline
{WINERED-HVS1} & {6434696676903108864} &  {$-0.147^{+0.046}_{-0.045}$} & {$-0.058^{+0.077}_{-0.082}$} & {$0.144^{+0.088}_{-0.091}$} & {$-0.001^{+0.175}_{-0.177}$} & {$0.008^{+0.140}_{-0.151}$} & 
{--} & 
{$0.02 \pm 0.07$} & 
{$-0.23^{+0.08}_{-0.08}$} & {$+0.06^{+0.04}_{-0.02}$} \\
{mrHVSxp2} & {4548672495942395264} & {--} & {--} & {--} & {--} & {--} & {--} & {--} & {$-0.40^{+0.11}_{-0.11}$} & {$+0.25^{+0.04}_{-0.03}$} \\
{mrHVSxp3} & {6173750574418020352} & {--} & {--} & {--} & {--} & {--} & {--} & {--} & {$-0.56^{+0.10}_{-0.12}$} & {$+0.15^{+0.11}_{-0.04}$} \\
{mrHVSxp4} & {4572287497444895104} & {--} & {--} & {--} & {--} & {--} & {--} & {--} & {$-0.59^{+0.10}_{-0.16}$} & {$+0.27^{+0.02}_{-0.06}$} \\
{mrHVSxp5} & {6307609625404677376} & {--} & {--} & {--} & {--} & {--} & {--} & {--} & {$-0.63^{+0.16}_{-0.15}$} & {$+0.26^{+0.04}_{-0.07}$} \\
{mrHVSxp6} & {5765626877690171904} & {--} & {--} & {--} & {--} & {--} & {--} & {--} & {$-0.69^{+0.19}_{-0.22}$} & {$+0.25^{+0.05}_{-0.07}$} \\
{mrHVSxp7} & {6773393846499438464} & {--} & {--} & {--} & {--} & {--} & {--} & {--} & {$-0.71^{+0.14}_{-0.31}$} & {$+0.28^{+0.05}_{-0.07}$} \\
{mrHVSxp8} & {6425246443181463680} & {--} & {--} & {--} & {--} & {--} & {--} & {--} & {$-0.73^{+0.17}_{-0.29}$} & {$+0.27^{+0.04}_{-0.08}$} \\
{mrHVSxp9} & {4501932497166815360} & {--} & {--} & {--} & {--} & {--} & {--} & {--} & {$-0.84^{+0.18}_{-0.25}$} & {$+0.07^{+0.08}_{-0.03}$} \\
{mrHVSxp10} & {1497611497903517056} & {--} & {--} & {--} & {--} & {--} & {--} & {--} & {$-0.85^{+0.11}_{-0.12}$} & {$+0.28^{+0.04}_{-0.04}$} \\
{mrHVSxp11} & {1329060763412103808} & {--} & {--} & {--} & {--} & {--} & {--} & {--} & {$-0.91^{+0.19}_{-0.27}$} & {$+0.25^{+0.05}_{-0.07}$} \\
{mrHVSxp12} & {5130376441238310016} & {--} & {--} & {--} & {--} & {--} & {--} & {--} & {$-0.93^{+0.28}_{-0.17}$} & {$+0.27^{+0.04}_{-0.07}$} \\
{mrHVSxp13} & {3929723865560075776} & {--} & {--} & {--} & {--} & {--} & {--} & {--} & {$-0.96^{+0.16}_{-0.15}$} & {$+0.17^{+0.07}_{-0.06}$} \\
\hline 
\enddata
\tablecomments{
(1) Name of the metal-rich HVS candidates. 
(2) Gaia DR3 source id. 
(3)-(9) Chemical abundances from spectroscopic observations. 
The spectroscopic data 
for mrHVSspec1-2 are taken from 
APOGEE DR17 \citep{Abdurrouf2022ApJS..259...35A}, 
those for mrHVSspec3-6 are taken from 
GALAH DR4 \citep{Buder2024arXiv240919858B}, 
those for mrHVSspec7-8 are taken from 
RAVE DR6 \citep{Steinmetz2020AJ....160...83S}, 
those for mrHVSspec9 are taken from 
LAMOST DR9 v2.0 \citep{Zhao2012}, 
and 
those for \targetstar are derived from our analysis in \autoref{section:WINERED}. 
(We note that mrHVSspec4 is flagged as a single-line spectroscopic binary in GALAH data.)
For mrHVSxp1-6, we adopt 
$\mathrm{[\alpha/Fe]_{spec}} = (
\mathrm{[Mg/Fe]_{spec}} + 
\mathrm{[Si/Fe]_{spec}} + 
\mathrm{[Ca/Fe]_{spec}} + 
\mathrm{[Ti/Fe]_{spec}})/4$ 
(see also \autoref{table:notation} 
for the definition of $\mathrm{[\alpha/Fe]_{spec}}$). 
For mrHVSxp7-8, 
we adopt $\mathrm{[\alpha/Fe]_{spec}}$ from RAVE DR6. 
(10)-(11) 
The metallicity [M/H]$_\mathrm{XP}$ and 
$\alpha$-abundance $[\alpha/\mathrm{M}]_\mathrm{XP}$ 
estimated from Gaia XP spectra \citep{Hattori2025ApJ...980...90H}.
}
\end{deluxetable*}



\section{Spectroscopic observation of a hyper-velocity star candidate 
WINERED-HVS1}
\label{section:WINERED}

\begin{figure*}
\centering 
\includegraphics[width=\textwidth , page=1]{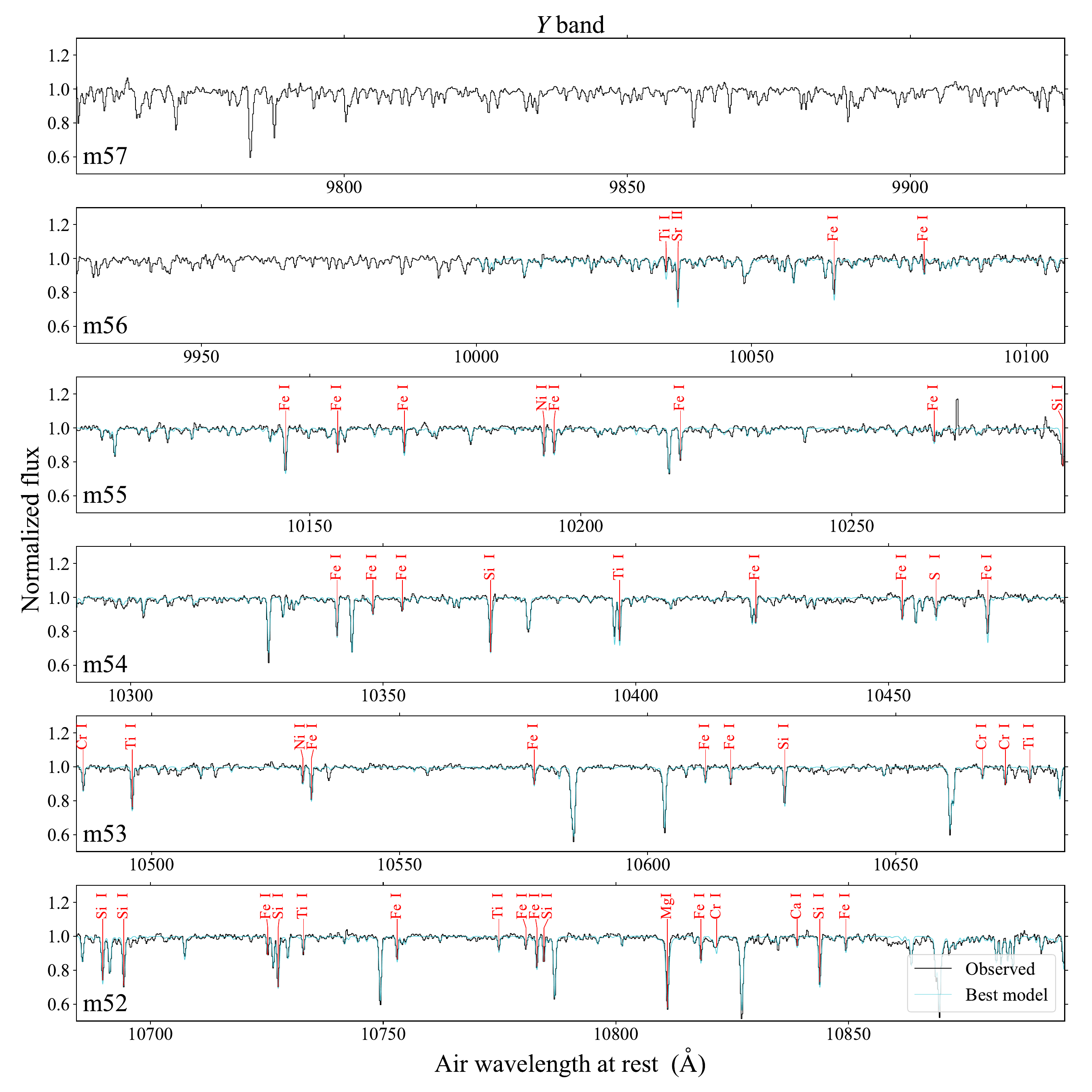}
\caption{\revise{WINERED near-infrared high-resolution spectrum of \targetstar in the \textit{Y} band (echelle orders 57-52). Black lines show the reduced spectrum of \targetstar , after telluric lines were removed. Cyan lines show the best-fit model spectrum synthesized with the MB99 line list. Red vertical lines near the top edge of each panel mark the wavelengths from the MB99 line list of the lines that were used for measuring [X/H]. We note that the MB99 list includes only lines with wavelengths longer than $10,000\,\text{\AA }$, and thus the synthetic spectrum is shown only for wavelengths greater than $10,000\,\text{\AA }$. (The observed and best-fit spectra are available as Data behind the Figure associated with this publication.)}}
\label{fig:specsample1}
\end{figure*}

\begin{figure*}
\centering 
\includegraphics[width=\textwidth , page=2]{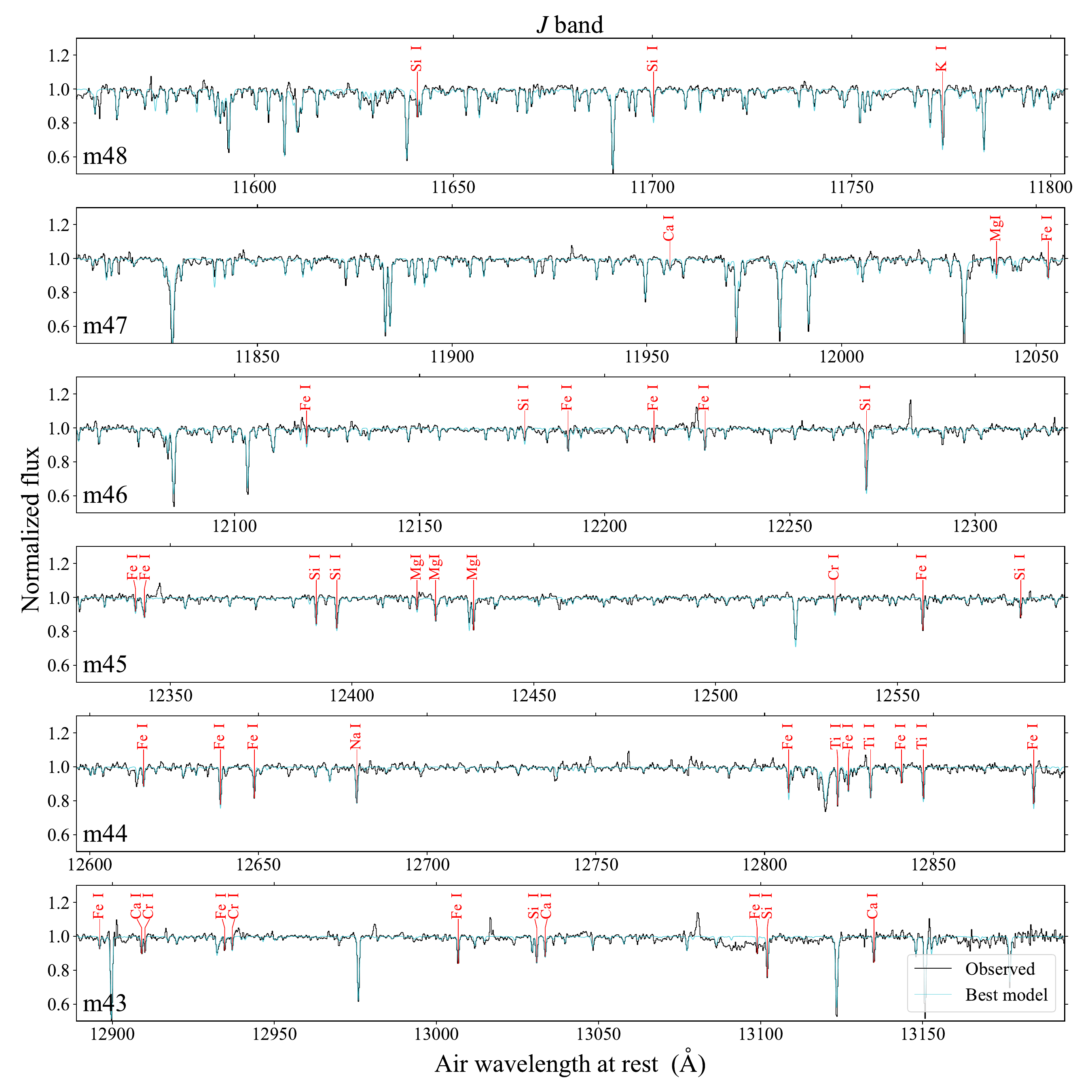}
\caption{\revise{Same as Fig.~\ref{fig:specsample1} but for the \textit{J} band (echelle orders 48--43). }}
\label{fig:specsample2}
\end{figure*}

We obtained a near-infrared \textit{YJ}-band high-resolution spectrum of \targetstar using the WINERED spectrograph~\citep[$0.90\text{--}1.35\,\micron $ with $R=28,000$;][]{Ikeda2022} attached to the Magellan Clay Telescope on June 6th 2023. 
We performed the ABBA nodding with $600\,\mathrm{sec}$ exposures. 
With the \textit{J}-band magnitude of $12.69$~\citep[2MASS;][]{Cutri2003}, the resultant signal-to-noise ratio (S/N) per pixel is $40\text{--}115$ for the \textit{Y} band ($0.97\text{--}1.09\,\micron $) and $25\text{--}90$ for the \textit{J} band ($1.15\text{--}1.32\,\micron $). 
We also observed a telluric-standard A0V star, HD~170115, with a sufficient S/N of $200\text{--}400$. 
The raw data was reduced using the WINERED Automatic Reduction Pipeline~(WARP version 3.8.7; \citealt{Hamano2024PASP..136a4504H}), and telluric lines were removed with the standard star using the method described by \citet{Sameshima2018}. 
Then the line-of-sight velocity was determined using the cross-correlation function between 
\revise{the observed spectrum and a radial velocity template spectrum}, 
and the wavelengths were adjusted to the standard-air scale at rest. 
\revise{The reduced spectrum, together with the best-fit synthetic spectrum whose parameters will be determined in this section, is shown in Figs.~\ref{fig:specsample1} and \ref{fig:specsample2}. }

After thorough examination, we concluded that the systematic bias in the determined line-of-sight velocity, summarized in \autoref{table:stellar_abn_all_res}, would be ${\sim }1.0\ur{\si{km.s^{-1}}}$ at most. 
\revise{Our line-of-sight velocity measurement for this star 
($34.79 \pm 0.13 \mathrm{(rand.)} \pm 1.0 \mathrm{(sys.)} \kms$) differs from the value reported in Gaia DR3 ($26.52 \pm 2.07 \kms$). 
To maintain consistency across all stars in our sample,
we adopt the Gaia DR3 line-of-sight velocity for this star in our orbital analysis. 
We have confirmed that adopting the value from our follow-up spectroscopic data has a minor effect on the orbital analysis for this star (e.g., reducing the value of $v_\mathrm{ej}$ by $\sim 25 \kms$), which does not affect the main conclusion of this paper. 
}

With the reduced spectrum, we determined chemical abundances of \targetstar with a method similar to those performed by 
\cite{Kondo2019} and \cite{Fukue2021} 
using the \textsc{Octoman} code~\citep{Taniguchi2025}, which is a wrapper of the spectral synthesis code MOOG~\citep{Sneden1973,Sneden2012}. 
Briefly, we first determined the effective temperature $T_{\mathrm{eff}}$, using the relations between $T_{\mathrm{eff}}$ and line-depth ratios (LDRs) of \ion{Fe}{i} lines calibrated using well-studied red giants by \citet{Taniguchi2021}. 
Following the procedure described by Taniguchi et al. (2025b, in prep.), we removed the systematic bias in the determined $T_{\mathrm{eff}}$ due to the metallicity effect on the $T_{\mathrm{eff}}$--LDR relations. 
In this step, we adopted the metallicity of \targetstar of $-0.2624\ur{dex}$, which is the value tabulated in the Gaia DR3 catalog (\texttt{mh\_gspphot}) after removing the systematic bias in it following the recipe by \citet{Andrae2023b}. 
The amount of the removed bias is $13\ur{K}$, which is much smaller than the error in the determined $T_{\mathrm{eff}}$, $76\ur{K}$. 
Next, we calculated the surface gravity $\log g$ using the Stefan-Boltzmann law with some stellar parameter values: \revise{the LDR effective temperature determined here, the mass $M=1\text{--}3M_{\odot }$ estimated in \autoref{section:mass_age}, and luminosity $L=51.06^{+8.22}_{-7.52}L_{\odot }$ tabulated in the Gaia DR3 catalog (\texttt{lum\_flame}). }
Then, by fitting some of the \ion{Fe}{i} lines that are relatively free from the contamination by surrounding lines, we determined $v_{\mathrm{micro}}$ and [Fe/H] that give no correlation between the line strengths and [Fe/H]s determined with the individual lines. 
Finally, we determined [X/H] of Na, Mg, Si, S, K, Ca, Ti, Cr, Ni, and Sr by fitting relatively uncontaminated individual lines of these elements. 
Throughout the analysis, we used the grid of ATLAS9-APOGEE model atmospheres~\citep{Meszaros2012}. 
We used two line lists, the list with astrophysical $\log gf$ values calibrated against the Sun by \citet[MB99]{Melendez1999} and the Vienna Atomic Line Database~\citep[VALD3;][]{Ryabchikova2015}, and compared the results to identify potential differences if any. 

In determining [Fe/H] and [X/H], we employed the so-called differential analysis method~\citep{Jofre2019} to reduce the systematic bias in the determined chemical abundances due to the inaccuracy in infrared line lists~\citep[see, e.g.,][]{Andreasen2016,Matsunaga2020}. 
As the reference stars of the differential analysis, we analyzed the WINERED spectra of two well-investigated thin-disk red giants, $\varepsilon $~Vir and Pollux, observed by \citet{Taniguchi2018}. 
The two stars have the stellar parameters and observed spectra similar to those of \targetstar . 
Stellar parameters ([Fe/H], $T_{\mathrm{eff}}$, $\log g$, and $v_{\mathrm{micro}}$) and chemical abundances of the two reference stars were mainly adopted from the Gaia FGK benchmark stars v2.1~\citep{Jofre2018}. 
They only determined [X/H] of Mg, Si, Ca, Ti, Cr, Fe, and Ni among the elements of our interest, and thus we adopted [X/H] of the other elements (Na, S, K, and Sr) from other literature or estimated with an assumption: [Na/H] was adopted from the study by \citet{Jofre2015b}, [S/H] and [K/H] were estimated assuming $\text{[S/H]}=\text{[K/H]}=(\text{[Mg/H]}+\text{[Si/H]}+\text{[Ca/H]})/3$, and [Sr/H] was assumed to have the same value as [Ba/H] determined by \citet{Jofre2015b}. 
Hence, the determined [X/H] of Na, S, K, and Sr for \targetstar are possibly subject to a systematic bias in the zero point. 

Two sources of errors were considered in the analysis: the standard error of line-by-line dispersion and the errors propagated from the errors in stellar parameters~\citep[see equations (9) and (10) in][]{Taniguchi2025}. 
The errors for most elements are dominated by the standard error of the line-by-line dispersion, but the error propagated from the error in $T_{\mathrm{eff}}$ also markedly contribute to some elements (Si, S, Ti, and Cr). 

\begin{figure}
\centering 
\includegraphics[width=\columnwidth ]{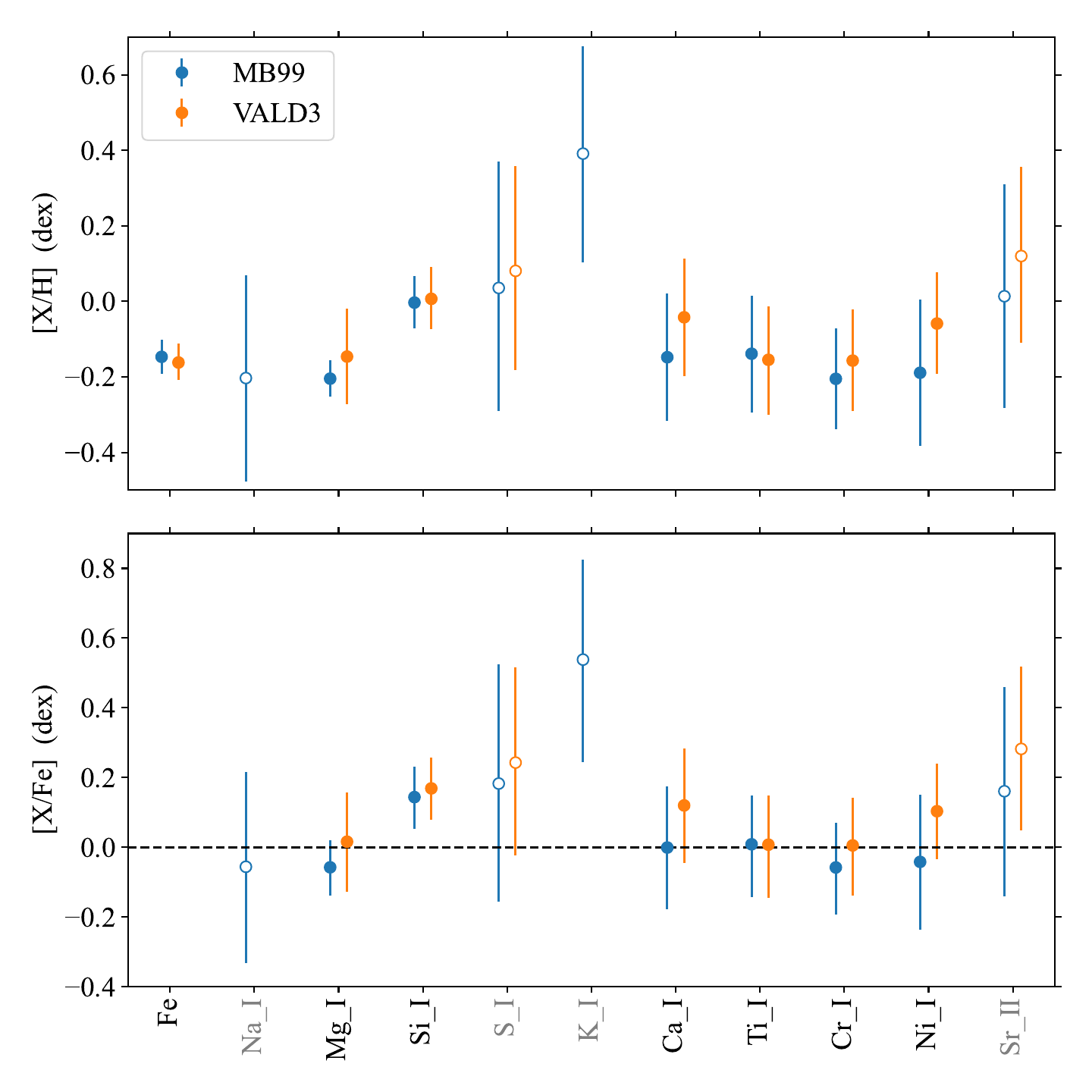}
\caption{Chemical abundances, [X/H] and [X/Fe], of \targetstar , summarized in \autoref{table:stellar_abn_all_res} (blue; fiducial result using the MB99 line list) and \autoref{table:stellar_abn_all_res_VALD3} (orange; using the VALD3 line list). We employed the differential analysis against two Gaia FGK benchmark stars, $\varepsilon $~Vir and Pollux. For elements indicated by filled circles, we adopted [X/H] of the reference stars determined by \citet{Jofre2018}. In contrast, reference [X/H] of elements indicated by open circles were taken from other literature or estimated with an assumption, and thus they are subject to a possible systematic bias in the zero point. See the main text for more detail. 
}
\label{fig:abn_all_res}
\end{figure}

Figure~\ref{fig:abn_all_res} compares the resultant chemical composition of \targetstar determined with the MB99 (\autoref{table:stellar_abn_all_res}) and VALD3 (\autoref{table:stellar_abn_all_res_VALD3}) line lists. 
We found an overall good agreement between the abundances determined with the two lists. 
Since the MB99 list generally better reproduces near-infrared spectra of late-type stars~\citep{Kondo2019,Taniguchi2025}, we used the results with the MB99 list, presented in \autoref{table:stellar_abn_all_res}, in the following discussion. 
For brevity, we only present [X/H] and [X/Fe] in \autoref{table:stellar_abn_all_res}. 
However, our analysis also allows us to obtain 
abundance ratio between $\alpha$ elements, \revise{such as 
$\mathrm{[Si/Mg]}= 0.201^{+0.073}_{-0.074}$, 
$\mathrm{[Ca/Mg]}= 0.057^{+0.176}_{-0.177}$, 
and 
$\mathrm{[Ti/Mg]}= 0.066^{+0.162}_{-0.170}$}. 
These [X/Mg] ratios will turn out to be informative 
in understanding the origin of \targetstar.

\begin{table}
\centering 
\caption{\revise{Spectroscopic results of \targetstar  }}
\label{table:stellar_abn_all_res}
\begin{tabular}{l rrc} \toprule 
$v_{\mathrm{helio}}$ [\si{km.s^{-1}}] & \multicolumn{3}{c}{$34.79\pm 0.13$\tablenotemark{a}} \\
\midrule 
$T_{\mathrm{eff}}$ [K] & \multicolumn{3}{c}{$4680\pm 76$} \\
$\log g$ & \multicolumn{3}{c}{$2.6\pm 0.2$} \\
$v_{\mathrm{micro}}$ [\si{km.s^{-1}}] & \multicolumn{3}{c}{$1.531^{+0.160}_{-0.150}$} \\
{[Fe/H]} [dex] & \multicolumn{3}{c}{$-0.147^{+0.046}_{-0.045}$ ($43$)} \\
\midrule 
 & \multicolumn{1}{c}{[X/H] [dex]} & \multicolumn{1}{c}{[X/Fe] [dex]} \\ \midrule 
\ion{Na}{i}\tablenotemark{b} & $-0.203^{+0.272}_{-0.274}$ & $-0.056^{+0.272}_{-0.276}$ & ($1$) \\
\ion{Mg}{i} & $-0.205^{+0.048}_{-0.047}$ & $-0.058^{+0.077}_{-0.082}$ & ($5$) \\
\ion{Si}{i} & $-0.003^{+0.071}_{-0.068}$ & $+0.144^{+0.088}_{-0.091}$ & ($17$) \\
\ion{S}{i}\tablenotemark{b} & $+0.036^{+0.334}_{-0.326}$ & $+0.182^{+0.342}_{-0.339}$ & ($1$) \\
\ion{K}{i}\tablenotemark{b} & $+0.391^{+0.284}_{-0.288}$ & $+0.538^{+0.286}_{-0.293}$ & ($1$) \\
\ion{Ca}{i} & $-0.148^{+0.170}_{-0.170}$ & $-0.001^{+0.175}_{-0.177}$ & ($5$) \\
\ion{Ti}{i} & $-0.139^{+0.154}_{-0.157}$ & $+0.008^{+0.140}_{-0.151}$ & ($9$) \\
\ion{Cr}{i} & $-0.205^{+0.134}_{-0.133}$ & $-0.058^{+0.129}_{-0.134}$ & ($7$) \\
\ion{Ni}{i} & $-0.189^{+0.194}_{-0.193}$ & $-0.042^{+0.194}_{-0.194}$ & ($2$) \\
\ion{Sr}{ii}\tablenotemark{b} & $+0.013^{+0.297}_{-0.296}$ & $+0.160^{+0.299}_{-0.301}$ & ($1$) \\
\bottomrule 
\end{tabular}
\flushleft{
\tablecomments{Chemical abundances, [X/H] and [X/Fe], were determined using the MB99 line list~\citep{Melendez1999}. The results using the VALD3 list are presented in \autoref{table:stellar_abn_all_res_VALD3}. Numbers in brackets show the number of lines analyzed. }
\tablenotetext{a}{With a possible systematic bias of ${\sim }1.0\ur{\si{km.s^{-1}}}$ at most.}
\tablenotetext{b}{[Na/H], [S/H], [K/H], and [Sr/H] are possibly subject to a systematic bias in the zero point. See text for details. }
}
\end{table}

\begin{figure*}
\centering
\includegraphics[width=0.7\textwidth]{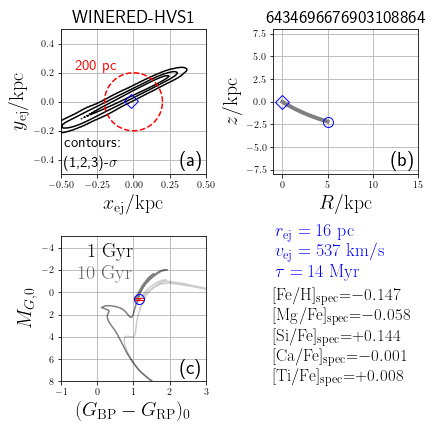}
\caption{
The orbital properties of \targetstar 
(\targetstarGaiaSID)
\revise{
under the fiducial assumptions (Method A) 
since the last disk crossing.} 
\revise{(a) 
The intersection of the Monte Carlo orbits with the Galactic disk plane at the last disk crossing, referred to as the ejection position.
The three contours enclose 39.3, 86.5, and 98.9 percent 
of the ejection positions.} 
The open 
\revise{blue diamond} 
indicates the ejection position of the best orbit---the orbit most consistent with a 
\revise{recently-ejected} 
HVS orbit (with the smallest $r_\mathrm{ej}$). 
\revise{The red dashed circle indicates a Galactocentric radius of $200 \pc$, which corresponds to the systematic uncertainty in 
the ejection position arising from the specific choice of the Galactic potential model.} 
\revise{(b)} 
The best orbit in the Galactocentric meridional plane. 
\revise{The current position and the ejection position are} 
marked by an open 
\revise{blue circle and diamond, respectively.} 
\revise{The ejection velocity of the best orbit is $v_\mathrm{ej} = 537 \kms$. 
If \targetstar was recently ejected from the Galactic center, 
its flight time is 14 Myr.}
In 5 Myr, the orbit is expected to reach the apocenter at $r=11.6 \kpc$. 
\revise{(c)} 
The location of the \targetstar in the color-magnitude diagram (CMD). 
We compute the extinction-corrected color $(G_\mathrm{BP} - G_\mathrm{RP})_0$ 
and absolute magnitude $M_{G,0}$ for each Monte Carlo orbit, 
as different stellar distance is assumed for each orbit. 
\revise{The blue circle corresponds to the location in the CMD when the best orbit is adopted. 
The red vertical error bar represents the range of $M_{G,0}$ 
if its ejection position is within 200 pc of the Galactic center. 
}
In correcting for the dust extinction, 
we use a 3D dust map Combined19 
from \texttt{mwdust} package \citep{Bovy2016ApJ...818..130B}, 
in which three 3D dust maps are combined \citep{Drimmel2003A&A...409..205D, Marshall2006A&A...453..635M, Green2019ApJ...887...93G}. 
We also plot 1 Gyr and 10 Gyr isochrones from the BaSTI model \citep{Hidalgo2018ApJ...856..125H}, 
by assuming the stellar metallicity closest to that of this star. 
\revise{The observed properties suggest that \targetstar 
is roughly 0.5--10 Gyr old (see \autoref{section:mass_age}).} 
}
\label{fig:WINEREDHVS1}
\end{figure*}

\begin{figure*}
\centering
\includegraphics[width=0.9\textwidth]{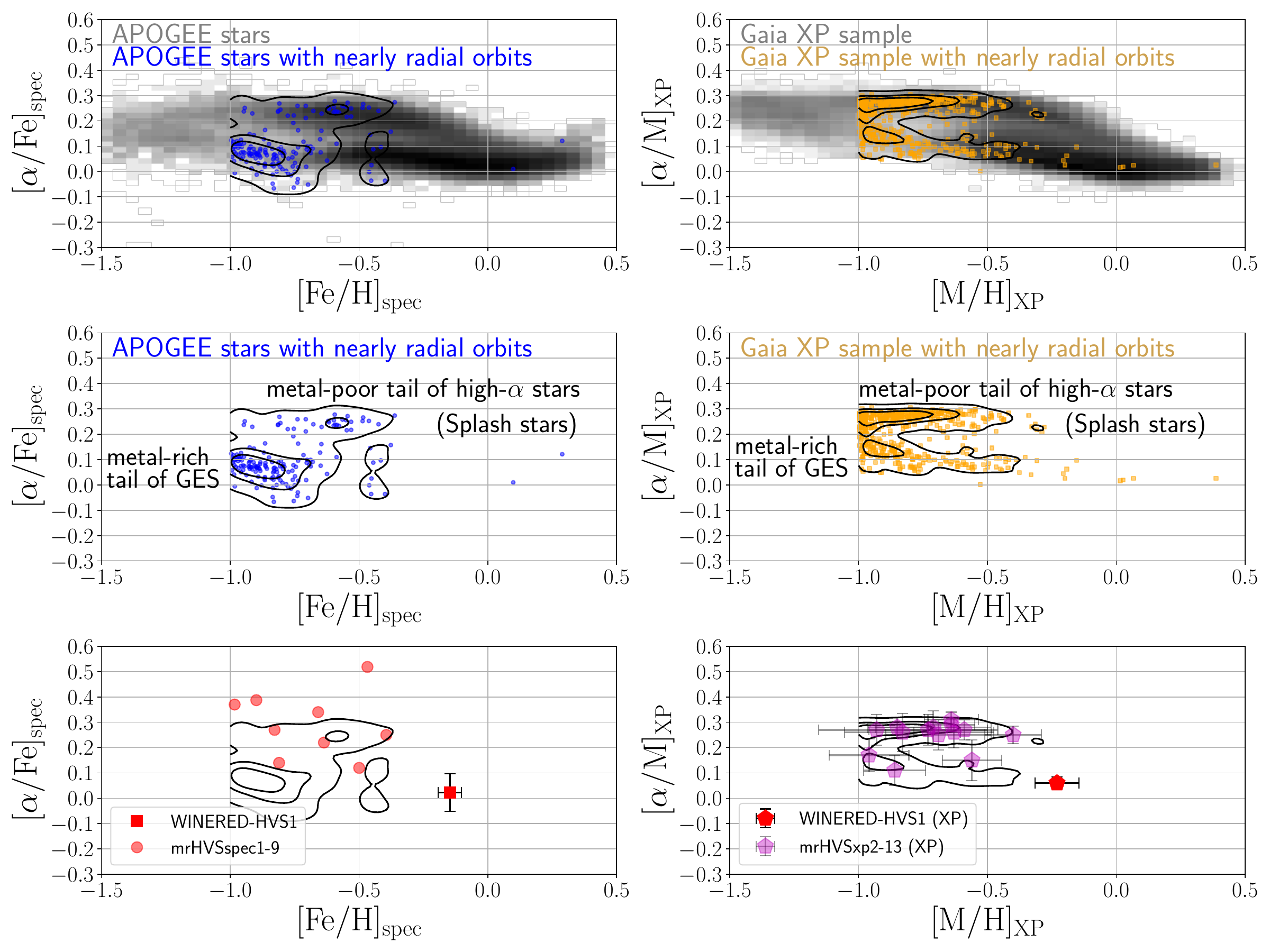}
\caption{
We compare the metallicity and $\alpha$-abundance of metal-rich stars that exhibit nearly radial orbits (which are presumably non-HVS stars; 
\revise{see \autoref{appendix:nearly_radial_orbits} for selection criteria}) and those of our HVS candidates.
(Left Column)  
We compare the APOGEE stars and our HVS candidates with known spectroscopic abundances (mrHVSspec1-9 and \targetstar). The distribution of the cleaned sample of APOGEE DR17 stars is shown by a grayscale image in the top panel. A subset of APOGEE stars with nearly radial orbits and $\mathrm{[Fe/H]_{spec}} > -1$ is marked with blue dots, accompanied by contours enclosing 25\%, 50\%, and 90\% of these stars. This group consists of two distinct sequences: the metal-rich tail of Gaia Enceladus Sausage (GES) and the metal-poor tail of high-$\alpha$ disk stars. Notably, 99.7\% of stars with nearly radial orbits and $\mathrm{[Fe/H]_{spec}} > -1$ satisfy $\mathrm{[Fe/H]_{spec}} < -0.3$, and our HVS candidates labeled as mrHVSspec1-9 also meet this criterion. \targetstar with 
\revise{$\mathrm{[Fe/H]_{spec}} = -0.147$}
emerges as the most promising candidate, distinguishing itself as a metal-rich outlier.
(Right Column)
Next, we compare Gaia XP stars with our XP-based metal-rich HVS candidates (mrHVSxp2-13 and \targetstar). The top and middle panels are the same as in the left column, but using a cleaned sample of stars with reliable Gaia XP chemistry. About 98\% of stars with nearly radial orbits and $\mhxp > -1$ have $\mhxp < -0.3$, and our HVS candidates labeled as mrHVSxp2-13 also meet this condition. \targetstar has a metallicity of $\mhxp = -0.23$, which is a clear metal-rich outlier, even when assessed through our XP-based metallicity measurements.
}
\label{fig:fe_alpha}
\end{figure*}



\section{Discussion} \label{section:discussion} 

In \autoref{section:search_HVS}, 
we selected 22 metal-rich HVS candidates from the Gaia DR3 catalog. 
For one of the candidates, \targetstar, 
we conducted \revise{follow-up} observations 
to obtain the detailed chemical abundances (see \autoref{section:WINERED}). 
In this Section, we will first 
discuss the kinematical (\autoref{table:kinematics}) 
and chemical properties 
(\autoref{table:chemistry}) 
of these 22 stars.  
Then we focus on the most-metal rich candidate, \targetstar, 
and discuss its origin. 


\subsection{Kinematical properties of metal-rich HVS candidates}

\label{section:kinematical_properties_of_HVSs}

\revise{
As mentioned in \autoref{section:kinematic_conditions}, 
we trace the stellar orbit in the last 200 Myr
and call the position and velocity at the last disk crossing 
as the {\it ejection} position and {\it ejection} velocity for brevity.  
While we use them to identify the HVS candidates, 
in general there is no guarantee 
that the time at the last disk crossing ($t=-\tau$) 
corresponds to the time when the star was ejected (from the SMBH). 
Still, the value of $v_\mathrm{ej}$ is useful 
to characterize the HVS ejection, 
and the orbit since the last disk crossing 
allows an intuitive understanding of the HVS orbit. 
In the following, 
we analyze the orbital properties 
at $-\tau < t < 0$ 
for each HVS candidate.}

For each metal-rich HVS candidate, 
we have $N_\mathrm{MC}$ orbit realizations. 
By using these Monte Carlo orbits in our fiducial search   
(i.e., Method A), 
we select the orbits that satisfy 
either (Condition 1) or (Condition 2), 
and then find the `best' orbit that has the smallest ejection radius 
$r_\mathrm{ej}$.

The properties of the best orbits of the metal-rich 
HVS candidates are summarized in  
\autoref{table:kinematics}.
In Fig.~\ref{fig:WINEREDHVS1}, 
we show the Monte Carlo orbits as well as the best orbit 
for \targetstar.\footnote{ 
Orbits for the other HVS candidates are shown in  
Figs.~\ref{fig:mrHVSspec1to5}, 
\ref{fig:mrHVSspec6to9},
\ref{fig:mrHVSxp2to6}, 
\ref{fig:mrHVSxp7to11}, and
\ref{fig:mrHVSxp12to13},
in \autoref{section:orbits_for_other_HVSs}. 
} 
Mainly due to the uncertainty in the heliocentric distance to this star, 
the ejection position of this star is uncertain, 
as shown in 
\revise{Fig.~\ref{fig:WINEREDHVS1}(a).} 
However, the best orbit 
(shown in \revise{Fig.~\ref{fig:WINEREDHVS1}(b))}
can be traced back to very close to the Galactic center, 
showing that this star is a candidate of an HVS.\footnote{
\revise{
The systematic uncertainty in $r_\mathrm{ej}$ 
arising from the specific choice of the Galactic potential is $\sim 200 \pc$ 
(see \autoref{section:justification_for_condition_1_2}). 
Thus, if $r_\mathrm{ej} \lesssim 200 \pc$, 
the orbits are consistent with the Galactic center origin.} 
}

\autoref{table:kinematics} indicates 
that all of the metal-rich HVS candidates have 
ejection velocities of 
$v_\mathrm{ej} \simeq 500$--$600 \kms$
in their best orbits. 
This result suggests that 
they are bound to the Milky Way, 
because their $v_\mathrm{ej}$ values 
are lower than the escape velocity 
near the Galactic center 
(ignoring the gravity from the SMBH Sgr A*), 
$v_\mathrm{esc, GC} \simeq 700$-$800 \kms$.\footnote{
We checked some of the widely used Galactic potential models 
from literature \citep{McMillan2017,Piffl2014MNRAS.445.3133P, Portail2017MNRAS.465.1621P}. 
Among these models, the one in \cite{Portail2017MNRAS.465.1621P} has the smallest value of $v_\mathrm{esc, GC}=684 \kms$. 
}
Indeed, one candidate (mrHVSxp7) 
\revise{has $v_r<0$ (see Fig.~\ref{fig:mrHVSxp7to11}), 
which implies a bound orbit.} 
\revise{
We stress that, due to the bound nature of these stars, 
we lack a clear understanding of when exactly these metal-rich HVS candidates were ejected by the SMBH, assuming they are indeed genuine HVSs.
}

The only unambiguous HVS known to date 
is S5-HVS1, 
whose ejection velocity is estimated to be $\sim 1800 \kms$ 
\citep{Koposov2020MNRAS.491.2465K}. 
Compared to the ejection velocity of S5-HVS1, 
the ejection velocities of our metal-rich HVS candidates 
are considerably smaller. 
In terms of ejection velocity, our sample does not contain 
\revise{any unambiguous HVS candidates.}
An independent analysis by \cite{Marchetti2022MNRAS.515..767M} 
also reached the same conclusion 
(i.e., they did not find unambiguous HVSs in their analysis). 
However, 
we note that the ejection velocities of our metal-rich HVS candidates, 500--600 \kms, 
are consistent with the theoretical predictions for the ejection velocities of HVSs 
\citep{Bromley2006ApJ...653.1194B, Brown2015ARAA, Marchetti2018MNRAS.476.4697M}.

\subsection{Chemical properties of metal-rich HVS candidates}
\label{section:chemistry_of_HVSs}

Since our HVS candidates are bound to the Milky Way 
(unlike the unambiguous HVS known as S5-HVS1), 
the orbital properties alone 
are not informative enough to 
constrain the origin of these stars. 
Rather, given the small ejection rate of HVSs 
(\citealt{Brown2015ARAA}; 
see also our discussion in \autoref{section:introduction_HVS}), 
it would not be surprising 
if most of our candidates were non-HVSs. 
In this regard, chemical properties of our candidates, 
which are summarized in \autoref{table:chemistry}, 
are important to understand the origin of these stars.

\subsubsection{Metallicity and $\alpha$-abundance}

In the left column of Fig.~\ref{fig:fe_alpha}, we compare the metallicity and $\alpha$-abundance of our HVS candidates with spectroscopic abundances 
(\targetstar and mrHVSspec1-9) 
with a sample of reference stars from APOGEE DR17 
\revise{(see \autoref{appendix:nearly_radial_orbits})}. 
The top-left and middle-left panels indicate that the APOGEE stars with radial orbits primarily originate from two populations of stars. 
The first population 
is the Gaia-Enceladus-Sausage (GES), 
which is the remnants of the last major merger 
\citep{Belokurov2018MNRAS.478..611B, Helmi2018Natur.563...85H}. 
Because of the near head-on merger of the GES 
and the ancient Milky Way, 
the stars originated from GES have nearly radial orbits. 
The second component 
is the metal-poor tail of the high-$\alpha$ (disk) stars. 
This population may be a so-called Splash component, 
which are the stars in the pre-existing high-$\alpha$ disk 
that were kicked out of the disk due to the GES merger 
\citep{Bonaca2017ApJ...845..101B, Belokurov2020MNRAS.494.3880B}. 
These stars 
may also be kinematically hot tail of the high-$\alpha$ disk stars 
(see Fig.~15(a)(b) in \citealt{Hattori2025ApJ...980...90H}). 
Due to their eccentric orbits, 
these two populations may contaminate our metal-rich HVS candidates. 
Most of the radial orbit contaminants have metallicities  $\mathrm{[Fe/H]_{spec}}<-0.3$. 
In our selection, we identified 628 APOGEE stars with radial orbits and $\mathrm{[Fe/H]_{spec}}>-1$. 
Notably, among these stars, 626 (99.7\%) have $\mathrm{[Fe/H]_{spec}}<-0.3$. 
Our candidates mrHVSspec1-9 exhibit $\mathrm{[Fe/H]_{spec}}<-0.3$, 
similar to the majority of the reference APOGEE sample. 
Among mrHVSspec1-9, 
seven of them display high $\mathrm{[\alpha/Fe]_{spec}}>0.2$, suggesting they may be metal-poor tail of high-$\alpha$ stars; 
while 
two of them show low $\mathrm{[\alpha/Fe]_{spec}}<0.2$, indicating that they may belong to the metal-rich tail of the Gaia Enceladus Sausage. 
(The latter view is supported by the low-[Al/Fe] of these two stars, 
as we will discuss in \autoref{section:AlFe}.)
These results indicate that 
our candidates mrHVSspec1-9 are consistent with 
the majority of non-HVS reference stars, 
possibly associated with GES (mrHVSspec2-3) or Splash (mrHVSspec1 and mrHVSspec4-9). 
Among the 628 APOGEE stars, only 2 stars (0.3\%) have $\mathrm{[Fe/H]_{spec}}>-0.3$,\footnote{These two stars do not exhibit HVS-like orbits.} 
which indicates that it is very rare to find a non-HVS with high metallicity and low $\alpha$-abundance. 
This result provides a supporting evidence that 
\targetstar, which shows an exceptionally high metallicity, 
is an HVS ejected from the Galactic center. 
(Indeed, as we will discuss in \autoref{section:compare_with_mw}, 
the detailed chemical abundances of this star are consistent with 
the Galactic center population.)

In the right column of Fig.~\ref{fig:fe_alpha}, we compare the chemistry of XP-based metal-rich HVS candidates with a sample of (non-HVS) stars with reliable (\mhxp, \amxp) 
\revise{(see \autoref{appendix:nearly_radial_orbits})}. 
The top-right and middle-right panels confirm the presence of the same two populations of non-HVS stars with nearly radial orbits in the Gaia XP sample.\footnote{
In the reference APOGEE sample 
(blue dots in the middle-left panel in Fig.~\ref{fig:fe_alpha}), 
we see more GES-like stars than Splash-like stars. 
In contrast, in the reference Gaia XP sample 
(orange dots in the middle-right panel in Fig.~\ref{fig:fe_alpha}), 
we see more Splash-like stars than GES-like stars. 
We do not fully understand the origin for this difference, 
but it may be due to the sample selection effect, 
such that 
we see low Galactic latitude region for the APOGEE sample and 
we see high Galactic latitude region (with low dust extinction; \citealt{Hattori2025ApJ...980...90H}) for the Gaia XP sample. 
In this regard, it is interesting to point out that 
our candidates mrHVSspec1-9 and mrHVSxp2-13 
are selected from 
the catalog of \cite{Hattori2025ApJ...980...90H}, 
and we see more Splash-like HVS candidates than GES-like HVS candidates, 
similar to the distribution of orange dots in 
the middle-right panel in Fig.~\ref{fig:fe_alpha}. 
}
Again, the XP-based metal-rich HVS candidates show 
$\mathrm{[Fe/H]_{spec}}<-0.3$, 
except for \targetstar. 
In terms of the \amxp\ value, 
mrHVSxp3,9,13 may be possibly associated with the GES component, 
while 
mrHVSxp2,4-8,10-12 may be possibly associated with the Splash component.

Based on the distribution of stars in 
\revise{the metallicity and $\alpha$-abundance space,}
we conclude that 
\begin{itemize}
\item 
\targetstar is the most promising candidate for an HVS ejected from the Galactic center (see also \autoref{section:compare_with_mw}).  
\end{itemize}
The other metal-rich HVS candidates (mrHVSspec1-9 and mrHVSxp2-13) are less attractive, 
but we do not have strong evidence to 
conclude that they are non-HVSs. 
However, if we assume---as a working hypothesis---that 
they are non-HVSs, then our best guess for their origins are as follows:
\begin{itemize}
\item 
mrHVSspec2-3 (see also \autoref{section:AlFe}) and \\
mrHVSxp3,9,13 are possibly metal-rich tail of GES stars;
\item 
mrHVSspec1,4-9 and mrHVSxp2,4-8,10-12 are possibly metal-poor tail of Splash stars. 
\end{itemize}
Under this working hypothesis, 
we only find one HVS (candidate) from our Gaia DR3 catalog with Gaia XP spectra. 
The statistical implication of this result 
is discussed in \autoref{section:statistics}.

\subsubsection{[Al/Fe]-abundance}
\label{section:AlFe}

In understanding the origin of HVS candidates, 
the abundance ratio of [Al/Fe] is also informative. 
As we discussed earlier, 
one of the sources of contamination to the HVSs 
is the GES, 
which dominates the stellar halo 
within 20 kpc from the Galactic center. 
A distinct feature of GES stars is that 
they typically show negative [Al/Fe]  
\citep{Feuillet2021MNRAS.508.1489F}, 
unlike other Milky Way stars 
such as thin disk, thick disk, or bulge stars \citep{Das2020MNRAS.493.5195D, Feuillet2022ApJ...934...21F, Barbuy2023MNRAS.526.2365B, Nandakumar2024ApJ...964...96N}. 
Therefore, the [Al/Fe] ratio is 
informative to 
flag 
possible GES stars.

Among 22 HVS candidates, 
we have [Al/Fe] measurements for 7 stars, 
and two stars (mrHVSspec2 and mrHVSspec3) 
have [Al/Fe] $\lesssim -0.2$ 
(see \autoref{table:chemistry}). 
The [Al/Fe] values of these stars indicate that 
mrHVSspec2 and mrHVSspec3 are 
consistent with GES stars.\footnote{
Since we do not know the [Al/Fe] 
of stars in the immediate neighbor of the SMBH, 
low-[Al/Fe] nature of these stars 
does not necessarily mean that they are non-HVSs. 
However, their low-[Al/Fe] nature does 
flag that they may be GES stars.
}
Intriguingly, these two stars have 
$\mathrm{[\alpha/Fe]_{spec}} = 0.14$ and $0.12$, 
respectively, 
and have the lowest $\mathrm{[\alpha/Fe]_{spec}}$ 
among mrHVSspec1-9. 
In the $(\mathrm{[Fe/H]_{spec}}, \mathrm{[\alpha/Fe]_{spec}})$-space, 
these stars lie almost along 
the sequence of the metal-rich tail 
of the GES stars 
(see the lower-left panel of Fig.~\ref{fig:fe_alpha}). 
Both [Al/Fe] and $\mathrm{[\alpha/Fe]_{spec}}$ 
indicate that mrHVSspec2 and mrHVSspec3 
are consistent with GES stars. 
However, we stress that this argument needs to be handled 
with care, 
because currently we do not know the [Al/Fe] ratio 
for any stars in the Galactic center. 
For example, if future observations 
indicate that Galactic center stars show low-[Al/Fe], 
each of mrHVSspec2 and mrHVSspec3 
can be considered a good candidate for either an HVS or a GES star.

\begin{figure*}
\centering
\includegraphics[width=0.7\textwidth]{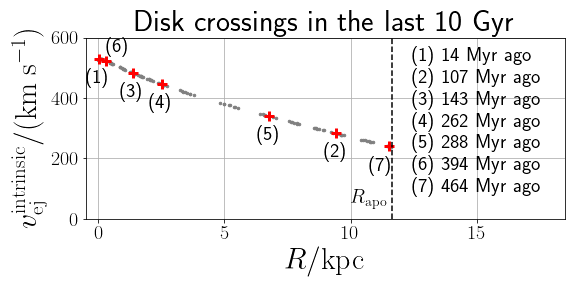}
\caption{
\revise{
Disk crossings of the best orbit of \targetstar\ in the last 10 Gyr. 
The $x$-axis shows the Galactocentric radius $R$ 
of the disk-crossing positions. 
The $y$-axis shows the disk-crossing velocity 
corrected for the streaming motion of the disk, $v_\mathrm{ej}^\mathrm{intrinsic}$. 
Gray dots represent disk crossings 
in the last 10 Gyr. 
Red pluses ($+$) represent disk crossings in the last 0.5 Gyr, 
enumerated according to the disk-crossing time. 
The vertical dashed line corresponds to the apocenter radius of the orbit, 
beyond which disk crossings do not occur. 
From this analysis, we see that 
low-velocity ejection 
($v_\mathrm{ej}^\mathrm{intrinsic}=200$--$400 \kms$) 
from the outer disk can kinematically explain the origin of \targetstar. 
(However, see also Fig.~\ref{fig:WINERED_and_disk_SiMg}.)
}
}
\label{fig:past_disk_crossings}
\end{figure*}

\begin{figure}
\centering
\includegraphics[width=0.48\textwidth]{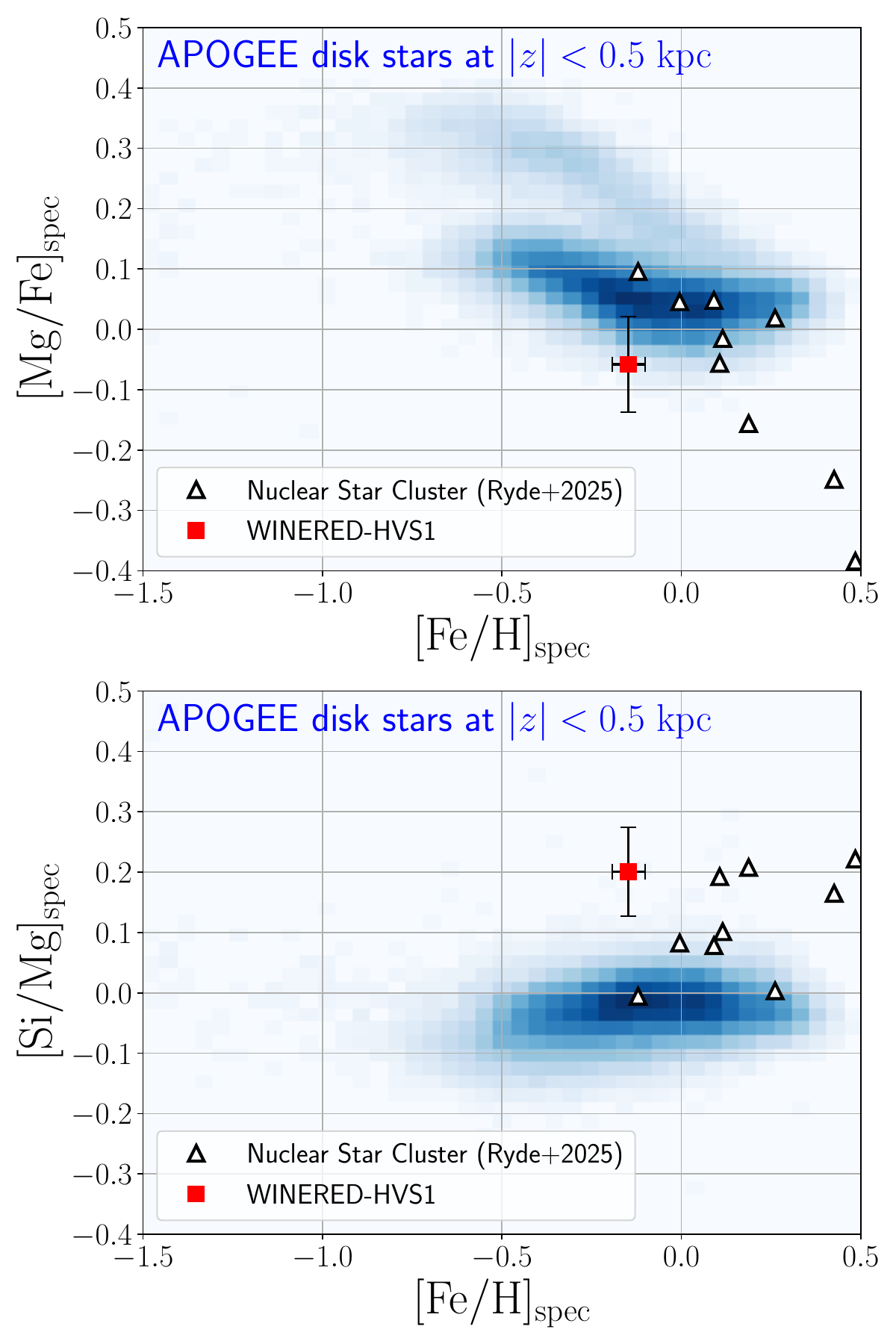}
%
\caption{
\revise{The chemistry of \targetstar\ 
(red data point) 
and the APOGEE red giant disk stars (blue density map; 
see \autoref{appendix:APOGEE_disk_stars}). 
For reference, the chemistry of NSC stars \citep{Ryde2025ApJ...979..174R} 
is shown by black triangles. 
The [Mg/Fe] abundance of \targetstar\ is consistent with the low-$\alpha$ disk, but its [Si/Mg] ratio differs significantly, indicating that this star was not ejected from the Galactic disk.
}
}
\label{fig:WINERED_and_disk_SiMg}
\end{figure}

\subsection{Orbital properties of WINERED-HVS1}

\revise{
As discussed in 
\autoref{section:chemistry_of_HVSs},
the high-metallicity and low-[$\alpha$/Fe] 
nature of \targetstar\ 
makes it the most promising HVS candidate. 
However, its [$\alpha$/Fe] is also similar 
to that of the thin disk (low-$\alpha$) stars 
(see top panels in Fig.~\ref{fig:fe_alpha}). 
Here we investigate the orbit of \targetstar\ 
to understand its origin. 
}

\subsubsection{Ejection of WINERED-HVS1 from the Galactic disk}
\label{section:disk_ejection}

\revise{
So far, 
we have focused on the orbits in the last 200 Myr 
and selected HVS candidates whose last disk crossings occurred 
near the Galactic center. 
In this section, 
we adopt a longer integration time of 10 Gyr 
and explore the past orbit of \targetstar, 
to assess the possibility 
that this star was not ejected by the SMBH 
but ejected elsewhere in the Galactic disk. 
To this end, 
we focus on the best orbit of \targetstar\ 
(evaluated with Method A) 
shown in Fig.~\ref{fig:WINEREDHVS1}(b). 
The adopted integration time (10 Gyr) 
is motivated by its stellar age
(0.5--10 Gyr; see \autoref{section:mass_age}).
}

\revise{
At each disk-crossing event, 
we record the disk-crossing time $t_\mathrm{ej}$, position $\vector{x}_\mathrm{ej}$ and velocity $\vector{v}_\mathrm{ej}$, 
as in \autoref{section:backward_orbit}. 
Since we are interested in the Galactic disk ejection, 
we derive the intrinsic ejection velocity 
\eq{
\vector{v}_\mathrm{ej}^\mathrm{intrinsic} = \vector{v}_\mathrm{ej}-\vector{v}_\mathrm{circ} (\vector{x}_\mathrm{ej}), 
}
where 
$\vector{v}_\mathrm{circ}$ 
corresponds to the velocity of a circular orbit at $\vector{x}_\mathrm{ej}$. 
This intrinsic ejection velocity 
represents the disk ejection velocity corrected for the streaming motion of the disk (see \citealt{Hattori2019ApJ...873..116H}). 
(For example, if the star was ejected from a star cluster in the disk, $\vector{v}_\mathrm{ej}^\mathrm{intrinsic}$ is the ejection velocity relative to the cluster.)}

\revise{
Fig.~\ref{fig:past_disk_crossings} 
shows the disk crossings of the best orbit of \targetstar\ in the last 10 Gyr. 
The vertical axis represents the intrinsic ejection velocity 
$v_\mathrm{ej}^\mathrm{intrinsic} = |\vector{v}_\mathrm{ej}^\mathrm{intrinsic}|$, 
and the horizontal axis represents the Galactocentric radius 
at each disk crossing ($R=|\vector{x}_\mathrm{ej}|$). 
Since the orbit has a nearly zero azimuthal angular momentum 
and we adopt a static and axisymmetric Galactic potential 
\citep{McMillan2017}, 
we see a clear sequence of $v_\mathrm{ej}^\mathrm{intrinsic}$ 
as a function of $R$. 
This sequence has a monotonically decreasing profile, 
which has a maximum near $R=0 \kpc$ 
and is truncated at 
the apocenter of this orbit, 
$R=R_\mathrm{apo}\simeq 11.6 \kpc$.  
At the last disk crossing event (14 Myr ago), 
the orbit crossed the disk plane near the Galactic center 
as described in Fig.~\ref{fig:WINEREDHVS1}. 
Over the past 10 Gyr, 
the orbit crossed within 200 pc from the Galactic center 15 times. 
At these disk crossings, 
the values of $v_\mathrm{ej}^\mathrm{intrinsic}$ 
(and $v_\mathrm{ej}$) are nearly identical, 
and satisfy 
$v_\mathrm{ej}^\mathrm{intrinsic} \simeq v_\mathrm{ej} \gtrsim 500 \kms$. 
}

\revise{
Since $v_\mathrm{ej}^\mathrm{intrinsic}(R)$ 
is well-behaved, 
we focus on the disk crossing events 
that occurred in the last 0.5 Gyr, marked by red pluses 
in Fig.~\ref{fig:past_disk_crossings}. 
We note that 0.5 Gyr corresponds to 
the lower bound for \targetstar's age. 
As shown in the figure, 
the orbit could have crossed the disk plane 
7 times in the past 0.5 Gyr 
(if the star was ejected more than 0.5 Gyr ago). 
Disk crossings (1) and (6) in the figure 
occurred near the Galactic center 
and show $v_\mathrm{ej}^\mathrm{intrinsic} > 500 \kms$. 
In contrast, 
disk crossings (2), (5), and (7) 
occurred at $6.8 \kpc \lesssim R \lesssim 11.5 \kpc$ 
and show 
$240 \kms \lesssim v_\mathrm{ej}^\mathrm{intrinsic} \lesssim 340 \kms$. 
These findings suggest that, 
if this star was ejected from the outer part of the disk, 
the required ejection velocity would be relatively small. 
}

\revise{
In general, 
a disk star can be ejected 
either by a `binary ejection mechanism' 
through the supernova explosion of the binary companion, 
or by the `dynamical ejection mechanism' 
through a few-body (i.e., 3- or 4-body) interaction. 
Although the ejection of a 1--3$\msun$ star 
(such as \targetstar) 
with $v_\mathrm{ej}^\mathrm{intrinsic} > 500 \kms$ 
is only possible in idealized environments, 
ejection with $v_\mathrm{ej}^\mathrm{intrinsic} = 200$--$400 \kms$ 
is viable through both binary ejection  
\citep{Tauris2015MNRAS.448L...6T} 
and dynamical ejection 
\citep{Leonard1991AJ....101..562L, Perets2012ApJ...751..133P}.}

\revise{
In this regard, the ejection of \targetstar\ 
from the (outer) Galactic disk might provide an alternative explanation for the origin of this star. 
If we hypothesize that 
the star was ejected from the Galactic disk, 
we would expect it to share chemical similarities with disk stars. 
To test this hypothesis, 
we compared the detailed chemical abundances of this star 
with those of disk stars. 
Fig.~\ref{fig:WINERED_and_disk_SiMg} 
shows the distribution of chemical abundances of 
red giant disk stars 
located at $|z|<0.5 \kpc$ from the disk plane. 
The top panel indicates that 
the [Fe/H] and [Mg/Fe] values of \targetstar\ 
are consistent with those of low-$\alpha$ disk stars, 
although its [Mg/Fe] is slightly lower 
than that of typical low-$\alpha$ disk stars. 
In contrast, the bottom panel shows that \targetstar\ has notably higher [Si/Mg] 
than the disk stars. 
This difference persists even when 
we further select disk stars based on their Galactocentric radius. 
These findings indicate that the chemical properties of \targetstar\ 
are distinct from those of disk stars, 
suggesting that it was not ejected from the Galactic disk. 
For reference, Fig.~\ref{fig:WINERED_and_disk_SiMg} also shows the chemical abundances of NSC stars. 
We find a chemical similarity between the NSC stars and \targetstar, 
which will be further discussed in \autoref{section:compare_with_mw} and Fig.~\ref{fig:WINERED_and_inner_mw}. 
}

\begin{figure}
\centering
\includegraphics[width=0.49\textwidth]{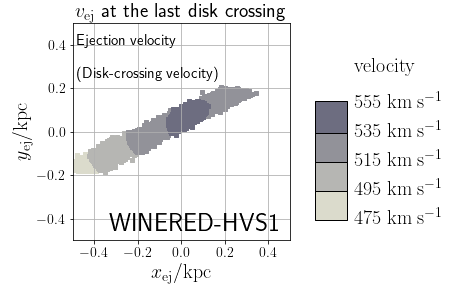}
\caption{
\revise{
The ejection velocity 
(disk-crossing velocity) $v_\mathrm{ej}$ of \targetstar\ 
as a function of the ejection position $(x_\mathrm{ej}, y_\mathrm{ej})$ 
for our $10^4$ Monte Carlo orbits 
evaluated with Method A. 
Here we only consider the last disk crossing, 
which occurred $\sim 14$ Myr ago. 
Based on these results, 
we infer that the ejection velocity of \targetstar\ 
is around $v_\mathrm{ej} \simeq 500$--$600 \kms$ 
if it originated from the Galactic center. 
}
}
\label{fig:vej_vint}
\end{figure}

\subsubsection{Ejection of WINERED-HVS1 from the Galactic center}

\revise{
Here we investigate the physical conditions 
under which \targetstar\ could have been ejected from the Galactic center.
Since the ejection velocity $v_\mathrm{ej}$ 
does not depend on the flight time 
(as long as the Galactic potential is nearly static), 
we once again use the $10^4$ Monte Carlo orbits in the last 200 Myr 
in this section. 
}

\revise{
Fig.~\ref{fig:vej_vint} shows 
the distribution of 
the ejection position 
$(x_\mathrm{ej},y_\mathrm{ej})$ 
of the Monte Carlo orbits of \targetstar, 
color-coded by 
the ejection velocity $v_\mathrm{ej}$. 
Since the Galactic potential is deeper 
at smaller Galactocentric radius, 
$v_\mathrm{ej}$ tends to be larger for 
smaller $r_\mathrm{ej}$. 
The value of $v_\mathrm{ej}$ 
range from $481 \kms$ to $539 \kms$; 
while the best orbit has an ejection velocity 
of $v_\mathrm{ej} = 537 \kms$. 
}

\revise{
In \autoref{section:simulation}, 
we conduct Monte Carlo simulations 
of HVSs with ejection velocities of 
$v_\mathrm{ej} \simeq 500$--$600 \kms$ 
arising from disrupted binary systems 
through Hills mechanism. 
We find that 
the ejection of \targetstar\ can be reproduced 
if the semi-major axis $a$ and 
the periapsis distance $\rperi$ 
of the progenitor binary 
fall within a reasonable range, 
namely 
$a \simeq 0.5$--$4 \AU$  
and $\rperi \simeq 30$--$300 \AU$ 
(see Fig.~\ref{fig:MonteCarlo}). 
}

\revise{
The ejection of 
\targetstar\ can be explained 
by two scenarios:
(i) it was ejected as a red giant; 
or 
(ii) it was ejected as a main-sequence star 
which later evolved into a red giant. 
In the case (i), 
it may or may not be a recently-ejected HVS. 
If this star is a recently-ejected HVS, 
its flight time is $\sim 14 \Myr$. 
In the case (ii), 
it may not be a recently-ejected HVS. 
This is because a main-sequence star with a mass of 1--3$\msun$ 
requires more than 20 Myr to evolve into a red giant 
similar to \targetstar. 
This timescale exceeds the time since the last disk crossing (14 Myr), 
implying that the star must have undergone at least one, 
and possibly multiple, disk crossings after ejection 
(see Appendix~\ref{section:simulation_summary}).}

\revise{
To summarize, the ejection of \targetstar\ 
by the SMBH is a viable scenario, 
which is reassuring since we are searching for HVSs ejected from the Galactic center. 
In the following section, 
we discuss the origin of \targetstar\ 
from a chemical perspective.}

\subsection{Chemistry of \targetstar compared to other stars 
in the Milky Way and beyond}
\label{section:compare_with_mw}

\revise{Here} we compare the chemistry of \targetstar\ 
with that of stars in 
the NSC, NSD, inner bulge, and dwarf galaxies. 
Throughout this section, 
we utilize 
[X/Y] instead of $\mathrm{[X/Y]_{spec}}$ for brevity, 
since we only deal with spectroscopically determined chemical abundances.

\noindent{(1) \it{Nuclear Star Cluster (NSC)}}

The SMBH at the center of the Milky Way is 
surrounded by a dense 
star cluster known as the NSC, 
whose radius is $\simeq 4 \pc$ \citep{Schodel2014CQGra..31x4007S}. 
According to the Hills mechanism, 
when the SMBH disrupts a binary star system 
at a typical radius of 
$\sim 100 \;\mathrm{AU} \simeq 5 \times 10^{-4} \pc$, 
it ejects one star as an HVS 
while capturing the companion star as an S-star. 
Since HVS ejections occur at the central part of the NSC, 
we expect a chemical similarity between HVSs and NSC stars.

\begin{figure*}
\centering
\includegraphics[width=0.24\textwidth]{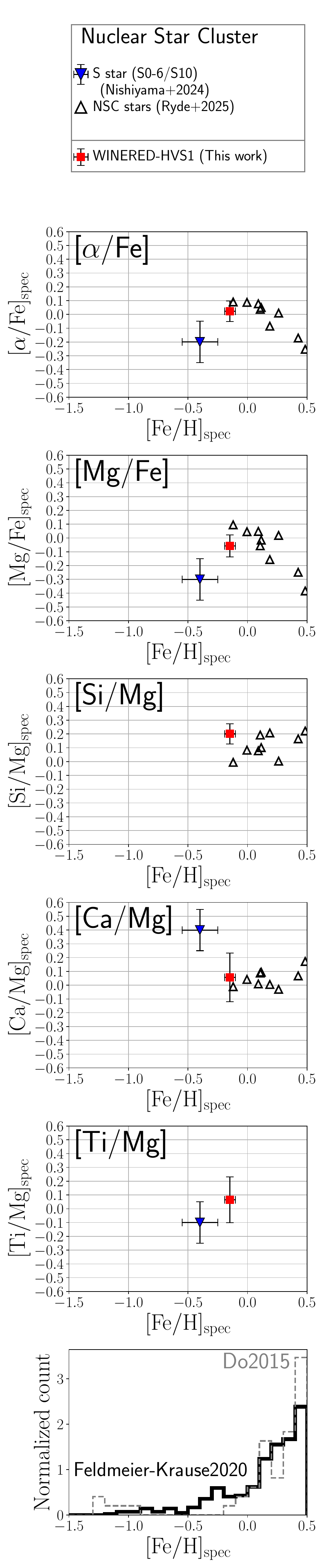}
\includegraphics[width=0.24\textwidth]{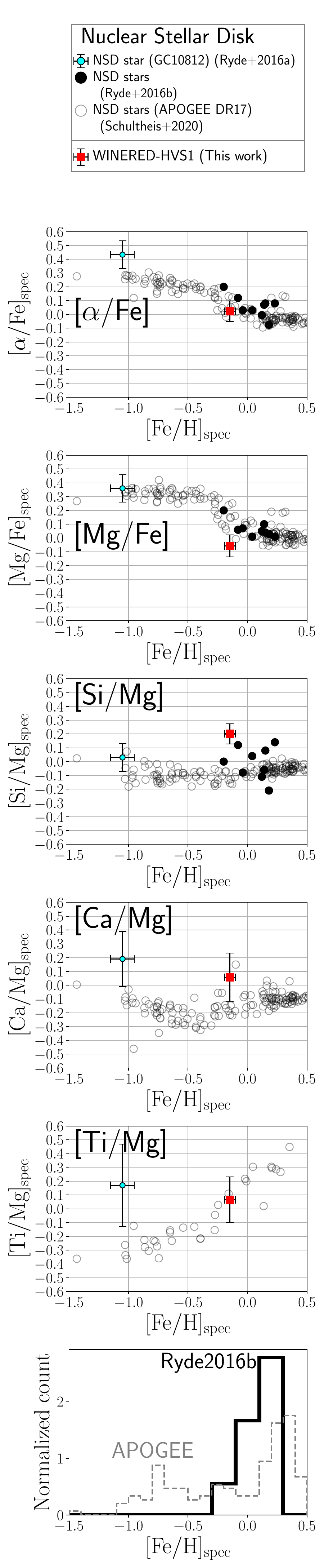}
\includegraphics[width=0.24\textwidth]{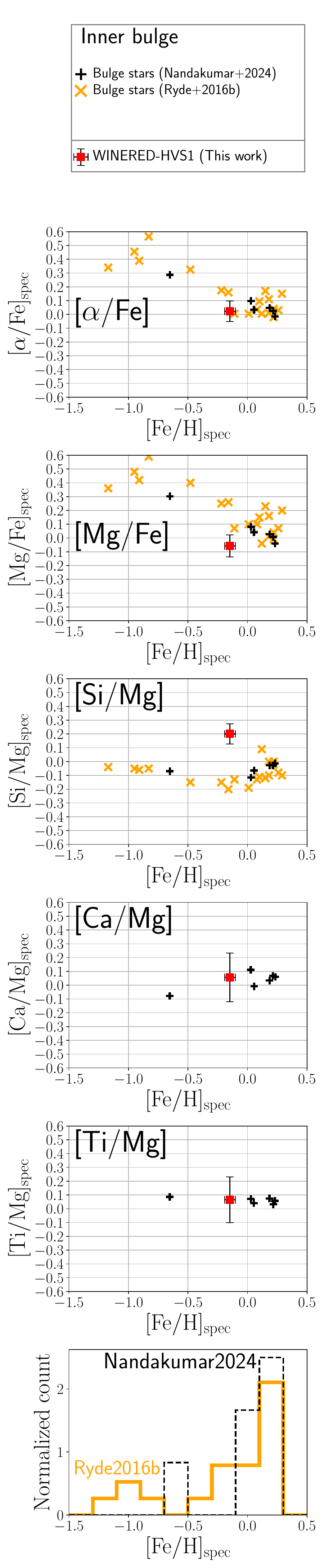}
\includegraphics[width=0.24\textwidth]{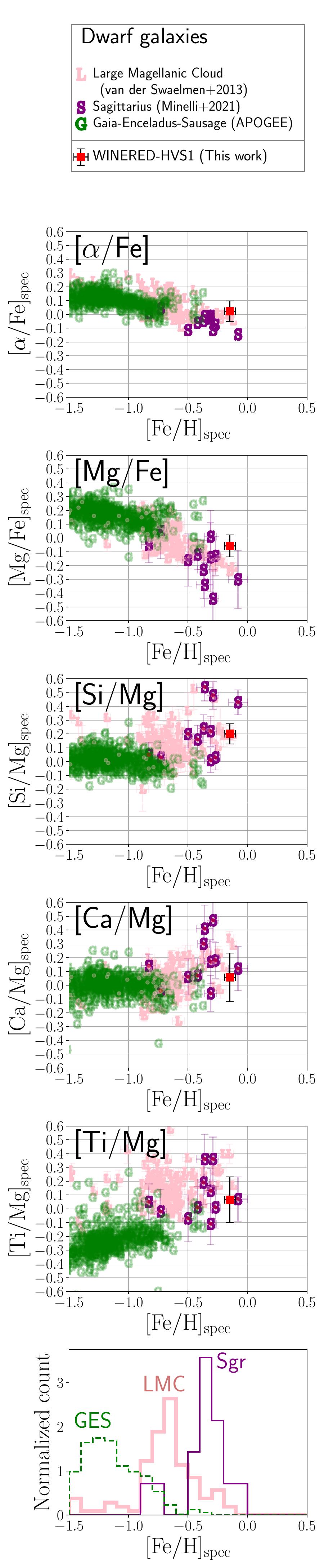} 
\caption{
Chemistry of \targetstar compared with 
stars in the NSC (left column), 
NSD (second column), 
inner bulge (third column), and 
dwarf galaxies (fourth column). 
The metal-rich, low-[Mg/Fe] and high-[Si/Mg] 
nature of \targetstar 
is most consistent with the NSC stars. 
Intriguingly, 
\targetstar is chemically similar to 
the metal-rich stars in massive dwarf galaxies; 
and it might have formed in the core of 
massive dwarf galaxy 
that merged to the Milky Way. 
}
\label{fig:WINERED_and_inner_mw}
\end{figure*}

The observed chemical abundances for NSC stars 
(including an S-star) 
are summarized in the left column of Fig. \ref{fig:WINERED_and_inner_mw}. 
As seen from the metallicity distribution of the NSC stars 
(bottom panel; \citealt{FeldmeierKrause2020MNRAS.494..396F,Do2015ApJ...809..143D}), 
the majority of stars in the NSC have super-solar metallicity, 
while more than $\sim 10$\% of the stars show sub-solar metallicity. 
The observed metallicity of 
\revise{$\mathrm{[Fe/H]} = -0.147$}
for \targetstar 
is within the metallicity range of NSC stars. 
Also, the values of [$\alpha$/Fe], [Mg/Fe], [Si/Mg], and [Ca/Mg] for \targetstar 
are within the range of the nine NSC red giants 
at 0.3--4 pc 
from the SMBH 
analyzed by \cite{Ryde2025ApJ...979..174R} 
(black open diamonds). 
Therefore, broadly speaking, 
the chemistry of \targetstar is consistent with 
that of NSC stars. 
Of course, a more detailed comparison with larger number of NSC stars 
is required to support this view. 
For example, 
the nine red giant stars analyzed by \cite{Ryde2025ApJ...979..174R} 
include two stars with solar or slightly sub-solar [Fe/H]. 
Although these two stars have similar metallicity to \targetstar, 
they have slightly higher [Mg/Fe] and slightly lower [Si/Mg] 
than \targetstar. 
It is unclear if these differences are serious or benign, 
especially because 
(i) we have only two NSC stars to make a comparison; 
(ii) [Mg/Fe] and [Si/Mg] for the NSC stars and \targetstar are associated with non-negligible 
uncertainties ($\sim 0.1$ dex); and 
(iii) these two stars and \targetstar are observed and analyzed 
in a different manner, and hence subject to a possible systematic bias in the zero point.

The blue upside-down triangle in 
the left column of Fig.~\ref{fig:WINERED_and_inner_mw} 
shows the chemistry of S0-6/S10. 
This star is the only S-star for which detailed chemistry is known 
\citep{Nishiyama2024PJAB..100...86N}, 
\revise{and its projected distance from the SMBH is only 0.012 pc.} 
The chemistry of the nine NSC stars in \cite{Ryde2025ApJ...979..174R} 
is not particularly similar to 
that of S0-6/S10, 
although these nine red giants, along with the S0-6/S10, 
are all located within the NSC. 
This difference might indicate the chemical inhomogeneity within the NSC \citep{FeldmeierKrause2020MNRAS.494..396F}. 
Intriguingly, in the ([Fe/H], [Mg/Fe])-space, 
\targetstar is located in between 
the S0-6/S10 and the sample in \cite{Ryde2025ApJ...979..174R}. 
Since we only have one S-star and one HVS candidate, 
we can not draw a useful conclusion 
as to whether the HVSs and S-stars have similar chemical properties 
as a population.

\noindent{(2) \it{Nuclear Stellar Disk (NSD)}}

NSD is a rotationally supported stellar disk 
\revise{at $R \lesssim 220 \pc$} 
\citep{Nishiyama2013ApJ...769L..28N, Schultheis2021A&A...650A.191S}. 
\cite{FeldmeierKrause2022MNRAS.513.5920F}
studied the chemical abundances of stars within the NSD 
to find that NSC and NSD show different chemistry, 
such that NSD stars are typically more metal-poor than NSC stars.

The chemical properties of NSD is summarized in the second column in 
Fig.~\ref{fig:WINERED_and_inner_mw}. 
The black filled circles show the NSD stars 
analyzed by \cite{Ryde2016AJ....151....1R}. 
The open gray circles show the NSD stars in APOGEE 
selected by \cite{Schultheis2020A&A...642A..81S}. 
In the original study of \cite{Schultheis2020A&A...642A..81S}, 
they used the chemical abundances in APOGEE DR16, 
while in this figure we adopt the more recent values from APOGEE DR17. 
The cyan circle is a metal-poor NSD star known as GC10812,\footnote{
The projected distance of this star from the SMBH is $1.5 \pc$, 
but \cite{Ryde2016ApJ...831...40R} and \cite{Ryde2025ApJ...979..174R} 
interpret this star as a NSD star, 
based on its kinematics and photometric information. 
} 
whose metallicity almost corresponds to the lowest metallicity of 
the NSD sample in \cite{Schultheis2020A&A...642A..81S}. 
The two NSD samples from 
\cite{Ryde2025ApJ...979..174R}
and 
\cite{Schultheis2020A&A...642A..81S} 
show a broadly consistent profiles of 
$\mathrm{[\alpha/Fe]}$, [Mg/Fe] and [Si/Mg] as a function of [Fe/H]. 
Notably, the value of [Si/Mg] for the sample in \cite{Schultheis2020A&A...642A..81S} 
is almost insensitive to [Fe/H]. 
This trend is also true for the sample of \cite{Ryde2016AJ....151....1R}, 
although we see a larger scatter for this sample. 
Interestingly, the low-[Si/Mg] nature of the NSD stars 
seems to be inconsistent with 
the high [Si/Mg] ratio observed for \targetstar. 
Also, 
the observed [Mg/Fe] ratio of \targetstar 
is slightly smaller than [Mg/Fe] 
of both \cite{Ryde2016AJ....151....1R} sample 
and \cite{Schultheis2020A&A...642A..81S} sample. 
The offset in [Si/Mg] (and a marginal offset in [Mg/Fe]) 
may indicate that the \targetstar is not chemically consistent with 
the stellar population in the NSD.

\noindent{(3) \it{Inner bulge}}

We further investigate the chemical properties of the inner bulge 
\revise{(at $R \lesssim 300 \pc$),} 
which are 
\revise{shown} 
in the third column of Fig.~\ref{fig:WINERED_and_inner_mw}. 
Here, the black plus and orange cross correspond to the inner bulge samples 
taken from 
\cite{Nandakumar2024ApJ...964...96N}
and 
\cite{Ryde2016AJ....151....1R}, 
respectively. 
Most notably, 
the inner bulge stars show low values of [Si/Mg], 
which is in contrast to the high-[Si/Mg] ratio seen in \targetstar. 
This discrepancy indicates that 
the \targetstar is chemically distinct from inner bulge stars.

\noindent{(4) \it{Dwarf galaxies}}

The [Fe/H] and [$\alpha$/Fe] of \targetstar 
resembles that of stars found at the metal-rich end of massive dwarf galaxies.
Motivated by this fact, we compare \targetstar 
with stars in dwarf galaxies and GES (remnants of a disrupted dwarf galaxy). 
In the right-most column in Fig.~\ref{fig:WINERED_and_inner_mw}, 
we show the chemical properties of stars in 
the LMC (pink `L'; \citealt{Swaelmen2013A&A...560A..44V}), 
Sagittarius dwarf galaxy (purple `S'; \citealt{Minelli2021ApJ...910..114M}), 
and 
the GES (green `G'; 
\revise{see \autoref{appendix:nearly_radial_orbits}}).

The stars in the GES are not as metal-rich as \targetstar, 
\revise{possibly} due to the quenched star formation in the GES 
when it merged to the Milky Way 
\citep{Ernandes2024A&A...691A.333E}. 
\revise{The chemistry of GES stars indicates} 
that \targetstar is not 
\revise{a GES star.}

The other massive, surviving dwarf galaxies in this plot 
(LMC and Sagittarius) 
are large enough to maintain star formation activity for a long duration of time. 
Therefore, some stars located in the central part of the LMC and Sagittarius dwarf galaxy are as metal-rich as \targetstar. 
For example, the Sagittarius dwarf galaxy has a NSC known as M54 \citep{AlfaroCuello2019ApJ...886...57A}, in which young ($\sim 1 \Gyr$ old) and metal-rich stars are confirmed. 
Although \targetstar is metal-rich, 
it is unclear if this star is as young as $\sim 1 \Gyr$ old. 
(Our current guess of its age is 
0.5--10 Gyr based on the CMD in Fig.~\ref{fig:WINEREDHVS1} and quite uncertain.) 
However, if \targetstar is indeed young, 
we see a similarity between \targetstar 
and the metal-rich stars in massive dwarf galaxies. 
In such a case, 
\targetstar might have formed in the NSC of a massive dwarf galaxy. 
If this star was formed in the NSC of a massive dwarf galaxy, and yet it was ejected from the Galactic center, we need to speculate a scenario to bring this star from its progenitor dwarf galaxy to the Galactic center.

It is unrealistic to suggest that this star was formed in the Sagittarius (or LMC) and was later brought to the Milky Way's NSC because the apocenter distance of this star is too small to have originated from these 
surviving 
dwarf galaxies. 
Therefore, it is more likely that \targetstar was formed in the NSC of a hypothetical massive dwarf galaxy that later merged with the Milky Way.  
If this hypothetical dwarf galaxy was massive enough to host an NSC, the dynamical friction might have brought the dwarf galaxy and its  NSC to the Galactic center.
Then the dwarf galaxy's NSC can merge with the Milky Way's NSC.\footnote{
If this scenario holds true, it is possible to propose an alternative scenario in which \targetstar was formed in a Young Massive Cluster \citep{PortegiesZwart2010ARAA..48..431P} within the stellar disk. 
If the progenitor cluster was later dragged to the Galactic center due to the dynamical friction and merged to the NSC, 
\targetstar could be ejected by the SMBH. 
However, this scenario cannot account for the origin of \targetstar, as stars in Young Massive Clusters typically exhibit chemistry similar to that of low-$\alpha$ disk stars, which is characterized by a low value of [Si/Mg] 
\revise{(see Fig.~\ref{fig:WINERED_and_disk_SiMg})} 
and inconsistent with the high value of 
\revise{$\mathrm{[Si/Mg]}= 0.201^{+0.073}_{-0.074}$} 
for this star.
}

If \targetstar was formed before the NSC merger, it could reflect the chemistry of the dwarf galaxy's NSC. 
Also, if \targetstar was brought to the center of the Milky Way's NSC, it could be ejected as a Galactic HVS. 
Indeed, the Milky Way's NSC exhibits an asymmetric spatial distribution of sub-solar metallicity stars \citep{FeldmeierKrause2020MNRAS.494..396F}, 
which can be interpreted as a result of the merger of NSCs within the last $\sim 3 \Gyr$---a time frame recent enough to retain the dynamical relics 
\citep{ArcaSedda2020ApJ...901L..29A}.

There are several caveats to our scenario. 
If the merger with this hypothetical dwarf galaxy occurred recently and if that dwarf galaxy was sufficiently massive to host an NSC, 
we should expect to see the stellar stream of its merger debris in the halo. 
Thus, the merger must have happened a long time ago ($\sim 10 \Gyr$). 
In this context, a hypothetical dwarf galaxy known as Kraken (also known as Heracles), 
which is a dwarf galaxy with a stellar mass of $\sim 10^9 \msun$ claimed to contribute to the central part of the Milky Way \citep{Kruijssen2019MNRAS.486.3180K, Horta2021MNRAS.500.1385H}, 
might be a good candidate for the progenitor dwarf galaxy of \targetstar. 
Although the Kraken dwarf galaxy is predominantly composed of 
metal-poor and high-$\alpha$ stars ($\mathrm{[Fe/H]}\simeq -1$ and $\mathrm{[Mg/Fe]}\simeq 0.3$; \citealt{Horta2023MNRAS.520.5671H}), 
its NSC could be as metal-rich and low-$\alpha$ as \targetstar. 

Assuming the progenitor dwarf galaxy (possibly Kraken) merged with the Milky Way $\sim 10 \Gyr$ ago, 
we need to explore a mechanism that would gradually bring its NSC to the Galactic center over a timescale of $\sim 7 \Gyr$ 
(if we presume the NSC merger occurred $3 \Gyr$ ago). 
According to the simulations in \cite{ArcaSedda2020ApJ...901L..29A}, 
some models classified as `late infalls' satisfy this condition. 
Therefore, while not entirely convincing, it may be possible for \targetstar to have formed in the NSC of a massive dwarf galaxy and been ejected from the Galactic center.

More detailed chemical abundances of \targetstar, 
such as its abundances of $r$-process elements, 
are required to fully understand 
\revise{its} birth environment.\footnote{
In this context, it is important to mention the work by \cite{Nishiyama2024PJAB..100...86N}. 
According to their study, 
the S-star known as S0-6/S10 is more than 10 Gyr old 
and its [Fe/H] and [$\alpha$/Fe]  
are similar to some metal-rich stars in massive dwarf galaxies 
such as LMC and Sagittarius. 
We note that metal-rich stars in massive dwarf galaxies 
have a relatively young age ($\sim 1 \Gyr$ old), 
and therefore stars like S0-6/S10, 
namely old and metal-rich stars, 
cannot be formed in large dwarf galaxies. 
For our HVS candidate \targetstar, 
the stellar age is around 0.5--10 Gyr. 
Thus, in case it is young ($\sim 1$ Gyr old), 
this star could be more or less similar to the metal-rich stars in massive dwarf galaxies. 
}

\subsection{Future prospect}

\subsubsection{Use of other large catalogs}

In this work, 
\revise{we use the catalog of chemical abundances 
in \cite{Hattori2025ApJ...980...90H} 
to search for metal-rich HVSs, 
because it is the largest catalog 
containing both \mhxp\ and \amxp.}
In the hindsight, however, 
the metallicity data 
is more important than $\alpha$-abundance data 
in identifying HVS candidates. 
Thus, it may be beneficial to utilize 
existing large catalogs of metallicity, 
such as those by \cite{Andrae2023ApJS..267....8A} and \cite{Zhang2023MNRAS.524.1855Z}. 
This approach could help in identifying metal-rich HVS candidates, even if the catalogs do not include $\alpha$-abundance information.

Several  
stellar surveys, 
including 
WEAVE \citep{Jin2023MNRAS.tmp..715J}, 
4MOST \citep{deJong2019Msngr.175....3D}, and 
PFS \citep{Takada2014PASJ...66R...1T}, 
are measuring the line-of-sight velocities for millions of stars. 
These new catalogs 
will allow us identify more HVS candidates.

\subsubsection{Longer integration time}

In this paper, we focused on 
\revise{the stellar orbits in the last 200 Myr 
to minimize orbital uncertainties. 
However, since we may have missed some HVSs 
whose last disk crossings occurred far from the Galactic center, it could be interesting to 
search for other bound HVSs by adopting a longer integration time.}

\subsubsection{Optical view of the Galactic center environment using hyper-velocity stars}

In \autoref{section:WINERED}, 
we use the near-infrared \textit{YJ}-band spectrum 
to derive the chemical abundances of \targetstar. 
As this star is moving in the halo region 
(see Fig.~\ref{fig:WINEREDHVS1}), 
where dust extinction is minimal, 
we can obtain an optical spectrum for this star. 
The optical spectrum contains 
absorption lines of several heavy elements 
(such as Eu and other $r$-process elements)  
that are not available in an infrared spectrum with measurable strengths.  
By analyzing the chemical abundances of \targetstar from its optical spectrum, 
we may gain insights into the chemical landscape near the Galactic center.

Traditionally, 
understanding the Galactic center environment 
has been challenging due to the high levels of dust in that direction, 
which demands extensive telescope time, especially in the near-infrared \textit{K} band. 
By searching for many more HVSs and 
by characterizing their chemical abundances 
with optical spectra, 
we will open a new window into the Galactic center environment 
without being hindered by dust obscuration.

\section{Summary and conclusion}
\label{section:conclusion}

By using astrometric data from Gaia DR3, 
photometric distance from StarHorse catalog \citep{Anders2022AandA...658A..91A}, 
and \revise{(\mhxp, \amxp)} 
from \cite{Hattori2025ApJ...980...90H}, 
we searched for candidates of metal-rich HVSs
\revise{ejected from the Galactic center.} 
After selecting stars with some quality cut, 
we searched for stars 
that are kinematically consistent with HVSs ejected from the Galactic center,  
by using 
four different sets of assumptions on 
the Galactic potential 
and on the Solar position/velocity. 
We find 74 kinematically selected 
metal-rich HVS candidates 
that are robust against the different assumptions.

We further pruned the sample by using the chemical \revise{data} 
to identify 22 metal-rich HVS candidates, 
all of which 
\revise{were} red giants (see Fig.~\ref{fig:flow_chart_chemistry}). 
Unlike the only unambiguous HVS, S5-HVS1 \citep{Koposov2020MNRAS.491.2465K}, 
our candidates are gravitationally bound to the Milky Way, 
and their ejection velocities are $500$--$600 \kms$ 
(see \autoref{table:kinematics}). 
Thus, chemical characterizations of these stars is essential for understanding their origins.

Among these 22 candidates, 9 HVS candidates (labeled as mrHVSspec1-9) 
have spectroscopic abundances derived from large surveys 
(APOGEE, GALAH, RAVE, and LAMOST; see \autoref{table:chemistry}). 
The remaining 13 HVS candidates (labeled as \targetstar and mrHVSxp2-13) 
lack existing spectroscopic abundances, 
but have relatively reliable estimates of (\mhxp, \amxp) 
from Gaia XP spectra \citep{Hattori2025ApJ...980...90H} 
(see \autoref{table:chemistry}).

Based on the metallicity and $\alpha$-abundance of these stars,
we cannot ignore the possibility that 21 stars out of our 22 candidates may be non-HVSs. 
Namely, these 21 stars might be part of 
either the metal-rich tail of the GES (which is the remnants of the ancient major merger; 
\citealt{Helmi2018Natur.563...85H, Belokurov2018MNRAS.478..611B}), 
or the metal-poor tail of the high-$\alpha$ disk 
(which may be so-called Splash stars 
kicked out of the pre-existing disk when GES merged to the ancient Milky Way; 
\citealt{Bonaca2017ApJ...845..101B, Belokurov2020MNRAS.494.3880B}). 
In contrast to these 21 stars, 
one candidate, \targetstar, 
has a near-solar value of \mhxp, 
making it unlikely to be classified as either a GES star or a Splash star.

To investigate this star further, 
we conducted a follow-up observation 
with Magellan/WINERED (see \autoref{section:WINERED}). 
Through a careful analysis of its high-resolution spectrum in the near-infrared \textit{YJ} bands, 
we determined the metallicity [Fe/H] as well as 
the abundances for Na, Mg, Si, S, K, Ca, Ti, Cr, Ni, and Sr 
(see \autoref{table:stellar_abn_all_res}). 
We obtained the metallicity and $\alpha$-abundance 
\revise{
($\mathrm{[Fe/H]}=-0.147^{+0.046}_{-0.045}$, 
$\mathrm{[Mg/Fe]}=-0.058^{+0.077}_{-0.082}$)
},
which are consistent with \revise{(\mhxp, \amxp).}

\revise{
Since \targetstar\ has a bound orbit  
and it is not necessarily young, 
there is a possibility that 
this star was ejected from the (outer) Galactic disk. 
However, the observed value of [Si/Mg] 
determined from our follow-up observations 
is distinct from that of Galactic disk stars, 
making this scenario unlikely (see Fig.~\ref{fig:WINERED_and_disk_SiMg}). 
This analysis supports our view that 
the star was ejected from the Galactic center. 
Indeed, through Monte Carlo simulations (see \autoref{section:simulation}), 
we find that the ejection of \targetstar\ 
with a velocity of $v_\mathrm{ej}=500$--$600 \kms$ 
via Hills mechanism involving the SMBH (Sgr A*) 
is dynamically viable. 
}

To better understand the origin of this star, 
we compared the chemical properties of \targetstar with those of other stars in the Milky Way and beyond (see Fig.~\ref{fig:WINERED_and_inner_mw}). 
We found that \targetstar is chemically consistent with stars in the NSC, 
but somewhat inconsistent with stars in the NSD or inner bulge, 
mainly due to its high [Si/Mg] ratio. 
This result is promising, 
as HVSs are ejected from the central region of the NSC, where the SMBH resides. 
Though the metallicity of \targetstar is lower than that of typical NSC stars, it remains consistent with the metal-poor tail of NSC stars.

Another interesting point regarding the chemistry of \targetstar 
is that the chemical abundance pattern of \targetstar 
resembles the abundance pattern of the metal-rich stars 
in massive dwarf galaxies, such as the LMC or the Sagittarius dwarf galaxy. 
We do not think that  \targetstar was formed in either of these dwarf galaxies, 
because bringing this star to the central part of the Milky Way 
is unrealistic from a dynamical point of view. 
However, this chemical similarity indicates that 
this star may have formed in an environment similar to the core of 
a massive dwarf galaxy. 
Further observational studies are awaited to better understand 
the birth environment of this star.

\acknowledgments

\revise{The authors thank the anonymous referee for thorough reading 
and constructive comments.}
K.H. thanks Sergey E. Koposov for the inspiring collaboration at CMU, which provided valuable insights. Some of the insights gained during that time form a foundation for this work. 
K.H. is supported by JSPS KAKENHI Grant Numbers JP24K07101, JP21K13965, and JP21H00053.
D.T. acknowledges financial support by the JSPS Research Fellowship for Young Scientists and the accompanying JSPS KAKENHI Grant Number 23KJ2149. 
T.T. acknowledges the support by JSPS KAKENHI Grant No. 23H00132.

%
This paper uses the WINERED data gathered with the 6.5 meter Magellan Telescope located at Las Campanas Observatory, Chile. 
We are grateful to Yuki Sarugaku, Tomomi Takeuchi, and the staff of Las Campanas Observatory for their support during the WINERED's observations. 
WINERED was developed by the University of Tokyo and the Laboratory of Infrared High-resolution Spectroscopy, Kyoto Sangyo University, under the financial support of KAKENHI (Nos.~16684001, 20340042, and 21840052) and the MEXT Supported Program for the Strategic Research Foundation at Private Universities (Nos.~S0801061 and S1411028). 
The observing run in 2023 June was partly supported by KAKENHI (grant No.~19KK0080) and JSPS Bilateral Program Number JPJSBP120239909. 

This work has made use of data from the European Space Agency (ESA) mission
{\it Gaia} (\url{https://www.cosmos.esa.int/gaia}), processed by the {\it Gaia}
Data Processing and Analysis Consortium (DPAC,
\url{https://www.cosmos.esa.int/web/gaia/dpac/consortium}). Funding for the DPAC
has been provided by national institutions, in particular the institutions
participating in the {\it Gaia} Multilateral Agreement.

\facility{Magellan/WINERED, Gaia}

\software{
Agama \citep{Vasiliev2019_AGAMA}, 
ATLAS9-APOGEE \citep{Meszaros2012},
BaSTI \citep{Pietrinferni2004ApJ...612..168P, Hidalgo2018ApJ...856..125H}, 
MOOG \citep{Sneden1973,Sneden2012},
\texttt{mwdust} \citep{Bovy2016ApJ...818..130B}, 
\textsc{Octoman} \citep{Taniguchi2025},
SAGA \citep{Suda2008PASJ...60.1159S}, 
WARP \citep{Hamano2024PASP..136a4504H},
matplotlib \citep{Hunter2007},
numpy \citep{vanderWalt2011},
scipy \citep{Virtanen2020SciPy-NMeth}
}


\bibliographystyle{aasjournal}
\bibliography{mybibtexfile}

\appendix

\section{Spectral analysis using a different line list}

In the main part of this paper, 
we use the MB99 line list \citep{Melendez1999} 
to derive the chemical abundances of \targetstar. 
\autoref{table:stellar_abn_all_res_VALD3} shows 
the same result but using the VALD3 line list \citep{Ryabchikova2015}.

\section{Mass and age of WINERED-HVS1 and other metal-rich HVS candidates}
\label{section:mass_age}

\subsection{WINERED-HVS1}
\revise{
Given the location of \targetstar\ in the CMD, 
it could be either a red clump star or a red giant branch star. 
If it is a red clump star, 
its mass can be $1$--$2 \msun$, 
implying an old age. 
In contrast, 
if it is a red giant branch star, 
its mass can be $2$--$3 \msun$, 
implying a younger age. 
We currently do not 
have enough data to determine whether this star 
is a red clump star or red giant branch star. 
Therefore, we estimate its stellar age using two ways, 
each based on the assumption that the star is in one of these two stages.
}

\revise{
From the APOKASC-3 catalog \citep{Pinsonneault2025ApJS..276...69P}, 
which provides reliable stellar ages from asteroseismology, 
we selected red giant branch stars and red clump stars 
whose astrophysical parameters are similar to those of \targetstar.
In the following analysis, we used the effective temperature $T_\mathrm{eff}^\mathrm{xgboost}$, surface gravity $\log g^\mathrm{xgboost}$, and metallicity $\mathrm{[M/H]}^{\mathrm{xgboost}}$ in \cite{Andrae2023ApJS..267....8A} estimated from Gaia XP spectra. 
First, by using the APOKASC-3 red giant branch sample, 
we estimated the relationship between  
$T_\mathrm{eff}^\mathrm{xgboost}$, $\log g^\mathrm{xgboost}$, and stellar age.  
We then used this relationship 
as well as 
the values $(T_\mathrm{eff}^\mathrm{xgboost}/\;\mathrm{K}, \log g^\mathrm{xgboost})=(4808.1, 2.416)$ of \targetstar\ 
to estimate its age in a Bayesian manner. 
We find that 
the stellar age is 0.5--3 \Gyr\ (95\% range of the posterior distribution) 
if this star is a red giant branch star. 
Secondly, 
we repeated the same analysis 
by using the APOKASC-3 red clump star sample. 
In this case, we find that 
the stellar age is 2--10 \Gyr\ (95\% range) 
if this star is a red clump star. 
Since we are not sure whether this star is a red giant branch 
or red clump star, 
we conclude that its age is likely $0.5$--$10 \Gyr$. 
Finally, by comparing $\mathrm{[M/H]}^{\mathrm{xgboost}}$, the evolutionary stage (i.e., red giant), and the estimated age 
of this star with the predictions from the PARSEC isochrone models version 1.2S~\citep{Bressan2012,Chen2015}, 
we estimate its mass to be 1--3$\msun$. 
}

\revise{
We note that Gaia DR3 estimated its age to be \texttt{age\_flame}$=$0.4--0.7$\Gyr$ 
and its mass to be \texttt{mass\_flame}$=$2.4--2.9$\msun$. 
However, their estimates should be treated with caution. 
As described in \cite{Fouesneau2023A&A...674A..28F}, 
when they estimated the stellar parameters by comparing the data and theoretical isochrone models, 
they only considered the stellar evolution stages from 
the zero-age-main-sequence to the tip of the giant branch, 
and 
they did not consider the possibility of the red clump stage. 
Indeed, their age estimate (0.4--0.7 \Gyr) 
is consistent with 
our estimate of $0.5$--$3 \Gyr$, 
under the assumption that it is a red giant branch star. 
We also note that StarHorse catalog (v1.2) estimated 
its mass to be 0.86--0.95$\msun$, which is consistent with the interpretation that this star is a red clump star.
}

\subsection{Other metal-rich HVS candidates}

\revise{
For the other metal-rich HVSs candidates, 
we roughly estimated their stellar masses 
to be $1$--$3\msun$ 
based on the CMD, 
although we did not analyze these stars as thoroughly 
as \targetstar. 
}

\begin{table}
\centering 
\caption{\revise{Chemical abundances of \targetstar measured with the VALD3 line list~\citep{Ryabchikova2015} }}
\label{table:stellar_abn_all_res_VALD3}
\begin{tabular}{l rrc} \toprule 
$v_{\mathrm{micro}}$ [\si{km.s^{-1}}] & \multicolumn{3}{c}{$1.453^{+0.188}_{-0.186}$} \\
{[Fe/H]} [dex] & \multicolumn{3}{c}{$-0.162^{+0.051}_{-0.047}$ ($46$)} \\
\midrule 
 & \multicolumn{1}{c}{[X/H] [dex]} & \multicolumn{1}{c}{[X/Fe] [dex]} \\ \midrule 
\ion{Na}{i} & --- & --- \\
\ion{Mg}{i} & $-0.146^{+0.127}_{-0.126}$ & $+0.016^{+0.140}_{-0.144}$ & ($2$) \\
\ion{Si}{i} & $+0.007^{+0.085}_{-0.080}$ & $+0.169^{+0.088}_{-0.091}$ & ($19$) \\
\ion{S}{i} & $+0.081^{+0.278}_{-0.262}$ & $+0.243^{+0.274}_{-0.267}$ & ($1$) \\
\ion{K}{i} & --- & --- \\
\ion{Ca}{i} & $-0.042^{+0.156}_{-0.156}$ & $+0.120^{+0.162}_{-0.165}$ & ($7$) \\
\ion{Ti}{i} & $-0.155^{+0.142}_{-0.145}$ & $+0.007^{+0.140}_{-0.153}$ & ($11$) \\
\ion{Cr}{i} & $-0.157^{+0.135}_{-0.134}$ & $+0.005^{+0.138}_{-0.143}$ & ($9$) \\
\ion{Ni}{i} & $-0.059^{+0.137}_{-0.134}$ & $+0.103^{+0.137}_{-0.137}$ & ($2$) \\
\ion{Sr}{ii} & $+0.120^{+0.236}_{-0.229}$ & $+0.282^{+0.236}_{-0.233}$ & ($1$) \\
\bottomrule 
\end{tabular}
\end{table}

\section{Selection of nearly radial orbit stars in APOGEE and Gaia XP samples} \label{appendix:nearly_radial_orbits}

\revise{
In Fig.~\ref{fig:fe_alpha}, 
we show the stars with nearly radial orbits 
in APOGEE sample  
and 
in Gaia XP sample. 
This APOGEE sample is also used to represent 
the GES stars in Fig.~\ref{fig:WINERED_and_inner_mw}. 
Here we describe how these samples are selected. 
}

\revise{
For APOGEE DR17 sample, we first select stars 
that satisfy:
SNR$> 80$, TEFF$< 6000$, LOGG$ < 3.5$, 
FE\texttt{\_}H$>-1$,
MG\texttt{\_}FE$>-0.5$,
SI\texttt{\_}FE$>-0.5$,
CA\texttt{\_}FE$>-0.5$,
TI\texttt{\_}FE$>-0.5$, 
MG\texttt{\_}FE\texttt{\_}ERR$<0.05$, 
STARFLAG is 0, 
ASPCAPFLAG is 0, 
MEMBERFLAG is 0, 
PROGRAMNAME is not `magclouds' (removing Magellanic Clouds regions), 
22nd and 23rd bits of APOGEE2\texttt{\_}TARGET1 is 0 (removing potential member stars of Sagittarius dwarf galaxy or LMC), 
and 
4th bit of EXTRATARG is 0 (removing duplicate observations).  
}

\revise{
For Gaia XP sample, we first select stars 
from the catalog of \cite{Hattori2025ApJ...980...90H} 
that satisfy:
\texttt{bool\_flag\_cmd\_good}=True, 
\texttt{mh\_50\_qrf}$>-1$, 
\texttt{mh\_84\_qrf}$-$\texttt{mh\_16\_qrf}$< 0.5$, 
\texttt{alpham\_84\_qrf}$-$\texttt{alpham\_16\_qrf}$< 0.16$, 
$G<13$, 
and 
RUWE $<1.4$. 
}

\revise{
After these selections, 
for both APOGEE DR17 sample and 
Gaia XP sample, we select stars with 
good parallax (\texttt{parallax\_over\_error}$>5$). 
We compute the stellar position and velocity 
as in \autoref{section:data_selection}. 
We select stars with bound orbit, 
and compute the radial and azimuthal 
orbital action $J_r$ and $J_\phi$. 
Finally, we select stars with radial orbits 
by imposing 
$30 < (J_r / \mathrm{(\kpc\kms)} )^{1/2} < 55$ 
and $|J_\phi| < 500 \kpc \kms$, 
following \cite{Feuillet2021MNRAS.508.1489F}. 
}

\section{Selection of disk stars in APOGEE sample} \label{appendix:APOGEE_disk_stars}

\revise{
The disk stars in Fig.~\ref{fig:WINERED_and_disk_SiMg} 
are selected in the following manner. 
First, from the APOGEE DR17 catalog, 
we select stars with:
SNR$>80$, TEFF$< 6000$, LOGG$ < 3.5$, 
FE\texttt{\_}H$>-1$,
MG\texttt{\_}FE$>-0.5$,
SI\texttt{\_}FE$>-0.5$,
FE\texttt{\_}H\texttt{\_}ERR$<0.1$, 
MG\texttt{\_}FE\texttt{\_}ERR$<0.05$, 
SI\texttt{\_}FE\texttt{\_}ERR$<0.05$, 
STARFLAG is 0, 
ASPCAPFLAG is 0, 
MEMBERFLAG is 0, 
PROGRAMNAME is not `magclouds', 
22nd and 23rd bits of APOGEE2\texttt{\_}TARGET1 is 0, 
and 
4th bit of EXTRATARG is 0. 
We then select stars with good parallax from Gaia DR3 
(\texttt{parallax\_over\_error}$>5$). 
Finally, we compute the Galactocentric position 
of the stars as in \autoref{section:data_selection} 
and select stars at $|z|< 0.5 \kpc$ from the disk plane. 
}

\section{Monte Carlo simulations of hyper-velocity stars}
\label{section:simulation}

\subsection{Empirical model of Hills mechanism}

According to the theoretical investigations by 
\cite{Hills1988Natur.331..687H} and \cite{Bromley2006ApJ...653.1194B}, 
the typical ejection velocity of an HVS due to the 
Hills mechanism---involving the 3-body interaction between a SMBH 
(with mass $M_\mathrm{BH}$) 
and a stellar binary 
(consisting of two stars separated by a semi-major axis $a$ 
with masses of $m_1$ and $m_2$)---is given by: 
\eq{ \label{eq:vej_Hills}
v_\mathrm{ej} = 
1370 \kms 
\left( \frac{a}{0.1 \AU} \right)^{-1/2} 
\left( \frac{m_1 + m_2}{M_\odot} \right)^{1/2} 
\left( \frac{M_\mathrm{BH}}{4 \times 10^6 M_\odot} \right)^{1/6} f_R . 
}
In this equation, $v_\mathrm{ej}$ represents the typical velocity of the escaping star once it has left the gravitational influence of the SMBH (at a distance of $\sim 1$ pc from the SMBH). 
In the main part of this paper, 
we employ two models of the Galactic gravitational potential that do not account for the SMBH's potential. 
Therefore, the ejection velocity $v_\mathrm{ej}$ computed in the main text for stars in the Gaia catalog can be directly compared with the value of 
$v_\mathrm{ej}$ provided in equation (\ref{eq:vej_Hills}).

We note that the empirical factor $f_R < 1$ is of order unity 
and depends on the dimensionless quantity 
\eq{
D = \frac{r_\mathrm{peri}}{a} \left( \frac{10^6 M_\odot}{M_\mathrm{BH}} \frac{m_1+m_2}{2 M_\odot} \right)^{1/3} . 
}
In this equation, $r_\mathrm{peri}$ represents the closest approach distance between the binary system and the SMBH. 
The approximate formula for $f_R (D)$ provided by \cite{Bromley2006ApJ...653.1194B} is 
\eq{
f_R (D) = \sum_{k=0}^{5} c_k D^k \;\;(\mathrm{for} \;0 \leq D \leq 175)
}
with coefficients given by 
$(c_0, c_1, c_2, c_3, c_4, c_5) = 
(0.774, 0.0204, -6.23\times 10^{-4}, 7.62\times 10^{-6}, -4.24\times10^{-8}, 8.62\times10^{-11})$. 
The quantity $D$ also governs the empirical ejection probability, 
which can be expressed as 
\eq{
P_\mathrm{ej} = 
\begin{cases}
1 - \frac{D}{175} \;\; & (\mathrm{for} \;0 \leq D \leq 175), \\
0 \;\; & (\mathrm{otherwise}). 
\end{cases}
}

\subsection{Simulations of HVS ejections}

In the main analysis of this paper, we find that  
\revise{
\targetstar\ and other metal-rich HVS candidates  
have ejection velocities of $500$--$600 \kms$. 
These stars have stellar masses of around $1$--$3 M_\odot$ 
(see \autoref{section:mass_age}). 
}
To understand the physical parameters that 
lead to the ejection of a \revise{$1$--$3 M_\odot$} star with an ejection velocity ranging from 500 to 600 \kms, 
we perform Monte Carlo simulations consisting of 
$n_\mathrm{MC}=5\times 10^5$ realizations. 
In the following analysis, we assume that 
$M_\mathrm{BH}=4\times10^6 M_\odot$. 
\revise{
In Appendix~\ref{section:simulation_3Msun}, 
we assume $m_1 = m_2 = 3 M_\odot$. 
In Appendix~\ref{section:simulation_2Msun}, 
we also perform similar simulations assuming 
$m_1 = m_2 = 2 M_\odot$ or 
$m_1 = m_2 = 1 M_\odot$. 
}

\subsubsection{A Monte Carlo simulation of 3-$M_\odot$ HVSs}
\label{section:simulation_3Msun}

Following \cite{Bromley2006ApJ...653.1194B}, 
we assume that the binary's closest approach to the SMBH satisfies the condition $1 \;\mathrm{AU} < r_\mathrm{peri} < 700 \;\mathrm{AU}$. 
To simplify our calculations, we assume that $r_\mathrm{peri}$ 
is equal to the impact parameter of the center-of-mass orbit of the binary at infinity. 
In this context, the probability of having 
the impact parameter between 
$r_\mathrm{peri}$ 
and $r_\mathrm{peri} + \mathrm{d}r_\mathrm{peri}$ 
is proportional to $2 \pi r_\mathrm{peri} \mathrm{d}r_\mathrm{peri}$. 
We randomly sample $n_\mathrm{MC}$ values of $r_\mathrm{peri}$ 
from the range $1 \;\mathrm{AU} < r_\mathrm{peri} < 700 \;\mathrm{AU}$  according to this probability distribution.

Additionally, following \cite{Bromley2006ApJ...653.1194B}, 
we assume that the logarithmic value of the binary's semi-major axis, 
$\ln a$, follows a uniform distribution. 
The lower limit for $a$ is set at \revise{$a_\mathrm{min} = 0.028$ AU,} 
which is the minimum allowed value of $a$ 
so that two \revise{$3$-$\msun$} main-sequence stars in a binary 
do not merge.\footnote{
The Roche radius of an equal-mass binary is given by $R_\mathrm{Roche}=0.38 a$. 
By equating the Roche radius and the radius of a \revise{$3$-$\msun$} main-sequence star \revise{($\sim 0.011$ AU),} 
we obtain the minimum allowed value of $a$. 
} 
\revise{
(Due to the assumption in $a_\mathrm{min}$, 
our simulation technically addresses 
the ejection of 3-$\msun$ main-sequence HVSs. 
However, this simulation can be also used to gain insights into 
the ejection of 3-$\msun$ red giant HVSs, as we will demonstrate later.) 
}
The upper limit for $a$ is set at 100 AU, 
which is sufficiently large that our analysis is not sensitive to this value. 
We also randomly sample $n_\mathrm{MC}$ values of $a$ according to this probability distribution.

For each pair $(a, r_\mathrm{{peri}})$, 
we calculate three quantities: $D$, the ejection velocity $v_\mathrm{ej}$, and the ejection probability $P_\mathrm{ej}$. 
If $P_\mathrm{ej}$ is non-zero, we store the corresponding $v_\mathrm{ej}$ 
\revise{with a probability of $P_\mathrm{ej}$}. 
Out of $n_\mathrm{MC} = 5 \times 10^5$ realizations, 
we find that 2457 realizations result in an ejection with 
$v_\mathrm{ej} = 500$--$600 \kms$. 
The fraction of ejections within this range of $v_\mathrm{ej}$ is small, 
mainly because the upper limit for $a$ is set very high. 
Under our current assumptions, no HVSs would be ejected with $v_\mathrm{ej} > 500 \kms$ if \revise{$a>5$ AU} is satisfied.

Fig.~\ref{fig:MonteCarlo}(a)--(c) illustrates the results of this analysis. 
For an HVS with a mass of \revise{$3 M_\odot$,} 
achieving an ejection velocity of 500--600 \kms\ typically requires 
\revise{$a \sim 3.1$ AU} (median) 
and 
\revise{$r_\mathrm{peri} \simeq 230$ AU} (median). 
The values of $a$ and $r_\mathrm{peri}$ can vary by a factor of around 2. 
\revise{
Our result indicates that  
a binary system 
consisting of 
two 3-$M_\odot$ main-sequence stars 
can interact with the Galactic SMBH to produce a main-sequence HVS with an ejection velocity of 500--600 \kms. 
}

\revise{
While this simulation focuses on the ejection of main-sequence HVSs, 
our simulation can also provide insights into the ejection of red giant HVSs of similar masses.
}
\revise{For $a>0.526$ AU,}
the Roche radius of an equal-mass binary satisfies 
\revise{
$R_\mathrm{Roche}=0.38a>0.2$ AU.}
Given that \revise{a 3-$\msun$ red giant star typically has a radius smaller than $0.2$ AU,} 
a condition of 
\revise{$a>0.526$ AU} 
is a sufficient condition 
for a binary system consisting of two \revise{3-$\msun$} red giants does not merge. 
In this regard, 
it is noteworthy that 
\revise{all} 
the Monte Carlo realizations 
that yielded $v_\mathrm{ej} = 500$--$600 \kms$ 
satisfy the condition 
\revise{$a>0.526$ AU}. 
Our result indicates that  
a binary system 
consisting of 
two 3-$M_\odot$ red giant stars 
can interact with the Galactic SMBH to produce a red-giant HVS with an ejection velocity of 500--600 \kms.

\subsubsection{Monte Carlo simulations of 2-$M_\odot$ HVSs and 1-$M_\odot$ HVSs}
\label{section:simulation_2Msun}

\revise{
We conduct the same Monte Carlo simulation 
but assuming $m_1 = m_2 = 2 M_\odot$ 
and $a_\mathrm{min}=0.026$ AU; 
or $m_1 = m_2 = 1 M_\odot$ 
and $a_\mathrm{min}=0.012$ AU. 
Here, $a_\mathrm{min}$ is set by the same requirement as in Appendix~\ref{section:simulation_3Msun}.  
Figs.~\ref{fig:MonteCarlo}(d)--(f) and (g)--(i) illustrate the results of these analyses. 
By using the fact that 2-$\msun$ (1-$\msun$) red giant stars 
typically have a radius smaller than $0.2$ AU ($0.1$ AU), 
and by repeating the same argument as in Appendix~\ref{section:simulation_3Msun}, 
we conclude that 
a binary system consisting of 
two 2-$\msun$ stars (or two 1-$\msun$ stars)---irrespective of the main-sequence/red-giant phase---can interact with the Galactic SMBH to produce an HVS with an ejection velocity of 500--600 \kms. 
}

\subsection{Ejection conditions of WINERED-HVS1 inferred from our simulations}
\label{section:simulation_summary}

\revise{
Based on the simulations described above, 
we obtain the following insights into the 
ejection environment of \targetstar, 
assuming that it belonged to an equal-mass binary 
just before it was ejected from the SMBH. 
\begin{itemize}
\item 
Judging from the ejection velocity 
of \targetstar\ ($500$--$600 \kms$), 
we see that the binary separation $a$ 
is $\sim 0.5$--$4$ AU 
and the binary's closest approach of the SMBH 
is $r_\mathrm{peri} \sim 30$--$300$ AU. 
\item 
When \targetstar\ was ejected from the Galactic center, 
it could be  
a main-sequence star or a red giant. 
\item 
The timescales for 
1-, 2-, and 3-$\msun$ stars 
to evolve from the main-sequence stage to the red giant stage 
(with a similar temperature to \targetstar) 
are roughly 
20 Myr, 30 Myr, and 500 Myr, respectively. 
These timescales are longer than the time since the last disk crossing (14 Myr). 
Thus, if it was ejected from the Galactic center as a main-sequence star, it must have experienced one or more disk crossings after the ejection. 
\item 
If \targetstar\ was recently ejected from the Galactic center  
and has never experienced disk crossings after the ejection, 
it must have been ejected as a red giant. 
\end{itemize}
}

\begin{figure*}
\centering
\includegraphics[width=0.95\textwidth]{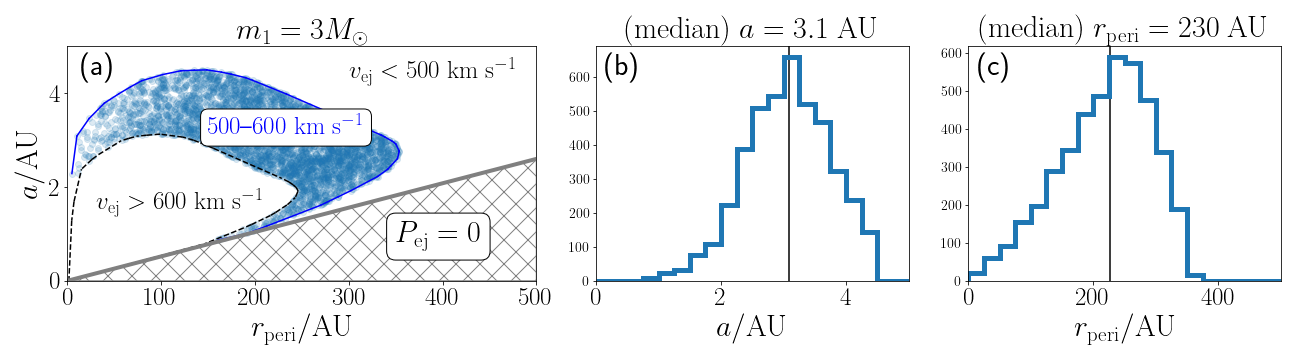}
\includegraphics[width=0.95\textwidth]{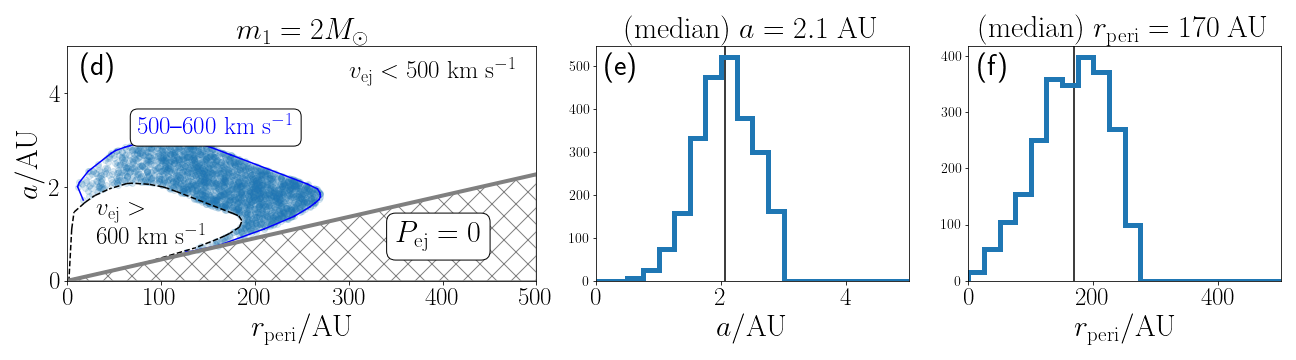}
\includegraphics[width=0.95\textwidth]{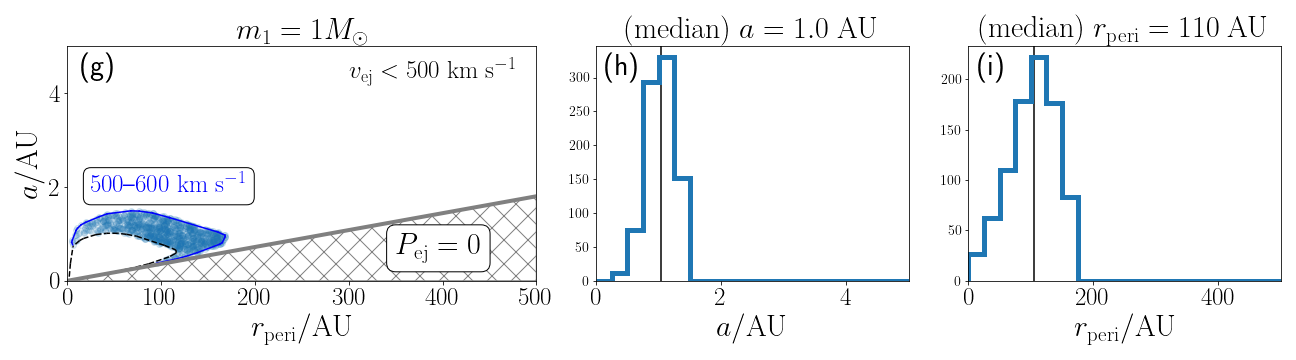}
\caption{
Results of Monte Carlo simulation of HVSs. 
\revise{On the top row (a--c), middle row (d--f), and bottom row (g--i), 
we show the results for simulated HVSs with masses of $3\msun$, $2 \msun$, and $1 \msun$, respectively.}
\revise{Panels (a), (d), and (e):} 
The blue dots shows the distribution of 
$(\rperi, a)$ 
that allows the ejection of an HVS to attain 
ejection velocity of 500--600 \kms. 
The hatched region corresponds to 
a parameter subspace where no ejection happens 
($P_\mathrm{ej}=0$; $D>175$). 
\revise{Panels (b), (e), and (h): 
The marginalized distribution of $a$ 
for HVSs with $v_\mathrm{ej} = 500$--$600 \kms$. 
The vertical line shows the median of $a$.
Panels (c), (f), and (i): 
The marginalized distribution of $r_\mathrm{peri}$ 
for HVSs with $v_\mathrm{ej} = 500$--$600 \kms$. 
The vertical line shows the median of $r_\mathrm{peri}$.}
}
\label{fig:MonteCarlo}
\end{figure*}

\section{Properties of our HVS candidate and the theoretical predictions}
\label{section:statistics}

In this paper, we identify 22 HVS candidates, 
and one of the candidates, \targetstar, 
is a promising candidate in terms of its chemistry. 
Here we investigate whether our discovery of one HVS (candidate) 
is statistically natural.

\subsection{Basic properties of \targetstar}
\label{section:basics_of_targetstar_in_Gaia}

First, we ignore the fact that this star is an HVS candidate and investigate the other observational properties (such as its magnitude) of \targetstar from a statistical perspective. 
Based on the Gaia DR3 catalog, its $G_\mathrm{RP}$, $G$, and $G_\mathrm{RVS}$ magnitudes are $G_\mathrm{RP}=13.58$, $G=14.28$, and $G_\mathrm{RVS}=13.26$. 
Among 1.7 billion stars in the Gaia DR3, only 1\% of them have $G_\mathrm{RP}$ or $G$ values that are brighter than \targetstar.
Also, among 219 million stars with Gaia XP spectra, only 9\% of them have $G_\mathrm{RP}$ or $G$ values that are brighter than \targetstar.
These numbers indicate that \targetstar is a bright object. 
However, except for its brightness, \targetstar 
is not particularly special. 
By randomly selecting stars with similar $G_\mathrm{RP}$ magnitudes 
($13.56<G_\mathrm{RP}<13.60$), 
we find 91\% of them have Gaia XP spectra, 
88\% have line-of-sight velocity ($\vlos$) measurements by Gaia, 
87\% have good parallax measurements (\texttt{parallax\_over\_error}$>5$), 
and 
74\% satisfy all the criteria above (as \targetstar does). 
The fractions are almost identical if we select stars based on $G$-magnitude ($14.26<G<14.30$) instead.

\subsection{Complications in the mock HVS catalog in Marchetti et al. (2018)}

\cite{Marchetti2018MNRAS.476.4697M} predicted the 
populations of HVSs observable in the Gaia catalog. 
Their HILLS mock catalog is a simulated catalog of both bound and unbound 
HVSs ejected from the Galactic center, 
assuming an HVS ejection rate of $2.8 \times 10^{-4} \yr^{-1}$. 
Since \targetstar is identified as a bound HVS candidate, their HILLS catalog is appropriate for comparison with our results. 
However, before using their results, we must clarify three key differences between their identification of HVSs in the simulation and our search for HVS candidates.

(Complication 1): 

We selected HVS candidates from a catalog of 48 million stars in  \cite{Hattori2025ApJ...980...90H}. 
In creating this catalog, \cite{Hattori2025ApJ...980...90H} did not use the full list of stars with Gaia XP spectra, which consists of 219 million stars. Instead, he focused on the 48 million stars located in regions with low dust extinction with $E(B-V)<0.1$.
Consequently, our search for HVSs is inevitably restricted to those HVSs in regions with low dust extinction. 
In contrast, \cite{Marchetti2018MNRAS.476.4697M} used a 3D dust extinction map to generate dust-obscured photometric data for their simulated HVSs. 
Their synthetic data included HVSs from both high and low dust extinction regions. 
If we crudely assume that the spatial distribution of stars with Gaia XP spectra is similar to that of the HVSs,\footnote{
This approximation may not be as inaccurate as it seems. Most of the stars with Gaia XP spectra are disk stars and have a flattened distribution. The spatial distribution of HVSs in the left panel of Fig.~10 from \cite{Marchetti2018MNRAS.476.4697M} also appears slightly oblate 
(even after accounting for the figure's aspect ratio). 
This flattened distribution may be due to bound HVSs being drawn toward the Galactic mid-plane because of the gravitational pull from the stellar disk.
}
we can estimate that we are searching for HVSs from approximately 22\% of the available data (48 million/219 million).

(Complication 2): 

Since \cite{Marchetti2018MNRAS.476.4697M} only estimated the expected number of main-sequence HVSs, we need to be careful when using their mock catalog to assess the likelihood of discovering a red-giant HVS, such as \targetstar. 
The key differences between main-sequence HVSs and red-giant HVSs (with the same initial masses) are that the main-sequence phase lasts longer, and a red giant is intrinsically brighter.

According to the Padova isochrone model \citep{Bressan2012}, a $\sim 2 \msun$ main-sequence star becomes approximately 1 mag brighter when it evolves into a red giant 
\revise{
(except at the very final stage of the red giant phase when it becomes $\sim 4$ mag brighter).}
Additionally, the duration of the red-giant phase is only 21\% of the main-sequence phase for a star with an initial mass of $2 \msun$. 
An increase of brightness by 1 mag allows a magnitude-limited survey to cover 2.5 times larger distances and $16 \; (= 2.5^3)$ times larger volumes. Fig.~8 of \cite{Marchetti2018MNRAS.476.4697M} predicted $\sim 95$ main-sequence HVSs with $G_\mathrm{RVS}<13.26$ (as bright as or brighter than \targetstar) and  $\sim 380$ main-sequence HVSs with $G_\mathrm{RVS}<14.26$ (which extends the upper limit by 1 mag). 
Thus, in their analysis, increasing the $G_\mathrm{RVS}$ by 1 mag results in a $4 \, (= 380/95)$ times larger number of main-sequence HVSs. Based on these arguments, we expect that an increase of 1 mag in HVS brightness may lead to an increase in its detectability by a factor of $4$ to $16$. 
Considering the short duration of the red-giant phase, we estimate that the detectability of a 2-$\msun$ red-giant HVS may be around $f_\mathrm{RG/MS} = 0.8$--$3.4\; (= 0.21 \times (4$--$16))$ times that of a main-sequence HVS with the same mass. 
\revise{
Similarly, 
the detectability of a 3-$\msun$ red-giant HVS is 
$f_\mathrm{RG/MS} = 0.3\; (= 0.15 \times (2.0$--$2.1))$ times that of a main-sequence HVS with the same mass; 
and the correction factor is 
$f_\mathrm{RG/MS} = 0.2$--$0.8\; (= 0.05 \times (4$--$16))$ for the case of 1-$\msun$ red-giant HVS. 
By combining these results, we infer that the detectability of a red-giant HVS with a mass of 1--3$\msun$ may be around $f_\mathrm{RG/MS} = 0.2$--$3.4$ times that of a main-sequence HVS with the same mass. 
}

(Complication 3): 

\revise{
The simulated HVSs described in 
\cite{Marchetti2018MNRAS.476.4697M} 
includes all the HVSs irrespective of the flight time, 
as long as the observational constraints 
(e.g., apparent magnitude) 
are satisfied. 
In contrast, 
in this paper, we aimed to identify HVSs 
by using the stellar orbits in the last 200 Myr. 
Our search is optimized for 
recently-ejected HVSs, 
and we may miss some fraction of 
HVSs ejected long time ago 
if their last disk crossings occurred at 
$R \gtrsim 200 \pc$. 
}

\revise{
In a pessimistic scenario, 
our search method may identify no HVSs 
that were ejected long time ago, 
and it may only identify recently-ejected HVSs. 
Based on our analysis in \autoref{table:kinematics}, the typical flight time for recently ejected HVSs is approximately 20 Myr. 
Therefore, in this pessimistic scenario, 
we might have excluded 
HVSs with flight times longer than $\sim 20$ Myr. 
Suppose that \targetstar\ is indeed a recently-ejected HVS (with no disk crossings after the ejection). Then, 
}
given the following points: (i) HVS ejections can occur at any time during a star's lifetime, (ii) the estimated lifetime of \targetstar is $\sim 0.5$--$10 \Gyr$, and (iii) the typical flight time of recently ejected HVSs is $\sim 20 \Myr$, it can be inferred that identifying one recently ejected HVS candidate (\targetstar) may suggest the existence of $\sim 25$--$500 \; (=0.5$--$10 \Gyr / 20 \Myr)$ bound HVSs with similar characteristics to \targetstar in our input catalog. 
\revise{
In reality, we cannot determine whether \targetstar\ was truly ejected recently; it may have been ejected in the more distant past. 
Thus, the correction factor ($\sim 25$--$500$) is an upper limit.  
}

\subsection{Assessment of the likelihood to find one red-giant HVS}

Despite the complications mentioned above, 
we dare to explore how our discovery of one HVS (candidate) 
can be interpreted in the context of the theoretical predictions 
made by \cite{Marchetti2018MNRAS.476.4697M}. 
Since \cite{Marchetti2018MNRAS.476.4697M} only considered 
main-sequence HVSs, we first pretend 
as if we found a `main-sequence' HVS with $G_\mathrm{RVS}=13.26$.

As predicted in Fig.~8 of \cite{Marchetti2018MNRAS.476.4697M}, 
we assume that there are $\sim 95$ main-sequence HVSs with $G_\mathrm{RVS}<13.26$ (i.e., at least as bright as \targetstar). 
Also, we assume that 74\% of Gaia stars with this same magnitude range possess Gaia XP spectra, $\vlos$, and reliable parallax measurements 
(see \autoref{section:basics_of_targetstar_in_Gaia}), 
and that 22\% of HVSs are found in low dust extinction regions 
(see (Complication 1)). 
Then we expect to identify $\sim 15 \; (=95 \times 0.74 \times 0.22$) main-sequence HVSs that meet the following criteria: 
(i) $G_\mathrm{RVS}<13.26$, 
(ii) located in low dust extinction regions ($E(B-V)<0.1$), and 
(iii) possessing the Gaia XP spectra, velocity measurements, and reliable parallax.

As noted in (Complication 2), 
what we found is 
\revise{
a red-giant HVS candidate with a mass of 1--3$\msun$}
(\targetstar), 
which is more likely to be detected 
by a factor of \revise{$0.2$}--$3.4$ than a main-sequence HVS with the same mass. 
\revise{
After correcting for this factor, 
the expected number of red-giant HVSs is 3--51 ($=15 \times $(0.2--3.4)). 
Although this number is associated with a large uncertainty, 
finding one HVS candidate from our input catalog is less than this theoretical prediction.
In this regard, it is worthwhile mentioning that 
this prediction includes all the observable HVSs irrespective of the flight time. 
For example, as discussed in (Complication 3), 
recently ejected HVSs might represent $\sim 1/500$--$1/25$ of the entire HVS population 
with similar properties. 
In this regard, the theoretical prediction for the recently ejected HVSs 
is $0.006$--$2$.
(Here, we adopt $3/500=0.006$ as the lower limit and $51/25\simeq 2$ as the upper limit.)
If we assume that \targetstar\ is truly a recently ejected HVS, 
this theoretical prediction may be consistent with our discovery 
of one HVS candidate. 
However, we need to stress that, 
even if \targetstar\ is a genuine HVS ejected from the Galactic center, 
we currently do not have a way to rigorously constrain the flight time of this star.}

Also, 
Fig.~11 (top panel) and Fig.~7 
\revise{
of \cite{Marchetti2018MNRAS.476.4697M}
}
indicate that the majority of HVSs with $G_\mathrm{RVS}<16$ have total velocities of $v<500 \kms$ and are located at the Galactocentric distances of $r<10 \kpc$. 
These findings are consistent with \targetstar, which has a velocity of $v=273 \kms$ and is located at $r=5.45 \kpc$. 
These characteristics suggest that the HVS candidate we identified may represent a typical bound HVS in the Gaia data.

\section{Orbital properties of metal-rich HVS candidates}
\label{section:orbits_for_other_HVSs}

\begin{figure*}
\centering
\includegraphics[width=0.9\textwidth]{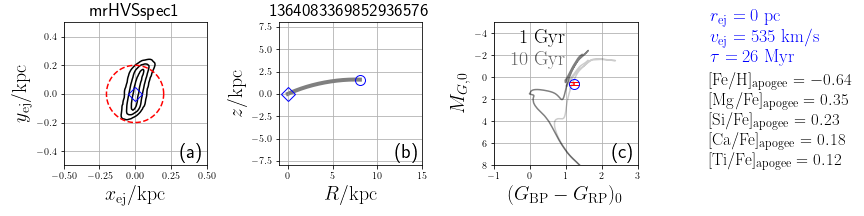}
\includegraphics[width=0.9\textwidth]{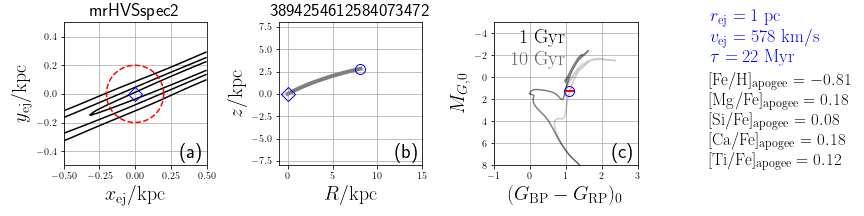}
\includegraphics[width=0.9\textwidth]{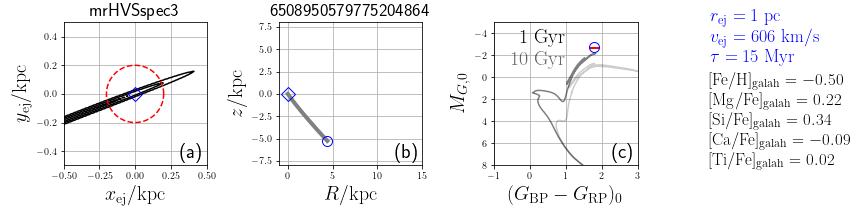}
\includegraphics[width=0.9\textwidth]{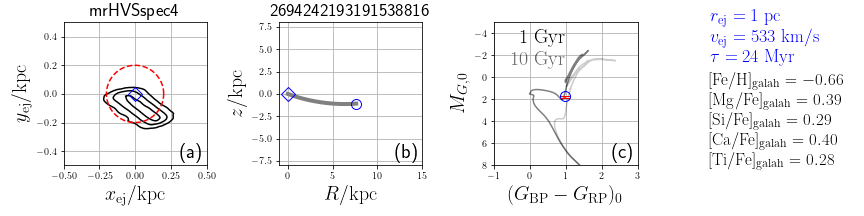}
\includegraphics[width=0.9\textwidth]{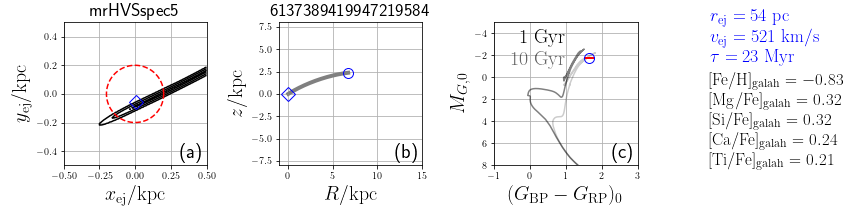}
\caption{
The same as Fig.~\ref{fig:WINEREDHVS1} but for mrHVSspec1-5.
}
\label{fig:mrHVSspec1to5}
\end{figure*}

\begin{figure*}
\centering
\includegraphics[width=0.9\textwidth]{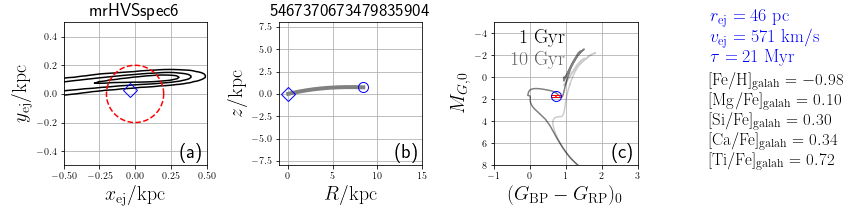}
\includegraphics[width=0.9\textwidth]{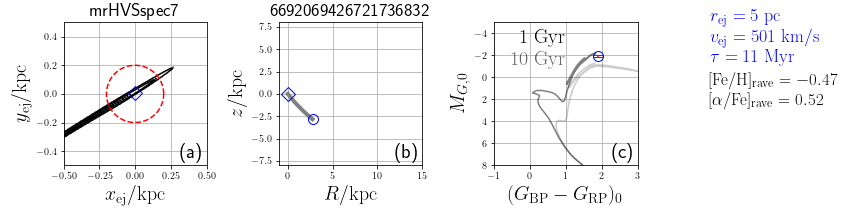}
\includegraphics[width=0.9\textwidth]{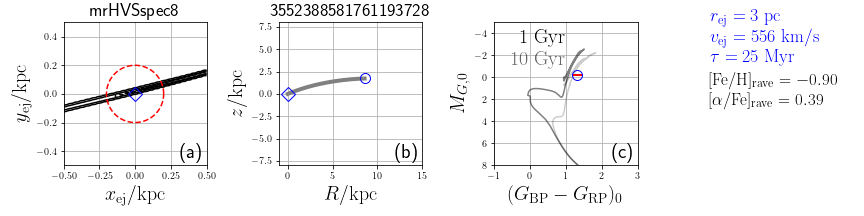}
\includegraphics[width=0.9\textwidth]{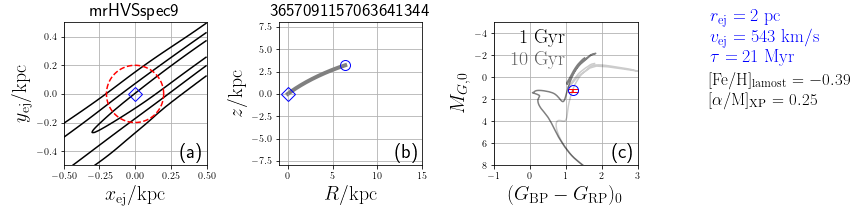}
\caption{
The same as Fig.~\ref{fig:WINEREDHVS1} but for mrHVSspec6-9.
}
\label{fig:mrHVSspec6to9}
\end{figure*}

\begin{figure*}
\centering
\includegraphics[width=0.9\textwidth]{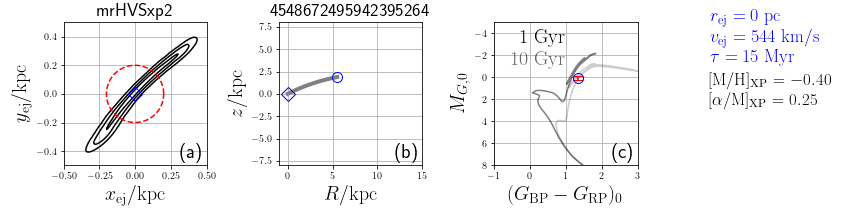}
\includegraphics[width=0.9\textwidth]{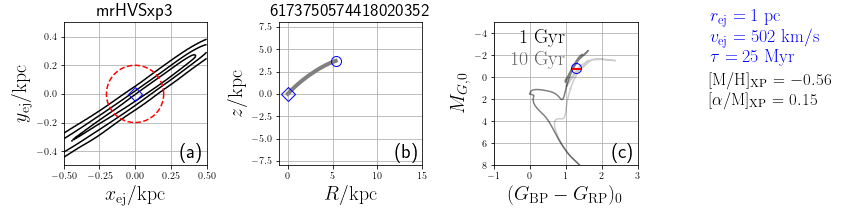}
\includegraphics[width=0.9\textwidth]{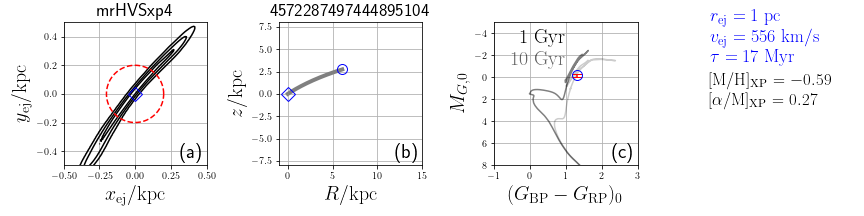}
\includegraphics[width=0.9\textwidth]{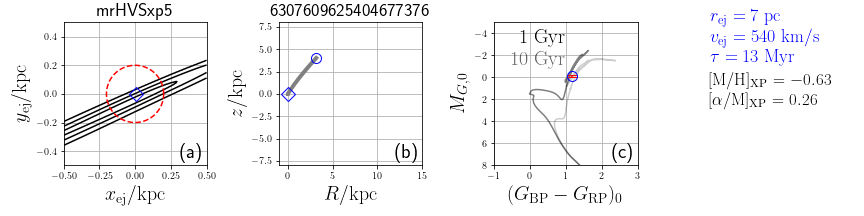}
\includegraphics[width=0.9\textwidth]{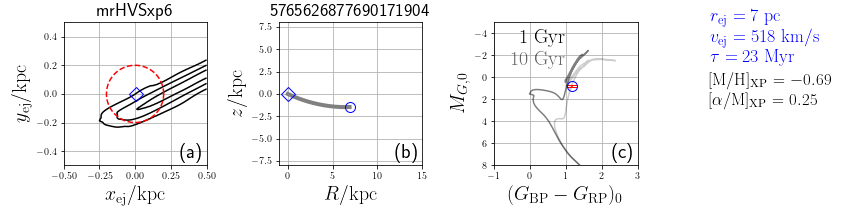}
\caption{
The same as Fig.~\ref{fig:WINEREDHVS1} but for mrHVSxp2-6.
}
\label{fig:mrHVSxp2to6}
\end{figure*}

\begin{figure*}
\centering
\includegraphics[width=0.9\textwidth]{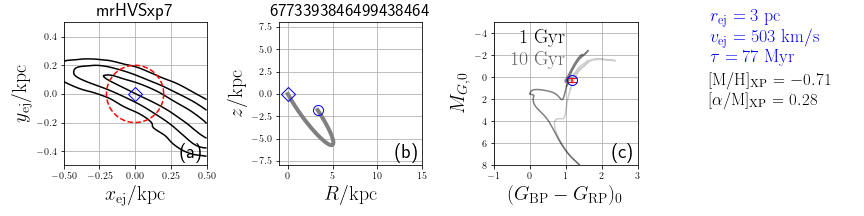}
\includegraphics[width=0.9\textwidth]{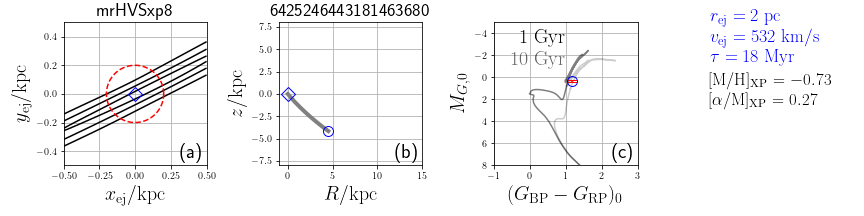}
\includegraphics[width=0.9\textwidth]{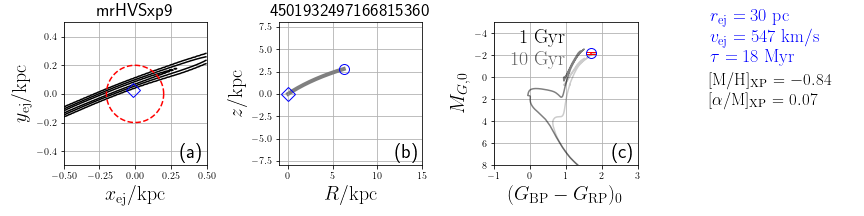}
\includegraphics[width=0.9\textwidth]{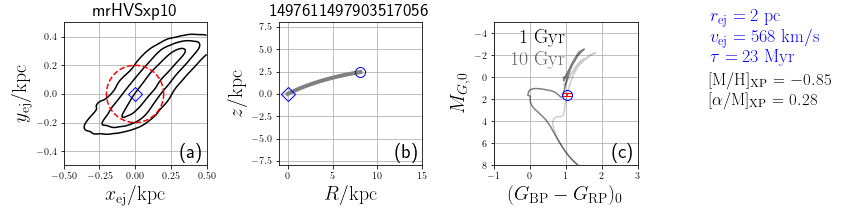}
\includegraphics[width=0.9\textwidth]{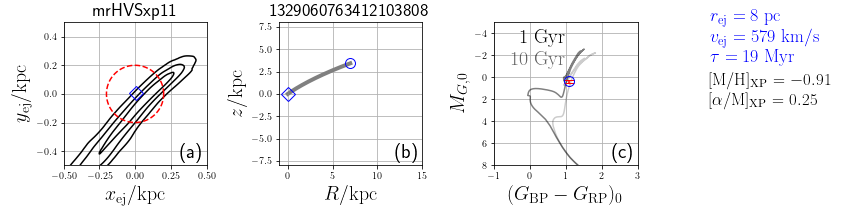}
\caption{
The same as Fig.~\ref{fig:WINEREDHVS1} but for mrHVSxp7-11.
}
\label{fig:mrHVSxp7to11}
\end{figure*}

\begin{figure*}
\centering
\includegraphics[width=0.9\textwidth]{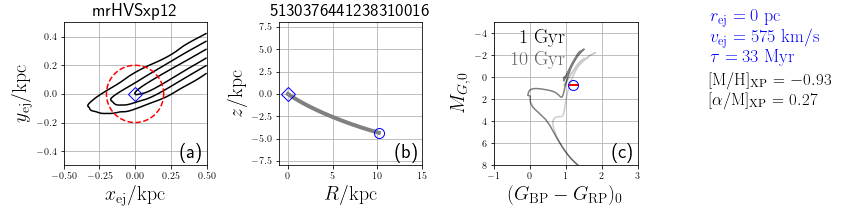}
\includegraphics[width=0.9\textwidth]{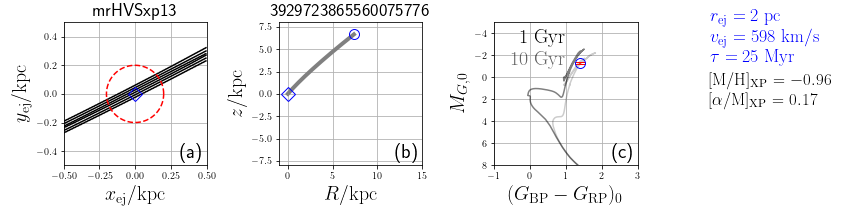}
\caption{
The same as Fig.~\ref{fig:WINEREDHVS1} but for mrHVSxp12-13.
}
\label{fig:mrHVSxp12to13}
\end{figure*}

In Fig.~\ref{fig:WINEREDHVS1}, we show the properties of \targetstar. 
Here we show the orbital properties for 
mrHVSsp1-9 
in Figs.~\ref{fig:mrHVSspec1to5} and \ref{fig:mrHVSspec6to9}. 
We show the plots for 
mrHVSxp2-13 
in Figs.~\ref{fig:mrHVSxp2to6}, \ref{fig:mrHVSxp7to11}, and \ref{fig:mrHVSxp12to13}.

\section{Notations in this paper} \label{appendix:notation}

\autoref{table:notation} 
summarizes the notations for astrometric quantities 
and chemical abundances 
used in this paper.

\begin{deluxetable}{l l }
\tablecaption{Notations in this paper 
\label{table:notation}}
\tablewidth{0pt}
\tabletypesize{\scriptsize}
\tablehead{
\colhead{Quantity} &
\colhead{Meaning} 
}
\startdata
$\alpha$ & Right ascension  \\ 
$\delta$ & Declination \\ 
$d$ & Heliocentric distance \\ 
$\varpi$ & Parallax \\ 
$\sigma_\varpi$ & Parallax error \\ 
$\pmra$ & Proper motion in $\alpha$  \\ 
$\pmdec$ &  Proper motion in $\delta$ \\ 
$\vlos$ & Heliocentric line-of-sight velocity \\ 
\hline
{$\mhxp$} & {Metallicity [M/H] determined from Gaia XP spectra \citep{Hattori2025ApJ...980...90H}} \\
{$\amxp$} & {$\alpha$-abundance [$\alpha$/M] determined from Gaia XP spectra \citep{Hattori2025ApJ...980...90H}} \\
{$\mathrm{[X/Y]_{spec}}$} & {Spectroscopically determined chemical abundance [X/Y]} \\
{$\mathrm{[\alpha/Fe]_{spec}}$} & {Spectroscopically determined $\alpha$-abundance.$^\mathrm{(a)}$} \\
\hline 
\enddata
\tablecomments{
(a) If $\mathrm{[\alpha/Fe]_{spec}}$ is in the literature, 
we use the literature value. 
If not, we compute $\mathrm{[\alpha/Fe]_{spec}}$ 
from available abundance ratios of $\alpha$ elements. 
If four abundance ratios [Mg/Fe], [Si/Fe], [Ca/Fe], and [Ti/Fe]
are available, we compute the average and standard deviation of these four values and use them as $\mathrm{[\alpha/Fe]_{spec}}$ and the associated uncertainty. 
(In computing these average, we ignore the uncertainties in each abundance ratios.) 
If only some of the four ratios are available, 
we take the average and standard deviation of the available ratios. 
}
\end{deluxetable}

\end{document}